\documentclass{elsart}
\usepackage{amsfonts}
\usepackage{natbib}
\usepackage{amssymb}
\usepackage{graphics}
\usepackage{graphicx}
\usepackage{epsfig}

\begin{document}


\begin{frontmatter}

\title{On the derivation of structural models with general thermomechanical prestress}

\author[a]{Silvano Erlicher\corauthref{cor1}}
\ead{erlicher@lami.enpc.fr}
\author[b]{, Fr\'{e}d\'{e}ric Bourquin}

\corauth[cor1]{Corresponding author. Tel: +33 1 64 15 37 80, Fax:
+33 1 64 15 37 41}

\corauth{Acknowledgements: this work has been partially supported by
the European RTN SMART SYSTEMS, while the first author was visiting
the LCPC}

\address[a]{UMR Navier (LAMI), \'{E}cole Nationale des Ponts et Chauss\'{e}es (ENPC),
6 et 8 av. B. Pascal, Cit\'{e} Descartes, Champs-sur-Marne, 77455
Marne-la-Vall\'{e}e, Cedex2, France}

\address[b]{Laboratoire Central des Ponts et Chauss\'{e}es (Public Works Research Laboratory)\\
58 boulevard Lefebvre, 75732, Paris, CEDEX 15, France}

\begin{abstract}
The vibrating behaviour of thin structures is affected by prestress
states. Hence, the effects of thermal prestress are important
research subjects in view of ambient vibration monitoring of civil
structures. The interaction between prestress, geometrically
non-linear behaviour,  as well as damping and its coupling with the
aforementioned phenomena has to be taken into account for a
comprehensive understanding of the structural behaviour. Since the
literature on this subject lacks a clear procedure to derive models
of thin prestressed and damped structures from 3D continuum
mechanics, this paper presents a new derivation of models for thin
structures accounting for generic prestress, moderate rotations and
viscous damping. Although inspired by  classical approaches, the
proposed procedure is quite different, because of (i) the definition
of a modified Hu-Washizu (H-W) functional, accounting for stress
constraints associated with Lagrange multipliers, in order to derive
lower-dimensional models in a convenient way; (ii) an original
definition of a (mechanical and thermal) strain measure and a
rotation measure enabling one  to identify the main terms in the
strain energy and to derive a cascade of lower-dimensional models
(iii) a new definition of "strain-rotation domains" providing a
clear interpretation of the classical assumptions of "small
perturbations" and "small strains and moderate rotations"; (iv) the
introduction of a pseudo-potential with stress constraints to
account for viscous damping. The proposed  procedure is applied to
thin beams.

\end{abstract}


\end{frontmatter}

\section{Introduction}

Ambient vibration monitoring (\emph{e.g.} \citet{wenzel-pichler-avm}, %
\citet{bassevillejsv04} ) has now become widely accepted as an important
tool for Structural Health Monitoring (SHM). But structural vibrations are
affected by prestress states. In particular, thermal variations may cause
very significant changes in a bridge spectrum %
\citep{Peeters-De-Roeck-Z24-01,farrar-I40-94}. Since temperature effects may
be orders of magnitude larger than the effect of a damage, overlooking them
prevents from a reliable damage detection based on vibration monitoring.
Most attempts to eliminate thermal effects have favoured blind approaches
not taking advantage of predictive models \citep{Farrarsohnworden03}.
However, successful endeavours to eliminate the temperature from
subspace-based damage detection algorithms prove the relevance of relying on
predictive thermomechanical models yielding the prestress state due to
temperature \citep{nasser-phd,basseville-ewshm06-adjust}. This suggests to
deeper understand the way temperature interacts with structural dynamics and
to revisit associated models. This paper steps forward in this direction.

On the other hand, identification of the geometrically non-linear behaviour
of beams or plates has recently given rise to numerous papers (\emph{e.g. } %
\citep{Kerschen2003}, \citep{ArgoulLe2003}, \citep{PerignonBellizzi}), often
using the notion of non-linear modes \citep{Rosenberg62} \citep{Vakakis97}.
Non-linear dynamics of prestressed beams or plates undergoing small strains
and moderate rotations is simulated in \citep{Ribeiro2001}, %
\citep{PerignonBellizzi}, \citep{Amabili2005}. However, the dynamics used in
these contributions seems to be based on some "historical" assumptions,
whose justification is often skipped.

The purpose of this paper is then to provide a new viewpoint on this
classical subject, where a very large number of contributions, sometimes not
clearly related, have been superposed over the years. We make clear the
series of simplifying assumptions leading from the 3D continuous
thermo-elasticity theory to the equations governing the dynamics of thin
structures with prestress and thermal field, under the assumption of "small
strains and moderate rotations". Attention is paid to the definition of the
range of validity of these classical equations, introducing the notion of
strain-rotation domains. More precisely: $\mathbf{(i)}$ a Hu-Washizu
functional including suitable stress constraints and the associated Lagrange
multipliers is defined: the constitutive law, the strain-displacement
relationship as well as the dynamic equilibrium equations are then derived
by imposing the stationarity conditions. Stress constraints characterizing
thin body theories are introduced in the 3D models in view of a convenient
derivation of 1D beam models.
$\mathbf{\ (ii)}$ The most important configurations characterizing the
structural behaviour are clearly identified. $\mathbf{(iii)}$ The assumption
that the \emph{\ prestressed} configuration has a known geometry, adopted as
\emph{Lagrangian reference} configuration (see \emph{e.g.} %
\citep{GeradinRixen}), is removed. When the pre-stress field and the
original geometry of the structure are very simple, this hypothesis is
convenient. However, for more general conditions, \emph{e.g.} when the
prestress state is time-dependent or related to a general thermal field,
this assumption appears to be over-simplified. Hence, a different analysis,
where prestressed and Lagrangian reference configurations are \emph{distinct}
, seems to be more suitable. The equilibrium equation governing the dynamics
of a structure subjected to a generic prestress is defined as the \emph{%
difference} between the Lagrangian equations at the dynamic configuration
and at the statically prestressed configuration. $\mathbf{(iv)}$ A rigorous
formalization of the assumption of "small strain and moderate rotations" is
provided. To this end, measures of the strain amplitude (symmetric part of
the displacement gradient) and of the rotations (skew-symmetric part) are
introduced in the 3D continuous framework, whereas standard approaches use
the aspect ratio as the governing parameter (\emph{e.g.} \citep{PGCvonkarman}%
). The new solution-dependent measures enables one to define strain-rotation
domains in which the leading terms of the strain energy are clearly
identified.

Damping, due to internal friction or other dissipative phenomena, should
carefully be taken into account in the structural analysis. Several damping
models exist, like general linear damping or viscous proportional or
non-proportional linear damping; see \emph{e.g.}
\citep{Adhikari Woodhouse
2001a}. However, the link between these structural damping models and the
corresponding dissipative material behaviour, described by a given
strain-stress law, does not seem to be clearly established in the
literature. Here the beam dynamic equations with linear viscous damping are
derived for the case of "small strains and moderate rotations", thus filling
the gap.

The different configurations characterizing a vibrating structure subjected
to static and dynamic loads and to a thermal field are defined in Section 2.
Then, the thermo-elastic constitutive rule is presented in Section 3.
Physical linearization is considered. Section 4 introduces a modified
Hu-Washizu functional accounting for stress constraints. The corresponding
stationarity conditions are discussed. In Section 5, two global measures for
the strains and the rotations are introduced and used to suggest
approximated expressions of the strain energy, each approximation being
valid in a strain-rotation domain. 2D finite element simulations
substantiate the approximation of the strain energy. Section 6 introduces a
dissipative stress tensor in view of modelling damping effect. Section 7
leads to a general expression of the weak equilibrium of a continuum
subjected to a general static prestress. In Section 8 the previous general
procedure is used to derive the Euler-Bernoulli beam equations for small
strains and moderate rotations in the undamped case, while the damped case
is treated in Section 9. After the Conclusions, the Appendix explains how to
compute the Lagrange multipliers.

\section{Configurations of a structure}

\begin{figure}[tbp]
\begin{center}
\includegraphics[width=11cm]{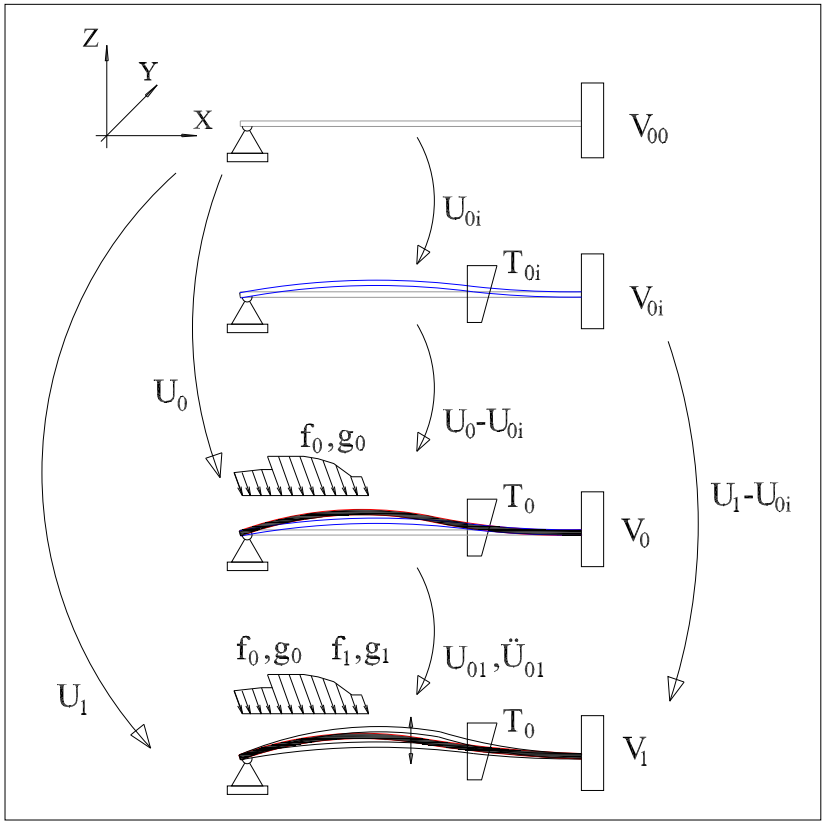}
\end{center}
\caption{The configurations of a structure.}
\label{Config}
\end{figure}

The following configurations (see Figure \ref{Config} and Table \ref{TabConf}%
) may be distinguished:

\begin{enumerate}
\item $V_{00}$ is the \emph{geometric reference configuration}, where the
\emph{displacements are assumed to vanish}. For a beam, $V_{00}$ is the
straight configuration. Let $\mathbf{X}$ be the reference position of a
material point.

\item $V_{0i}$ is the initial configuration of the body, where the
temperature field $T_{0i}$ is constant and where no external force is
applied. This configuration is relevant in the case of geometric
imperfections, where $V_{0i}$ differs from $V_{00}$. In this paper, however,
we assume $V_{0i}=V_{00}$, \emph{i.e.} $\mathbf{U}_{0i}=\mathbf{x}_{0i}-
\mathbf{X}=\mathbf{0}$. The stress may not vanish since a self-equilibrated
stress $\mbox{\boldmath ${\Pi }$}_{0i}$ may exist.

\item $V_{0}$ is the equilibrium configuration under \emph{static external}
loads $\mathbf{f}_{0}$, $\mathbf{g}_{0}$ and a temperature field $T_{0}$
resulting in a displacement $\mathbf{U}_0$, a relative displacement between $%
V_{0i}$ and $V_{0}$ equal to $\mathbf{U}_{0}-\mathbf{U}_{0i}$ and a stress
field $\mbox{\boldmath$\Pi$}_{0}- \mbox{\boldmath$\Pi$}_{0i}$.

\item $V_{1}$ is the instantaneous configuration of a vibrating structure
subjected to the static loads $\mathbf{f}_{0}$, $\mathbf{g}_{0}$, to the
temperature field $T_{0}$ and to \emph{dynamic} volume and surface forces $%
\mathbf{f}_{1} $ and $\mathbf{g}_{1}$. The relative displacement field
between $V_{0} $ and $V_{1}$ is indicated by $\mathbf{U}_{01}=\mathbf{U}_{1}-%
\mathbf{U}_{0}= \mathbf{x}_{1}-\mathbf{x}_{0}$. In practical situations $%
\mathbf{f}_{0}$ \textbf{, }$\mathbf{g}_{0}$ and $T_{0}$ may vary with time,
e.g. on a daily period, but their variations are supposed to be very slow
with respect to free vibrations of the structure. \emph{This decomposition
of a given external force into a static part and dynamic part} is convenient
but \emph{not unique}, \emph{e.g.} it may depend on the time interval
considered if two different static loads are successively imposed.
Therefore, the total \emph{response} of the structure does not split
uniquely in a static component and a dynamic response.
\end{enumerate}

\section{Constitutive law and physical linearization}

Based the objectivity principle, (see \emph{e.g.} \citet[p.
602]{Mandel66}), the Helmholtz free energy $\Psi $ can be defined as a
function of the Green-Lagrange strain tensor
\begin{equation}
\begin{array}{l}
\mathbf{E\left( \mathbf{U}\right) =}\frac{1}{2}\left( \frac{\partial \mathbf{%
U}}{\partial \mathbf{X}}+\left( \frac{\partial \mathbf{U}}{\partial \mathbf{%
X }}\right) ^{T}\right) +\frac{1}{2}\left( \frac{\partial \mathbf{U}}{%
\partial \mathbf{X}}\right) ^{T}\cdot \left( \frac{\partial \mathbf{U}}{%
\partial \mathbf{X}}\right) \mathbf{=\nabla }^{s}\left( \mathbf{U}\right) +%
\frac{1}{2}\mathbf{\nabla } \left( \mathbf{U}\right) ^{T}\cdot \mathbf{%
\nabla }\left( \mathbf{U}\right)%
\end{array}
\label{GreenLagr}
\end{equation}
and of the temperature $T$, viz. $\Psi =\Psi \left( \mathbf{E},T\right) $.
As usual, $\cdot $ indicates the dot product and $\mathbf{U}$ is the
displacement measured with respect to the given reference configuration $%
V_{00}$. When $\left\Vert \mathbf{E}\right\Vert \ll 1$ and the temperature
variations are small, $\Psi \left( \mathbf{E,}T\right) $ can be approximated
by a truncated series expansion. 
Truncating at the second order around the given initial configuration $%
V_{00} $, characterized by $\mathbf{E=0}$ and $T=T_{0i}$ (physical
linearization) leads to
\begin{equation}
\begin{array}{l}
\Psi \left( \mathbf{E,}T\right) =\mbox{\boldmath$\Pi$}_{0i}:\mathbf{E+}\frac{%
1}{2}\mathbf{E}:\mathbf{D:E-E:A}\left( T-T_{0i}\right) \\
-s_{0i}\left( T-T_{0i}\right) -\frac{1}{2}\frac{c_{\varepsilon }}{T_{0i}}%
\left( T-T_{0i}\right) ^{2}%
\end{array}
\label{helm2}
\end{equation}%
where $:$ denotes the doubly contracted inner product, $\mathbf{D}$ the
fourth order tensor of the elastic constants, $T-T_{0i}$ the temperature
variation and $\mathbf{A}$ a diagonal second order tensor accounting for
thermal expansion, $s_{0i}$ the initial volume entropy and $c_{\varepsilon }$
the specific volume heat [$J$ $m^{-3}K^{-1}$]. When a \emph{St.
Venant-Kirchhoff} material is considered, one has
\begin{equation}
\mathbf{D=}\lambda \mathbf{1}\otimes \mathbf{1+}2\mu
\mathbf{I}\textrm{ \ \ \ \ and \ \ }\mathbf{A=}\alpha \left(
3\lambda +2\mu \right) \textrm{ }\mathbf{1} \label{ADisotr}
\end{equation}
where $\otimes $ is the outer tensor product; $\mathbf{1}$ is the second
order identity tensor; $\mathbf{I}$ is the fourth order identity tensor; $%
\lambda $ and $\mu $ are the Lam\'{e} coefficients, $\alpha $ is the thermal
dilation coefficient. As it is well-known, the following identities hold:
\begin{equation}
\lambda =\frac{E\nu }{\left( 1+\nu \right) \left( 1-2\nu \right) },\textrm{ \ }%
\mu =\frac{E}{2\left( 1+\nu \right) },\textrm{ }
\label{lameYoungPoiss}
\end{equation}%
where $E$ is the Young modulus, $\nu $ is the Poisson ratio. 
In this case, the Helmholtz energy (\ref{helm2}) becomes
\begin{equation}
\begin{array}{l}
\Psi \left( \mathbf{E,}T\right) =\mbox{\boldmath$\Pi$}_{0i}:\mathbf{E+}\frac{%
1}{2}\lambda \textrm{ }\left( tr\left( \mathbf{E}\right) \right)
^{2}+\mu \textrm{ }\mathbf{E:E-}\alpha \left( 3\lambda +2\mu \right)
\left(
T-T_{0i}\right) tr\left( \mathbf{E}\right) \\
\textrm{ \ \ \ \ \ \ \ \ \ \ \ \ \ \ \ \ \ \ \ }-s_{0i}\left( T-T_{0i}\right) -%
\frac{1}{2}\frac{c_{\varepsilon }}{T_{0i}}\left( T-T_{0i}\right) ^{2}%
\end{array}
\label{helm2iso}
\end{equation}

Eq. (\ref{helm2}) leads to a constitutive law linear with respect to $%
\mathbf{E}$ and $T-T_{0i}$: 
\begin{equation}
\mbox{\boldmath$\Pi$}=\frac{\partial \Psi }{\partial \mathbf{E}}=%
\mbox{\boldmath$\Pi$}^{nd}\mathbf{=\Pi }_{0i}+\mathbf{D:E-A}\left(
T-T_{0i}\right)  \label{GenCostLaw}
\end{equation}%
where $\mbox{\boldmath$\Pi$}$ is the \emph{(second) Piola-Kirchhoff
symmetric stress tensor}, the index $nd$ means non-dissipative. Eq. (\ref%
{GenCostLaw}) also reads%
\begin{equation}
\mbox{\boldmath$\Pi$}=\mbox{\boldmath$\Pi$}^{nd}=\mbox{\boldmath$\Pi$}%
_{0i}+\lambda \textrm{ }tr\left( \mathbf{E}\right) \mathbf{1+}2\mu \textrm{ }%
\left( \mathbf{E}\right) -\alpha \left( 3\lambda +2\mu \right) \textrm{ }%
\left( T-T_{0i}\right) \mathbf{1}  \label{GenCostLawIso}
\end{equation}

For isotropic materials having elastic non-linear constitutive behaviour, a
direct generalization of (\ref{GenCostLawIso}) can be defined. It is called
\emph{second order elasticity} \cite[p. 607]{Mandel66}, supplemented here by
the self-equilibrated stress $\mbox{\boldmath$\Pi$}_{0i} $:
\begin{equation}
\begin{array}{l}
\mbox{\boldmath$\Pi$}=\mbox{\boldmath$\Pi$}^{nd}=\mbox{\boldmath$\Pi$}_{0i}+%
\left[ \lambda \textrm{ }tr\left( \mathbf{E}\right)
+\frac{A}{2}tr\left( \mathbf{E}^{2}\right) +3B\left( tr\left(
\mathbf{E}\right) \right) ^{2}-\alpha \left( 3\lambda +2\mu \right)
\left( T-T_{0i}\right) \right]
\mathbf{1} \\
+\left[ \lambda ^{\prime }tr\left( \mathbf{E}\right) \left( T-T_{0i}\right)
+a^{\prime }\left( T-T_{0i}\right) ^{2}\right] \mathbf{1}+2\mu \textrm{ }%
\mathbf{E}+A\textrm{ }tr\left( \mathbf{E}\right) \mathbf{E}+2\mu ^{\prime }%
\mathbf{E}\left( T-T_{0i}\right) +C\mathbf{\ \ E}\cdot \mathbf{E}%
\end{array}
\label{constQuad}
\end{equation}
where $A,B,C,\lambda ^{\prime },\mu ^{\prime }$ and $a^{\prime }$ are the
material parameters introduced in addition to the usual ones $\lambda $, $%
\mu $ and $\alpha $.

Remark: the temperature variations can be considered small when
\begin{equation}
\left\Vert \mathbf{D}^{-1}:\mathbf{A}\textrm{ }\left(
T-T_{0i}\right) \right\Vert \ll 1  \label{smallT}
\end{equation}%
viz. when strains associated with them are small with respect to
unity, similarly to strains fulfilling condition $\|\mathbf{E}\|\ll
1$. For isotropic materials, the temperature variations are small if
$\left\vert \alpha \textrm{ }\left( T-T_{0i}\right) \right\vert \ll
1$. This condition is consistent with the assumption of physical
linearization.

\section{A Hu-Washizu functional with additional stress constraints}

In this section, a special representation of the problem at hand is
introduced, in view of taking into account stress constraints in a
three-dimensional framework\ without forgetting about compatibility issues.
A Hu-Washizu (H-W) functional depending on the stress $\mbox{\boldmath$\Pi$}%
, $ the strain measure $\mathbf{\mathbf{\bar{E}}}$ and the displacement $%
\mathbf{U}$, considered independent is supplemented with constraints on the
stress, in order to model thin bodies. Let $\mathbb{U}$ , $\mathbb{T}$ and $%
\mathbb{F}$ denote spaces of smooth enough displacement fields, of symmetric
second order tensor fields and scalar fields of Lagrange multipliers,
respectively. The proposed H-W functional reads
\begin{equation}
\begin{array}{l}
J_{H-W}\left( \mbox{\boldmath$\Pi$}^{\ast }\mathbf{,\mathbf{\bar{E}}^{\ast
},U}^{\ast },\mbox{\boldmath$\Lambda$}^{\ast }\right) =\int_{V_{00}}\Psi
\left( \mathbf{\mathbf{\bar{E}}}^{\ast },T\right) \textrm{ }dV-\int_{V_{00}} %
\mbox{\boldmath$\Pi$}:\left( \mathbf{\mathbf{\bar{E}}}^{\ast }-\mathbf{E}
\left( \mathbf{U}^{\ast }\right) \right) dV \\
-\int_{V_{00}}\mathbf{f\cdot U}^{\ast }dV-\int_{\partial V_{00,\sigma }}
\mathbf{g\cdot U}^{\ast }dA-\int_{\partial V_{00,u}}\left( \left[ \left(
\mathbf{1}+\nabla \left( \mathbf{U}^{\ast }\right) \right) \mathbf{\cdot \Pi
^{\ast }}\right] \mathbf{\cdot N}\right) \mathbf{\cdot }\left( \mathbf{\
U^{\ast }-\bar{U}}\right) dA \\
-\int_{V_{00}}\sum_{k=1}^{n_{\Lambda }}\Lambda _{k}^{\ast }\mathbf{R}_{k}: %
\mbox{\boldmath$\Pi$}^{\ast }dV%
\end{array}
\label{Hu-Washizu-Gen}
\end{equation}
with $\mbox{\boldmath$\Pi$}^{\ast }\in \mathbb{T}$, $\mathbf{\mathbf{\bar{E}}
^{\ast }}\in \mathbb{T},\mathbf{U}^{\ast }\in \mathbb{U}$ and $%
\mbox{\boldmath$\Lambda$}^{\ast }\in \mathbb{F}^{n_{\Lambda}}$, where $%
n_{\Lambda }$ is the number of stress constraints and $\mathbf{\bar{U}}$ is
the given displacement on the boundary $\partial V_{00,u}=\partial
V_{00}-\partial V_{00,\sigma }$. Note that $J_{H-W} $ is a functional
defined on $3+n_{\Lambda }$ fields. At a solution $\left( %
\mbox{\boldmath$\Pi$},\mathbf{\bar{E}},\mathbf{U},\mbox{\boldmath$\Lambda$}%
\right)$ of the static problem, the functional above satisfies the
stationarity with respect to the $3+n_{\Lambda }$ fields.
The last term leads to $n_{\Lambda }$ \emph{linear} stress constraints:
\begin{equation}
\mathbf{R}_{k}:\mbox{\boldmath$\Pi$}:=\left(\mathbf{N}_{k}\otimes \mathbf{N}%
_{k}\right): \mbox{\boldmath$\Pi$}=\left( \mbox{\boldmath$\Pi$}\cdot \mathbf{%
N} _{k}\right) \cdot \mathbf{N}_{k}=0  \label{Nn}
\end{equation}%
where $\mathbf{N}_{k}$ is a vector and $\mathbf{R}_{k}$ is a second order
constant tensor. Eq. (\ref{Nn}) imposed on $\mbox{\boldmath$\Pi$}$ carries
over to an equivalent condition on the Cauchy stress $%
\mbox{\boldmath
$\sigma$}$. The use of this kind of constraints to derive beam equations
from 3D elasticity is illustrated in Section 7. The case of a general 3D
structure with no constraint can be addressed by formally setting $\Lambda
_{k}=0$ and $\mathbf{R}_{k}=\mathbf{0}$. The well-known impossibility to
derive beam or plate equations from purely kinematical assumptions in the
three-dimensional equations of elasticity in pure displacement motivate the
introduction of internal constraints as in \citep{PPGMindlin}, %
\citep{PPGTimoshenko}. These constraints enable one to use a pure
displacement approach while mimicking the averaging process underlying the
convergence of the equations of elasticity when the aspect ratio tends to
zero. The derivation of lower-dimensional models without assumptions can
rely on $\Gamma$-convergence \citep{acerbibuttazzoARMA86}, asymptotic
analysis \citep{PGCvonkarman}, or energy methods \citep{Babuska92}. This
paper aims at deriving the equations governing the evolution of thin
structures subject to prestress states without above mathematical apparatus.

\subsection{Action functional and stationarity conditions}

While the stationarity of the H-W functional leads to statics, the action
functional leads to elastodynamics over a time interval $\left[ 0,t_{f}%
\right] $. Let us set $\mathbb{V=}\left\{ \mathbf{U}\left( \mathbf{X}%
,t\right) :\mathbf{U}\left( \mathbf{\cdot },t\right) \in \mathbb{U}\textrm{ \ }%
\forall t\in \left[ 0,t_{f}\right] \right\} $. The kinetic energy $\mathcal{T%
}\left( \mathbf{\dot{U}^{\ast }}\right) =\frac{1}{2}\int_{V_{00}}\rho _{00}%
\mathbf{\dot{U}^{\ast }\cdot \dot{U}^{\ast }}dV$, combined with the H-W
functional enables one to define the action
\begin{equation}
\mathcal{D}\left( \mbox{\boldmath$\Pi$}^{\ast }\mathbf{,\mathbf{\bar{E}}%
^{\ast },U}^{\ast },\mbox{\boldmath$\Lambda$}^{\ast }\right) \mathcal{=}%
\int_{0}^{t_{f}}\left( \mathcal{T}\left( \mathbf{\dot{U}^{\ast }}\right)
\mathcal{-}J_{H-W}\left( \mbox{\boldmath$\Pi$}^{\ast }\mathbf{,\bar{E}^{\ast
},U}^{\ast },\mbox{\boldmath$\Lambda$}^{\ast }\right) \right) dt
\label{Action}
\end{equation}

Hamilton's principle is given by $\delta \mathcal{D}=0$, where small
variations $\left( \delta \mbox{\boldmath$\Pi$},\delta \mathbf{\bar{E},}%
\delta \mathbf{U,}\delta \mbox{\boldmath$\Lambda$}\right) \in \mathbb{%
T\times T\times V\times F}^{n_{\Lambda }}$ are considered. The stationarity
operator $\delta \left( \cdot \right) $ is intended as \emph{isochronous},
viz. $\delta t=0$ and $\delta \int_{0}^{t_{f}}\mathcal{L}$ $%
dt=\int_{0}^{t_{f}}\delta \mathcal{L}$ $dt$ (see also %
\citet{QuadrelliAtluri99}). Hamilton's principle leads to the stationarity
conditions

\begin{equation}
\begin{array}{l}
1.\textrm{ \ \ \ }\forall k=1,n_{\Lambda }\textrm{\ \ \ \ \ \ \ \ }\mathbf{R}%
_{k}:\mbox{\boldmath$\Pi$}=0\textrm{ \ \ \ \ \ in }V_{00} \\
2.\textrm{ \ \ \ }\left\{
\begin{array}{ll}
\mathbf{\bar{E}}=\mathbf{E}\left( \mathbf{U}\right) -\sum_{k=1}^{n_{\Lambda
}}\Lambda _{k}\mathbf{R}_{k} & \textrm{\ \ \ in }V_{00} \\
\mathbf{U}\mathbf{=\bar{U}} & \textrm{\ \ \ on }\partial V_{00,u}%
\end{array}%
\right. \\
3.\textrm{ \ \ \ }\mbox{\boldmath$\Pi$}=\left. \frac{\partial \Psi
\left(
\mathbf{\bar{E}}^{\ast },T\right) }{\partial \mathbf{\bar{E}}^{\ast }}%
\right\vert _{\mathbf{\bar{E}}^{\ast }=\mathbf{\bar{E}}}\textrm{ \ \
\ \ \ \ \
\ in }V_{00}%
\end{array}
\label{Strong123-gen}
\end{equation}%
In the physically linear case, Eq. (\ref{Strong123-gen}-3) corresponds to (%
\ref{GenCostLaw}). If isotropy is assumed, Eq. (\ref{Strong123-gen}-3)\ is
equal to (\ref{GenCostLawIso}). The first equation expresses the stress
constraints, the second one defines the strain $\mathbf{\bar{E}}$, which
differs from $\mathbf{\ E}\left( \mathbf{U}\right) $ due to the stress
constraints. Moreover, $\mbox{\boldmath$\Pi$}$ depends on the strain $%
\mathbf{\bar{E}}$ accounting for the Lagrange multipliers and not on the
standard strain measure $\mathbf{\ E}\left( \mathbf{U}\right) $. The
stationarity condition of the action $\mathcal{D}$ with respect to
displacements leads to
\begin{equation}
\begin{array}{l}
\forall \delta \mathbf{U}\in \mathbb{V}\textrm{, \ \
}\int_{0}^{t_{f}}\left[ \int_{V_{00}}\mbox{\boldmath$\Pi$}:\delta
\mathbf{E}\left( \mathbf{U,}\delta
\mathbf{U}\right) dV-\int_{V_{00}}\mathbf{f\cdot }\delta \mathbf{U}%
dV-\int_{\partial V_{00,\sigma }}\mathbf{\ g\cdot }\delta \mathbf{U}dA\right]
dt \\
-\int_{0}^{t_{f}}\left[ \int_{\partial V_{00,u}}\left( \left[ \left( \mathbf{%
1}+\nabla \left( \mathbf{U}\right) \right) \mathbf{\ \cdot }%
\mbox{\boldmath$\Pi$}\right] \mathbf{\cdot N}\right) \mathbf{\cdot }\delta
\mathbf{U}dA+\int_{V_{00}}\rho _{00}\mathbf{\dot{U}\cdot }\delta \mathbf{%
\dot{U}}dV\right] dt=0%
\end{array}
\label{Virtual-Work-Gen1}
\end{equation}%
where it has been used the identity $\mathbf{U}\mathbf{=\bar{U}}$\ on $%
\partial V_{00,u}$, given in (\ref{Strong123-gen}). Moreover, the virtual
strain is defined by
\begin{equation}
\begin{array}{l}
\delta \mathbf{E}\left( \mathbf{U,}\delta \mathbf{U}\right) =\left. \frac{%
\partial \mathbf{E}\left( \mathbf{U^{\ast }}\right) }{\partial \mathbf{U}%
^{\ast }}\right\vert _{\mathbf{U}^{\ast }\mathbf{=U}}\cdot \delta \mathbf{U}
\\
=\mathbf{\nabla }^{s}\left( \delta \mathbf{U}\right) +\frac{1}{2}\mathbf{%
\nabla }\left( \delta \mathbf{U}\right) ^{T}\cdot \mathbf{\nabla }\left(
\mathbf{U}\right) +\frac{1}{2}\mathbf{\nabla }\left( \mathbf{U}\right)
^{T}\cdot \mathbf{\nabla }\left( \delta \mathbf{U}\right)%
\end{array}
\label{deltaEbar}
\end{equation}%
Integrating by parts in time leads to the virtual works principle at every $%
t $
\begin{equation}
\mathcal{W}_{i}\left( \mbox{\boldmath$\Pi$},\mathbf{U},\delta \mathbf{U}%
\right) +\mathcal{W}_{e}\left( \mathbf{f,g,}\delta \mathbf{U}\right) =%
\mathcal{W}_{a}\left( \mathbf{\ddot{U},}\delta \mathbf{U}\right)
\textrm{\ \ \ \ \ \ \ \ \ \ \ }\forall \delta \mathbf{U}\in
\mathbb{V} \label{Virtual-Work-Gen2}
\end{equation}%
where
\begin{equation}
\begin{array}{l}
\mathcal{W}_{i}\left( \mbox{\boldmath$\Pi$},\mathbf{U},\delta \mathbf{U}%
\right) :=-\int_{V_{00}}\mbox{\boldmath$\Pi$}:\delta
\mathbf{E}\left( \mathbf{U,}\delta \mathbf{U}\right) dV,\textrm{ \ \
}\mathcal{W}_{a}\left(
\mathbf{\ddot{U},}\delta \mathbf{U}\right) :=\int_{V_{00}}\rho _{00}\mathbf{%
\ddot{U}\cdot }\delta \mathbf{U}dV \\
\mathcal{W}_{e}\left( \mathbf{f,g,}\delta \mathbf{U}\right) :=\int_{V_{00}}%
\mathbf{f\cdot }\delta \mathbf{U}dV+\int_{\partial V_{00,\sigma }}\mathbf{\
g\cdot }\delta \mathbf{U}dA+\int_{\partial V_{00,u}}\left( \left[ \left(
\mathbf{1}+\nabla \left( \mathbf{U}\right) \right) \mathbf{\cdot \Pi }\right]
\mathbf{\cdot N}\right) \mathbf{\cdot }\delta \mathbf{U}dA%
\end{array}
\label{Wdef}
\end{equation}%
denote the virtual work of internal, inertia and external forces,
respectively. Finally, by integration by parts, one obtains the
corresponding strong form equation and the boundary conditions on $\partial
V_{00,\sigma }$:
\begin{equation}
4.\textrm{ \ \ \ }\left\{
\begin{array}{ll}
div\left( \left( 1+\nabla \left( \mathbf{U}\right) \right) \cdot %
\mbox{\boldmath$\Pi$}\right) +\mathbf{f=}\rho _{00}\mathbf{\ddot{U}}
& \textrm{
\ \ \ in }V_{00} \\
\left( 1+\nabla \left( \mathbf{U}\right) \right) \cdot \mbox{\boldmath$\Pi$}%
\cdot \mathbf{N}=\mathbf{g} & \textrm{ \ \ \ on }\partial V_{00,\sigma }%
\end{array}%
\right.  \label{Strong0}
\end{equation}%
with the initial conditions $\mathbf{U}\left( t=0\right) =\mathbf{\tilde{U}}$
and $\mathbf{\dot{U}}\left( t=0\right) =\mathbf{\tilde{V}}$ in $V_{00}$.

\section{Physical linearization, strain-rotation domains and dominant terms
in the strain energy}

Let $\mathbf{U,}$ $\mathbf{\bar{E},}$ $\mbox{\boldmath$\Pi$}$ and $\Lambda
_{k}$ denote the fields defining the equilibrium at time $t$. The strain
energy density associated with the strain field $\mathbf{\bar{E}=\bar{E}}%
\left(t\right)$ reads
\begin{equation}
\mathcal{F}\left( \mathbf{\bar{E}},T\right)
\mathcal{=}\int_{V_{00}}\Psi \left( \mathbf{\bar{E}},T\right)
\textrm{ }dV  \label{StrainEnergy}
\end{equation}%
where $\Psi \left( \mathbf{\bar{E}},T\right) $ is given by (\ref{helm2iso}).
The displacement gradient writes
\begin{equation}
\mathbf{\nabla }\left( \mathbf{U}\right) =\mathbf{\nabla }^{s}\left( \mathbf{%
\ U}\right) +\mathbf{\nabla }^{sk}\left( \mathbf{U}\right) :=%
\mbox{\boldmath$\varepsilon$}\left( \mathbf{U}\right) \mathbf{+r}\left(
\mathbf{U}\right) =\mbox{\boldmath$\varepsilon$}+\mathbf{r}  \label{EdeltaE}
\end{equation}%
where the symmetric tensor $\mbox{\boldmath$\varepsilon$}$ is called \emph{%
small strain tensor}, while the skew-symmetric part $\mathbf{r}$ is called
the \emph{small rotation tensor}. For brevity, hereinafter they will be
referred to as \emph{strain tensor} and \emph{rotation tensor}. In the same
way, the strain increment (\ref{deltaEbar}) writes
\begin{equation}
\delta \mathbf{E}\left( \mathbf{U,}\delta \mathbf{U}\right) =\mathbf{\
\nabla }^{s}\left( \delta \mathbf{U}\right) +\frac{1}{2}\mathbf{\nabla }%
\left( \delta \mathbf{U}\right) ^{T}\cdot \left(
\mbox{\boldmath$
\varepsilon$}+\mathbf{r}\right) +\frac{1}{2}\left(
\mbox{\boldmath$
\varepsilon$}+\mathbf{r}\right) ^{T}\cdot \mathbf{\ \nabla }\left( \delta
\mathbf{U}\right)  \label{deltaEUdotdt}
\end{equation}

\subsection{A global measure for strains and for rotations and a hierarchy
for the strain energy terms}

In view of finding out the dominant terms in (\ref{StrainEnergy}), let us
introduce the solution-dependent scalar parameters $\eta $ and $p$ as
follows:
\begin{equation}
\begin{array}{l}
\eta =\eta _{\varepsilon }+\eta _{\Delta T},\textrm{ \ \ }\eta _{\varepsilon }=%
\frac{1}{\sqrt{V_{00}}}\left\Vert \mbox{\boldmath $\varepsilon$
}\right\Vert _{L_{2}\left( V_{00}\right) },\textrm{ \ \ } \eta
_{\Delta T}=\frac{1}{\sqrt{V_{00}}}\left\Vert \alpha \left(
T-T_{0i}\right)
\mathbf{1}\right\Vert _{L_{2}\left( V_{00}\right) } \\
\eta ^{p}:=\frac{1}{\sqrt{2}}\frac{1}{\sqrt{V_{00}}}\left\Vert \mathbf{r}
\right\Vert _{L_{2}\left( V_{00}\right) }=\left( \frac{1}{V_{00}}
\int_{V_{00}}\frac{1}{2}r_{ij}r_{ij}\textrm{ }dV\right) ^{\frac{1}{2}}%
\end{array}
\label{etap-def}
\end{equation}
where $\left\Vert \cdot \right\Vert _{L_{2}\left( V_{00}\right) }$ is the $%
L_{2}$-norm associated with the domain $V_{00}$. The strain measure $\eta $
is supposed to be small. Observe that: (i) $p$ is only defined whenever $%
\eta >0$. Thus infinitesimal rigid body motions at initial temperature are
excluded from the following analysis; (ii) If the rotations are not large, $%
\mathbf{r}$ is a rotation tensor and the quantity $\eta ^{p}$ is a global
measure of the rotations. Hence, $p$ can be interpreted as a global index of
the relative amplitude of rotations with respect to strains. As a result
(see Eq. (\ref{GreenLagr}))
\begin{equation}
\mathbf{E}\left( \mathbf{U}\right) =O\left( \eta \right) +O\left( \eta
^{2p}\right) +O\left( \eta ^{1+p}\right) +O\left( \eta ^{2}\right)
\label{Ebar-orders}
\end{equation}%
Under dynamic conditions and provided that the vibration frequency is
bounded, one also has $\left\Vert \frac{d\mbox{\boldmath$\varepsilon$}}{dt}%
\right\Vert _{L_{2}\left( V_{00}\right) }=O\left( \eta \right) \ $and\ \ $%
\left\Vert \frac{d\mathbf{r}}{dt}\right\Vert _{L_{2}\left( V_{00}\right)
}=O\left( \eta ^{p}\right) $. In addition, let us make a consistent
hypothesis on the self-equilibrated prestress $\mbox{\boldmath$\Pi$}_{0i}$:
\begin{equation}
\begin{array}{l}
\mbox{\boldmath$\Pi$}_{0i}=\mbox{\boldmath$\Pi$}_{0i}^{\left( 1\right) }+%
\mbox{\boldmath$\Pi$}_{0i}^{\left( 2p\right) }+ \mbox{\boldmath$\Pi$}%
_{0i}^{\left( 2\right) }+\mbox{\boldmath$\Pi$}_{0i}^{\left( 1+p\right) }
=O\left( \eta \right) +O\left(\eta^{2p}\right) +O\left( \eta ^{1+p}\right)
+O\left(\eta ^{2}\right)%
\end{array}
\label{T-P0i-order}
\end{equation}%
Since the Lagrange multipliers $\Lambda _{k}$ fulfill the relationship for
all $k=1,n_{\Lambda }$
\begin{equation}
\mathbf{R}_{k}:\mbox{\boldmath$\Pi$}=\mathbf{R}_{k}:\left[ %
\mbox{\boldmath$\Pi$}_{0i}+\mathbf{D}:\left( \mathbf{E}\left( \mathbf{U}%
\right) -\sum_{l=1}^{n_{\Lambda }}\Lambda _{l}\mathbf{R}_{l}\right) -\mathbf{%
A}\textrm{ }\left( T-T_{0i}\right) \right] =0  \label{RkPi}
\end{equation}%
then, provided the stress constraints $\mathbf{R}_{k}$ are linearly
independent, $\mbox{\boldmath$\Lambda$}\mathbf{=}\left( \Lambda _{k}\right),$
$k=1,n_{\Lambda }$, is the solution of a linear Gram system and splits as
follows (see Appendix):
\begin{equation}
\mbox{\boldmath$\Lambda$}\mathbf{=}\mbox{\boldmath$\Lambda$}^{\left(
1\right) }\mathbf{+}\mbox{\boldmath$\Lambda$}^{\left( 2p\right) }\mathbf{+}%
\mbox{\boldmath$\Lambda$}^{\left( 1+p\right) }\mathbf{+}%
\mbox{\boldmath$
\Lambda$}^{\left( 2\right) }=O\left( \eta \right) +O\left( \eta ^{2p}\right)
+O\left( \eta ^{1+p}\right) +O\left( \eta ^{2}\right)  \label{lambda-orders}
\end{equation}%
where $\mbox{\boldmath$\Lambda$}^{\left( 1\right) }$, $%
\mbox{\boldmath$
\Lambda$}^{\left( 2p\right) },$ $\mbox{\boldmath$\Lambda$}^{\left(
1+p\right) }$ and $\mbox{\boldmath$\Lambda$}^{\left( 2\right) }$ can be
evaluated analytically in simple cases. This implies that the Lagrange
multipliers, $\mathbf{\bar{E}}$ (Eq. (\ref{Strong123-gen}-2)) and $\mathbf{E}
\left( \mathbf{U}\right)$ split in terms of the same order.

\begin{figure}[tbp]
\begin{center}
\includegraphics[width=10cm]{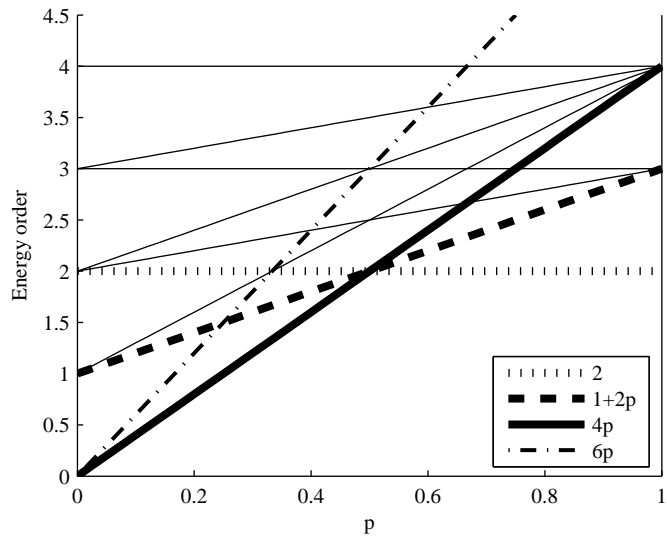}
\end{center}
\caption{Orders of the strain energy terms. The dominant terms ($2$, $1+2p$
and $4p$) are highlighted. The line corresponding to the order $6p$
represents the most important energy term associated with the non-linear
constitutive law (\protect\ref{constQuad}).}
\label{Exponents}
\end{figure}

From Eqs. (\ref{Strong123-gen}) and (\ref{lambda-orders}), the stress splits
in four terms:
\begin{equation}
\mbox{\boldmath$\Pi$}=\mbox{\boldmath$\Pi$}^{\left( phys.lin.\right) }=%
\mbox{\boldmath$\Pi$}^{\left( 1\right) }+\mbox{\boldmath$\Pi$}^{\left(
2p\right) }+\mbox{\boldmath$\Pi$}^{\left( 1+p\right) }+\mbox{\boldmath$\Pi$}%
^{\left( 2\right) }  \label{Pi-phys.lin}
\end{equation}%
where
\begin{equation}
\begin{array}{l}
\mbox{\boldmath$\Pi$}^{\left( 1\right)
}=\mbox{\boldmath$\Pi$}_{0i}^{\left( 1\right) }+\mathbf{D:}\left(
\mbox{\boldmath $\varepsilon$ -}\textrm{\textbf{\
}}\sum_{l=1}^{n_{\Lambda }}\Lambda _{l}^{\left( 1\right) }\mathbf{R}%
_{l}\right) \mathbf{-A}\textrm{ }\left( T-T_{0i}\right) =O\left(
\eta \right)
\\
\mbox{\boldmath$\Pi$}^{\left( 2p\right) }=\mbox{\boldmath$\Pi$}^{\left(
2p\right) }+\mathbf{D}:\frac{1}{2}\mathbf{r}^{T}\cdot \mathbf{r}-\mathbf{D}%
:\sum_{l=1}^{n_{\Lambda }}\Lambda _{l}^{\left( 2p\right) }\mathbf{R}_{l}%
\mathbf{=}O\left( \eta ^{2p}\right) \\
\mbox{\boldmath$\Pi$}^{\left( 1+p\right) }=\mbox{\boldmath$\Pi$}%
_{0i}^{\left( 1+p\right) }+\mathbf{D}:\left( \frac{1}{2}\mathbf{r}^{T}\cdot
\mbox{\boldmath $\varepsilon$
}\textrm{ }+\frac{1}{2}\mbox{\boldmath $\varepsilon$ }^{T}\cdot \mathbf{r}%
\right) -\mathbf{D}:\sum_{l=1}^{n_{\Lambda }}\Lambda _{l}^{\left( 1+p\right)
}\mathbf{R}_{l}=O\left( \eta ^{1+p}\right) \\
\mbox{\boldmath$\Pi$}^{\left( 2\right) }=\mbox{\boldmath$\Pi$}_{0i}^{\left(
2\right) }+\mathbf{D}:\frac{1}{2}\mbox{\boldmath $\varepsilon$ }^{T}\cdot %
\mbox{\boldmath $\varepsilon$ }-\mathbf{D}:\sum_{l=1}^{n_{\Lambda }}\Lambda
_{l}^{\left( 2\right) }\mathbf{R}_{l}\mathbf{=}O\left( \eta ^{2}\right)%
\end{array}
\label{Pi-orders}
\end{equation}%
Finally, assume that
\begin{equation}
0<\eta <1\textrm{ \ \ \ \ \ \ \ \ \ }p>0  \label{etap-intervals}
\end{equation}
Since $\eta <1$ , the terms with the smallest exponent are the largest ones.
The comparison between different terms must be done for a given $p$-value.
If $p\geq 1$, viz. when $\mathbf{r}$ is of the same order of magnitude as,
or smaller than, $\mbox{\boldmath
$\varepsilon$}$ (case of \emph{small} rotations), all non-linear terms in
the Green-Lagrange strain $\mathbf{E}$ and in the Lagrange multipliers $%
\Lambda _{k}$ can be discarded and the standard condition of small
transformation is retrieved. The exponent $p<1$ is introduced in order to
account for the possibility of having rotations \emph{larger} than strains.
The complete expression of the strain energy reads (recall Eq. (\ref{helm2}%
), where the terms that depends \emph{only} on the temperature are omitted
for brevity):
\[
\begin{array}{l}
\mathcal{F}\left( \mathbf{\bar{E}},T\right) \mathcal{=}\int_{V_{00}}\Psi
\left( \mathbf{\bar{E}},T\right) \textrm{ }dV=\int_{V_{00}}\left( %
\mbox{\boldmath$\Pi$}_{0i}-\mathbf{A}\left( T-T_{0i}\right) +\frac{1}{2}%
\mathbf{D:\bar{E}}\right) \mathbf{:\bar{E}}dV \\
=\int_{V_{00}}\left( \mbox{\boldmath$\Pi$}_{0i}-\mathbf{A}\left(
T-T_{0i}\right) +\frac{1}{2}\mathbf{D:}\left( \mathbf{E-}\sum_{l=1}^{n_{%
\Lambda }}\Lambda _{l}\mathbf{R}_{l}\right) \right) \mathbf{:}\left( \mathbf{%
\ E-}\sum_{l=1}^{n_{\Lambda }}\Lambda _{l}\mathbf{R}_{l}\right) dV%
\end{array}%
\]%
i.e. $\mathcal{F}\left( \mathbf{\bar{E}},T\right) \mathcal{=F}\left( \mathbf{%
E},T\right) \mathcal{+F}_{\Lambda }\left( \mathbf{E},T\right) $\ , where
\[
\begin{array}{l}
\mathcal{F}\left( \mathbf{E},T\right) =\int_{V_{00}}\left( %
\mbox{\boldmath$\Pi$}_{0i}-\mathbf{A}\left( T-T_{0i}\right) +\frac{1}{2}%
\mathbf{D:E}\right) \mathbf{:E}dV \\
\mathcal{F}_{\Lambda }\left( \mathbf{E},T\right) =-\int_{V_{00}}\left( %
\mbox{\boldmath$\Pi$}_{0i}-\mathbf{A}\left( T-T_{0i}\right) +\frac{1}{2}%
\mathbf{D:E}\right) \mathbf{:}\left( \sum_{l=1}^{n_{\Lambda }}\Lambda _{l}%
\mathbf{R}_{l}\right) dV \\
-\int_{V_{00}}\frac{1}{2}\mathbf{D:}\left( \sum_{l=1}^{n_{\Lambda }}\Lambda
_{l}\mathbf{R}_{l}\right) \mathbf{:E}dV+\int_{V_{00}}\frac{1}{2}\mathbf{D:}%
\left( \sum_{l=1}^{n_{\Lambda }}\Lambda _{l}\mathbf{R}_{l}\right) \mathbf{:}%
\left( \sum_{l=1}^{n_{\Lambda }}\Lambda _{l}\mathbf{R}_{l}\right) dV%
\end{array}%
\]%
More precisely,
\begin{equation}
\begin{array}{l}
\mathcal{F}\left( \mathbf{E},T\right) \mathcal{=}\int_{V_{00}}\left( \mathbf{%
\ \Pi }_{0i}^{\left( 1\right) }-\mathbf{A}\left( T-T_{0i}\right) +\frac{1}{2}%
\mathbf{D:}\mbox{\boldmath $\varepsilon$ }\right) \mathbf{:}\left(
\mbox{\boldmath $\varepsilon$
+}\frac{1}{2}\mathbf{r}^{T}\cdot \mathbf{r}\right) \textrm{ }dV \\
+\int_{V_{00}}\left( \mbox{\boldmath$\Pi$}_{0i}^{\left( 1\right) }-\mathbf{A}%
\left( T-T_{0i}\right) +\frac{1}{2}\mathbf{D:}%
\mbox{\boldmath $\varepsilon$
}\right) \mathbf{:}\left( \frac{1}{2}\mathbf{r}^{T}\cdot \mbox{\boldmath$%
\varepsilon$}+\frac{1}{2}\mbox{\boldmath$\varepsilon$}^{T}\cdot \mathbf{r+}%
\frac{1}{2}\mbox{\boldmath$\varepsilon$}^{T}\cdot
\mbox{\boldmath
$\varepsilon$ }\right) dV \\
+\int_{V_{00}}\left( \mbox{\boldmath$\Pi$}_{0i}^{\left( 2p\right) }+\frac{1}{%
2}\mathbf{D:}\left( \frac{1}{2}\mathbf{r}^{T}\cdot \mathbf{r}\right) \right)
\mathbf{:}\left( \mbox{\boldmath
$\varepsilon$
+}\frac{1}{2}\mathbf{r}^{T}\cdot \mathbf{r+}\frac{1}{2}\mathbf{r}^{T}\cdot %
\mbox{\boldmath$\varepsilon$}+\frac{1}{2}\mbox{\boldmath$\varepsilon$}%
^{T}\cdot \mathbf{r+}\frac{1}{2}\mbox{\boldmath$\varepsilon$}^{T}\cdot %
\mbox{\boldmath$\varepsilon$}\right) \textrm{ }dV \\
+\int_{V_{00}}\left( \mbox{\boldmath$\Pi$}_{0i}^{\left( 1+p\right) }+\frac{1%
}{2}\mathbf{D:}\left( \frac{1}{2}\mathbf{r}^{T}\cdot \mbox{\boldmath
$\varepsilon$ }\textrm{ }+\frac{1}{2}\mbox{\boldmath
$\varepsilon$ }^{T}\cdot \mathbf{r}\right) \right) \mathbf{:}\left( %
\mbox{\boldmath $\varepsilon$ +}\frac{1}{2}\mathbf{r}^{T}\cdot \mathbf{r}%
\right) \textrm{ }dV \\
+\int_{V_{00}}\left( \mbox{\boldmath$\Pi$}_{0i}^{\left( 1+p\right) }+\frac{1%
}{2}\mathbf{D:}\left( \frac{1}{2}\mathbf{r}^{T}\cdot
\mbox{\boldmath
$\varepsilon$ }+\frac{1}{2}\mbox{\boldmath
$\varepsilon$}^{T}\cdot \mathbf{r}\right) \right) \mathbf{:}\left( \frac{1}{2%
}\mathbf{r}^{T}\cdot \mbox{\boldmath$ \varepsilon$
}+\frac{1}{2}\mbox{\boldmath $\varepsilon$ }^{T}\cdot \mathbf{r+}\frac{1}{2}%
\mbox{\boldmath $\varepsilon$ }^{T}\cdot \mbox{\boldmath $\varepsilon$
}\right) dV \\
+\int_{V_{00}}\left( \mbox{\boldmath$\Pi$}_{0i}^{\left( 2\right) }+\frac{1}{2%
}\mathbf{D:}\left( \frac{1}{2}\mbox{\boldmath
$\varepsilon$}^{T}\cdot \mbox{\boldmath $\varepsilon$}\right) \right)
\mathbf{:}\left( \mbox{\boldmath $\varepsilon$
+}\frac{1}{2}\mathbf{r}^{T}\cdot \mathbf{r+}\frac{1}{2}\mathbf{r}^{T}\cdot
\mbox{\boldmath
$\varepsilon$ }+\frac{1}{2}\mbox{\boldmath
$\varepsilon$ }^{T}\cdot \mathbf{r+}\frac{1}{2}%
\mbox{\boldmath
$\varepsilon$ }^{T}\cdot \mbox{\boldmath $\varepsilon$ }\right) \textrm{ }dV%
\end{array}
\label{strain-energy-complete}
\end{equation}%
Eq. (\ref{strain-energy-complete}) shows all terms of the strain energy not
depending on $\mbox{\boldmath $\Lambda$}$. The exponents $s_{i}$ of these
energy terms are indicated in Figure \ref{Exponents} and in Table \ref%
{TabOrders}. The strain and rotation measures $\eta $ and $\eta ^{p}$ lead
to a hierarchy between the different terms. For every $p-$value, the
smallest exponents are $s_{1}=2$ , $s_{2}=1+2p,$ $s_{3}=4p$ (see Fig. \ref%
{Exponents}). Then, the dominant terms are
\begin{equation}
\begin{array}{l}
\mathcal{F}^{\left( 2\right) }\left( \mathbf{E},T\right)
=\int_{V_{00}}\left( \mbox{\boldmath$\Pi$}_{0i}^{\left( 1\right) }-\mathbf{A}%
\left( T-T_{0i}\right) +\frac{1}{2}\mathbf{D:}%
\mbox{\boldmath
$\varepsilon$ }\right) \mathbf{:}\mbox{\boldmath $\varepsilon$ }\textrm{ }%
dV=O\left( \eta ^{2}\right) \\
\mathcal{F}^{\left( 1+2p\right) }\left( \mathbf{E},T\right)
=\int_{V_{00}}\left( \mbox{\boldmath$\Pi$}_{0i}^{\left( 1\right) }-\mathbf{A}%
\left( T-T_{0i}\right) +\frac{1}{2}\mathbf{D:}%
\mbox{\boldmath
$\varepsilon$ }\right) \mathbf{:}\left( \frac{1}{2}\mathbf{r}^{T}\cdot
\mathbf{r}\right) \textrm{ }dV \\
\textrm{ \ \ \ \ \ \ \ \ \ \ \ \ \ \ \ \ \ }+\int_{V_{00}}\left( %
\mbox{\boldmath$\Pi$}_{0i}^{\left( 2p\right) }+\frac{1}{2}\mathbf{D:}\left(
\frac{1}{2}\mathbf{r}^{T}\cdot \mathbf{r}\right) \right) \mathbf{:}%
\mbox{\boldmath $\varepsilon$
}\textrm{ }dV=O\left( \eta ^{1+2p}\right) \\
\mathcal{F}^{\left( 4p\right) }\left( \mathbf{E},T\right)
=\int_{V_{00}}\left( \mbox{\boldmath$\Pi$}_{0i}^{\left( 2p\right) }+\frac{1}{%
2}\mathbf{D:}\left( \frac{1}{2}\mathbf{r}^{T}\cdot \mathbf{r}\right) \right)
\mathbf{:}\left( \frac{1}{2}\mathbf{r}^{T}\cdot \mathbf{r}\right) \textrm{ }%
dV=O\left( \eta ^{4p}\right)%
\end{array}
\label{StrEn-3approx}
\end{equation}%
Likewise, the dominant contribution to $\mathcal{F}_{\Lambda }$ splits in
three terms:
\[
\begin{array}{l}
\mathcal{F}_{\Lambda }^{\left( 2\right) }\left( \mathbf{E},T\right)
=-\int_{V_{00}}\left( \mbox{\boldmath$\Pi$}_{0i}^{\left( 1\right) }-\mathbf{A%
}\left( T-T_{0i}\right) +\frac{1}{2}\mathbf{D:}\mbox{\boldmath$\varepsilon$}%
\right) \mathbf{:}\sum_{l=1}^{n_{\Lambda }}\Lambda _{l}^{\left( 1\right) }%
\mathbf{R}_{l}dV \\
-\int_{V_{00}}\frac{1}{2}\left( \mathbf{D:}\sum_{l=1}^{n_{\Lambda }}\Lambda
_{l}^{\left( 1\right) }\mathbf{R}_{l}\right) \mathbf{:}%
\mbox{\boldmath
$\varepsilon$ }dV+\int_{V_{00}}\frac{1}{2}\left( \mathbf{D:}%
\sum_{l=1}^{n_{\Lambda }}\Lambda _{l}^{\left( 1\right) }\mathbf{R}%
_{l}\right) \mathbf{:}\sum_{l=1}^{n_{\Lambda }}\Lambda _{l}^{\left( 1\right)
}\mathbf{R}_{l}dV=O\left( \eta ^{2}\right)%
\end{array}%
\]%
\[
\begin{array}{l}
\mathcal{F}_{\Lambda }^{\left( 1+2p\right) }\left( \mathbf{E},T\right)
=-\int_{V_{00}}\left( \mbox{\boldmath$\Pi$}_{0i}^{\left( 1\right) }-\mathbf{A%
}\left( T-T_{0i}\right) +\frac{1}{2}\mathbf{D:}\mbox{\boldmath$\varepsilon$ }%
\right) \mathbf{:}\sum_{l=1}^{n_{\Lambda }}\Lambda _{l}^{\left( 2p\right) }%
\mathbf{R}_{l}dV \\
-\int_{V_{00}}\left( \mbox{\boldmath$\Pi$}_{0i}^{\left( 2p\right) }+\frac{1}{%
2}\mathbf{D:}\frac{1}{2}\mathbf{r}^{T}\cdot \mathbf{r}\right) \mathbf{:}%
\sum_{l=1}^{n_{\Lambda }}\Lambda _{l}^{\left( 1\right) }\mathbf{R}_{l}dV \\
-\int_{V_{00}}\frac{1}{2}\left( \mathbf{D:}\sum_{l=1}^{n_{\Lambda }}\Lambda
_{l}^{\left( 1\right) }\mathbf{R}_{l}\right) \mathbf{:}\frac{1}{2}\mathbf{r}%
^{T}\cdot \mathbf{r}dV-\int_{V_{00}}\frac{1}{2}\left( \mathbf{D:}%
\sum_{l=1}^{n_{\Lambda }}\Lambda _{l}^{\left( 2p\right) }\mathbf{R}%
_{l}\right) :\mbox{\boldmath$\varepsilon$}dV \\
+\int_{V_{00}}\frac{1}{2}2\left( \mathbf{D:}\sum_{l=1}^{n_{\Lambda }}\Lambda
_{l}^{\left( 1\right) }\mathbf{R}_{l}\right) \mathbf{:}\sum_{l=1}^{n_{%
\Lambda }}\Lambda _{l}^{\left( 2p\right) }\mathbf{R}_{l}dV=O\left( \eta
^{1+2p}\right)%
\end{array}%
\]%
\[
\begin{array}{l}
\mathcal{F}_{\Lambda }^{\left( 4p\right) }\left( \mathbf{E},T\right)
=-\int_{V_{00}}\left( \mbox{\boldmath$\Pi$}_{0i}^{\left( 2p\right) }+\frac{1%
}{2}\mathbf{D:}\frac{1}{2}\mathbf{r}^{T}\cdot \mathbf{r}\right) \mathbf{:}%
\sum_{l=1}^{n_{\Lambda }}\Lambda _{l}^{\left( 2p\right) }\mathbf{R}_{l}dV \\
-\int_{V_{00}}\frac{1}{2}\left( \mathbf{D:}\sum_{l=1}^{n_{\Lambda }}\Lambda
_{l}^{\left( 2p\right) }\mathbf{R}_{l}\right) \mathbf{:}\frac{1}{2}\mathbf{r}%
^{T}\cdot \mathbf{r}dV \\
+\int_{V_{00}}\frac{1}{2}\left( \mathbf{D:}\sum_{l=1}^{n_{\Lambda }}\Lambda
_{l}^{\left( 2p\right) }\mathbf{R}_{l}\right) \mathbf{:}\sum_{l=1}^{n_{%
\Lambda }}\Lambda _{l}^{\left( 2p\right) }\mathbf{R}_{l}dV=O\left( \eta
^{4p}\right)%
\end{array}%
\]%
In summary, one can write
\begin{equation}
\mathcal{F}\left( \mathbf{\bar{E}},T\right) =\mathcal{F}^{\left( b\right)
}\left( \mathbf{E},T\right) +\mathcal{F}_{\Lambda }^{\left( b\right) }\left(
\mathbf{E},T\right) +\sum_{i=4}^{n}O\left( \eta ^{s_{i}}\right)
\label{StrainEnergy-eta-p}
\end{equation}%
with%
\[
\begin{array}{c}
\mathcal{F}^{\left( b\right) }\left( \mathbf{E},T\right) =\mathcal{F}%
^{\left( 2\right) }\left( \mathbf{E},T\right) +\mathcal{F}^{\left(
1+2p\right) }\left( \mathbf{E},T\right) +\mathcal{F}^{\left( 4p\right)
}\left( \mathbf{E},T\right) \\
\mathcal{F}_{\Lambda }^{\left( b\right) }\left( \mathbf{E},T\right) =%
\mathcal{F}_{\Lambda }^{\left( 2\right) }\left( \mathbf{E},T\right) +%
\mathcal{F}_{\Lambda }^{\left( 1+2p\right) }\left( \mathbf{E},T\right) +%
\mathcal{F}_{\Lambda }^{\left( 4p\right) }\left( \mathbf{E},T\right)%
\end{array}%
\]%
A strain energy containing only the three dominant terms, viz. $\mathcal{F=F}%
^{\left( b\right) }\left( \mathbf{E},T\right) +\mathcal{F}_{\Lambda
}^{\left( b\right) }\left( \mathbf{E},T\right) $, can be directly derived
starting from the Hu-Washizu functional (\ref{Hu-Washizu-Gen}), provided
that the strain measure is chosen as follows:
\begin{equation}
\mathbf{E}\left( \mathbf{U}\right) \rightarrow \mathbf{E}^{\left( b\right)
}\left( \mathbf{U}\right) =\mbox{\boldmath$\varepsilon$}\left( \mathbf{U}%
\right) +\mbox{\boldmath$\chi$}\left( \mathbf{U}\right) ,\textrm{ \ \ }%
\mbox{\boldmath$\chi$}\left( \mathbf{U}\right) =\frac{1}{2}\mathbf{r}\left(
\mathbf{U}\right) ^{T}\cdot \mathbf{r}\left( \mathbf{U}\right)
\label{Eb-def}
\end{equation}%
The virtual works of the internal forces (see (\ref{Wdef}), (\ref%
{deltaEUdotdt}) and (\ref{Pi-orders})) read
\begin{equation}
\begin{array}{l}
-\mathcal{W}_{i}^{\left( 2\right) }\left( \mbox{\boldmath $\Pi$},\mathbf{U}%
,\delta \mathbf{U}\right) =\int_{V_{00}}\mbox{\boldmath$\Pi$}^{\left(
1\right) }\mathbf{:}\nabla ^{s}\left( \delta \mathbf{U}\right) \textrm{ }dV \\
-\mathcal{W}_{i}^{\left( 1+2p\right) }\left( \mbox{\boldmath $\Pi$},\mathbf{U%
},\delta \mathbf{U}\right) =\int_{V_{00}}\mbox{\boldmath$\Pi$}^{\left(
1\right) }\mathbf{:}\delta \mbox{\boldmath$\chi$}\left( \mathbf{U,}\delta
\mathbf{U}\right) dV+\int_{V_{00}}\mbox{\boldmath$\Pi$}^{\left( 2p\right) }%
\mathbf{:}\nabla ^{s}\left( \delta \mathbf{U}\right) \textrm{ }dV \\
-\mathcal{W}_{i}^{\left( 4p\right) }\left( \mbox{\boldmath $\Pi$},\mathbf{U}%
,\delta \mathbf{U}\right) =\int_{V_{00}}\mbox{\boldmath$\Pi$}^{\left(
2p\right) }\mathbf{:}\delta \mbox{\boldmath$\chi$}\left( \mathbf{U,}\delta
\mathbf{U}\right) dV%
\end{array}
\label{Wi2Wi1+2pWi4p}
\end{equation}%
where $\delta \mbox{\boldmath$\chi$}\left( \mathbf{U,}\delta \mathbf{U}%
\right) =\frac{1}{2}\left( \mathbf{r}\left( \delta \mathbf{U}\right)
^{T}\cdot \mathbf{\ r}\left( \mathbf{U}\right) +\mathbf{r}\left( \mathbf{U}%
\right) ^{T}\cdot \mathbf{r}\left( \delta \mathbf{U}\right) \right) $.

\subsection{Definition of "strain-rotation domains" and of approximated
expressions of the strain energy}

Eq. (\ref{StrainEnergy-eta-p}) reveals the dominant energy terms. However,
it is also interesting to find the \emph{strain-rotation domains} (depending
on $\eta $ and $p$)\ where the approximation of the exact strain energy by a
simplified expression is acceptable. \emph{Three} domains are of special
interest:

\begin{description}
\item[(a)] The domain $\mathbb{H}_{a}$ where
\begin{equation}
\mathcal{F}\left( \mathbf{\bar{E}},T\right) \simeq \mathcal{F}^{\left(
2\right) }\left( \mathbf{E},T\right) +\mathcal{F}_{\Lambda }^{\left(
2\right) }\left( \mathbf{E},T\right) =O\left( \eta ^{2}\right)  \label{Wi_a}
\end{equation}%
Eq. (\ref{Wi_a}) represents the situation usually called \emph{of small
perturbations}. The corresponding virtual work of internal forces reduces to
$\mathcal{W}_{i}^{\left( 2\right) }\left( \mbox{\boldmath $\Pi$},\mathbf{U}%
,\delta \mathbf{U}\right) $; see Eq. (\ref{Wi2Wi1+2pWi4p}). As it is seen
hereafter, a suitable name for $\mathbb{H}_{a}$ is \emph{\ domain of small
strains and relatively small squared rotations (the ratio }$\eta^{2p-1}$
\emph{is small)}.

\item[(b)] The domain $\mathbb{H}_{b}$ where
\begin{equation}
\mathcal{F}\left( \mathbf{\bar{E}},T\right) \simeq \mathcal{F}^{\left(
b\right) }\left( \mathbf{E},T\right) +\mathcal{F}_{\Lambda }^{\left(
b\right) }\left( \mathbf{E},T\right) =O\left( \eta ^{2}\right) +O\left( \eta
^{1+2p}\right) +O\left( \eta ^{4p}\right)  \label{Wi_b}
\end{equation}%
\emph{i.e. }three energy terms are considered together, of order $2$, $1+2p$
and $4p$, after the discussion of the previous Section. The virtual work
reduces to
\[
\mathcal{W}_{i}^{\left( b\right) }\left( \mbox{\boldmath
$\Pi$},\mathbf{U},\delta \mathbf{U}\right) =\mathcal{W}_{i}^{\left( 2\right)
}\left( \mbox{\boldmath $\Pi$},\mathbf{U},\delta \mathbf{U}\right) +\mathcal{%
W}_{i}^{\left( 1+2p\right) }\left( \mbox{\boldmath $\Pi$},\mathbf{U},\delta
\mathbf{U}\right) +\mathcal{W}_{i}^{\left( 4p\right) }\left(
\mbox{\boldmath
$\Pi$},\mathbf{U},\delta \mathbf{U}\right)
\]
This region is the \emph{domain of small strains and relatively moderate
squared rotations (the ratio }$\eta^{2p-1}\simeq 1 $\emph{)}, called \emph{%
moderate rotation domain} for brevity.

\item[(c)] The domain $\mathbb{H}_{c}$\ where
\begin{equation}
\mathcal{F}\left( \mathbf{\bar{E}},T\right) \simeq \mathcal{F}^{\left(
4p\right) }\left( \mathbf{E},T\right) +\mathcal{F}_{\Lambda }^{\left(
4p\right) }\left( \mathbf{E},T\right) =O\left( \eta ^{4p}\right)
\label{Wi_c}
\end{equation}%
and $\mathcal{W}_{i}\left( \mbox{\boldmath $\Pi$},\mathbf{U},\delta \mathbf{U%
}\right) \rightarrow \mathcal{W}_{i}^{\left( 4p\right) }\left( %
\mbox{\boldmath $\Pi$},\mathbf{U},\delta \mathbf{U}\right) =O\left( \eta
^{4p}\right) ,$ \emph{i.e.} the energy term of order $4p$ is larger than all
the others (\emph{domain of small strains and relatively large squared
rotations, i.e. the ratio }$\eta^{2p-1} $\emph{\ is large}).
\end{description}

In order to draw these domains, let us introduce a small positive number $%
\zeta \ll 1$, for instance $\zeta =0.01$. Then, the situations $p\leq \frac{1%
}{2}$ and $p\geq \frac{1}{2}$ can be distinguished, since for $p\leq \frac{1%
}{2}$ the dominant term is of order $4p$ and in the other case the term of
order $2$ is the largest (Figure \ref{Exponents}). When $p\geq \frac{1}{2}$,
look for the conditions on $\eta $ and $p$ \emph{such that all energy terms
are small compared with the second order term}. The following system of
inequalities defines $\mathbb{H}_{a}$:
\begin{equation}
\mathbb{H}_{a}:\textrm{ \ \ \ \ \ }\frac{\eta ^{s_{i}}}{\eta
^{2}}\leq \zeta \textrm{ \ \ \ \ \ \ \ \ \ \ \ for all }i=2,n
\label{condH1}
\end{equation}

\begin{figure}[tbp]
\begin{center}
\includegraphics[width=14cm]{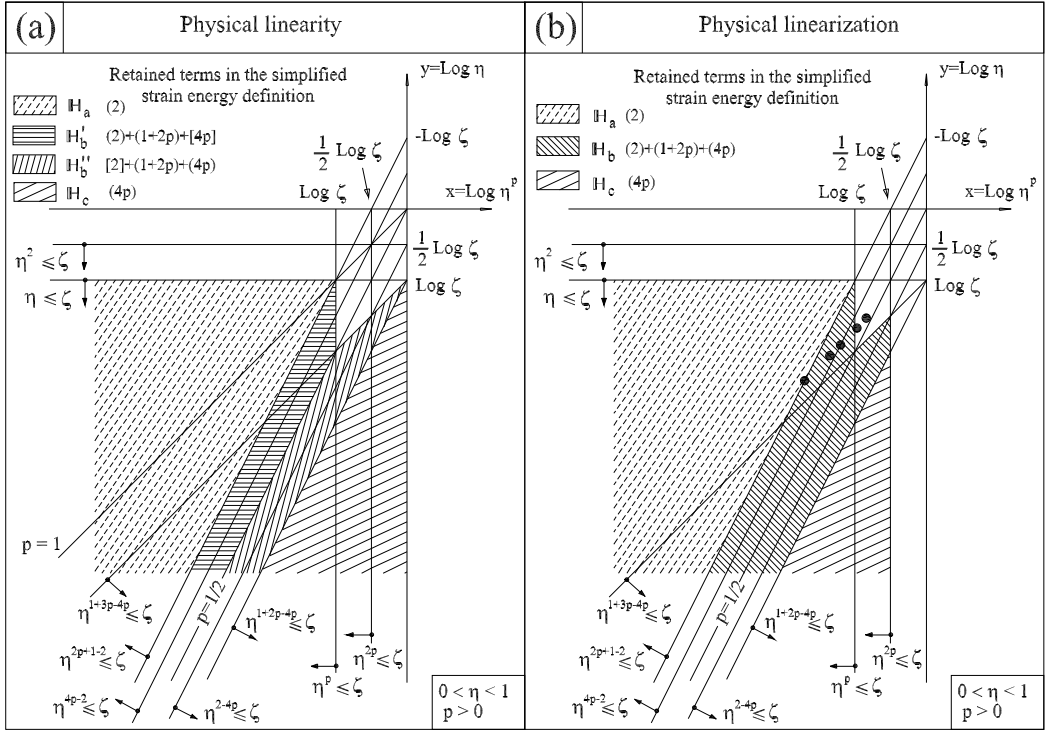}
\end{center}
\caption{(a) Strain-rotation domains for a physically linear material. (b)
Strain-rotation domains under the assumption of physical linearization. The
sets $\mathbb{H}_{b}^{\prime }$ and $\mathbb{H}_{b}^{\prime \prime }$ are
merged into the set $\mathbb{H}_{b}$. The "strain-rotation points" represent
the deformed configurations of the structure of the numerical example
discussed in Section 5.3 . In both Figures (a) and (b), the dominant terms
of the strain energy are indicated. }
\label{Str-rot-domains-gen}
\end{figure}

Every condition enjoys a simple log-log representation, $\left( x=\log \eta
^{p}, y=\log \eta \right) $, since $\eta ^{s_{i}-2}=\eta
^{c_{i}p+d_{i}}\rightarrow \log \eta ^{s_{i}-2}=\log\eta ^{c_{i}p+d_{i}}
=c_{i}x+d_{i}y\leq \log \zeta$, where $c_{i}$ and $d_{i}$ are integer
numbers. For instance, when $s_{i}=s_{2}=1+2p\leq \log \zeta$, one has $%
2x-y\leq \log \zeta$, which relates $y=\log \eta $ with $x=\log \eta ^{p}$.
This condition is associated with a line which is the bottom limit of the
domain $\mathbb{H}_{a}$ ; see Figure \ref{Str-rot-domains-gen}a. If $s_{i}=3$%
, one has $c_{i}=0,$ $d_{i}=1$, leading to the condition $y=\log \eta \leq
\log \zeta $, associated with an horizontal line as shown in same Figure.
This means that the ratio between the energy term of order 3 and that of
order 2 is smaller or equal to $\zeta $, provided that $\eta \leq \zeta $.
Still assuming $p\geq \frac{1}{2}$ , one can define the conditions such that
all the energy terms are small compared with that of order $2$, except those
of order $1+2p$ and $4p$:
\begin{equation}
\mathbb{H}_{b}^{\prime }:\frac{\eta ^{s_{i}}}{\eta ^{2}}\leq \zeta
\textrm{\ \ \ }\forall i=4,n\textrm{ , \ }\zeta < \frac{\eta
^{1+2p}}{\eta ^{2}}:=\rho _{1}\leq 1,\textrm{ \ }\zeta \lesseqgtr
\frac{\eta ^{4p}}{\eta ^{2}}:=\rho _{2}\leq 1  \label{Hb-rho12}
\end{equation}%
The third relationship shows that inside this domain, the ratio between the
term of order $4p$ and that of order $2$ may be either small, equal or
greater than $\zeta $. When $p\leq \frac{1}{2}$, the conditions such that
all the energy terms are small compared with that of order $4p$ read:
\begin{equation}
\mathbb{H}_{c}:\textrm{ \ \ \ \ \ }\frac{\eta ^{s_{i}}}{\eta
^{4p}}\leq \zeta \textrm{ \ \ \ \ \ \ \ \ \ \ \ for all
}i=1,2\textrm{ and }4,n
\end{equation}%
This set is associated with the approximated energy (\ref{Wi_c}). If the
terms of order $1+2p$ and $2$ are not small, one has
\begin{equation}
\mathbb{H}_{b}^{\prime \prime }:\frac{\eta ^{s_{i}}}{\eta ^{4p}}\leq \zeta
\textrm{ \ }\forall i=4,n\textrm{ , \ }\zeta < \frac{\eta ^{1+2p}}{\eta ^{4p}}=%
\frac{1}{\rho _{1}}\leq 1,\textrm{ \ }\zeta \lesseqgtr \frac{\eta
^{2}}{\eta ^{4p}}=\frac{1}{\rho _{2}}\leq 1  \label{Hb-rho12bis}
\end{equation}%
The set $\mathbb{H}_{b}=\mathbb{H}_{b}^{\prime }\cup \mathbb{H}_{b}^{\prime
\prime }$ is associated to the approximated energy (\ref{Wi_b}). Inside this
domain, three energy terms (of order $2$, $1+2p$ and $4p$) are retained. The
ratios $\rho _{1}$ and $\rho _{2}$ and their inverses (Eqs. (\ref{Hb-rho12}%
), (\ref{Hb-rho12bis})) give an estimation of the relative amplitude of
these dominant terms.

Observe that $\mathbb{H}_{c}$ is extended to large rotations ($\eta ^{4p}$
close to $1$) because we have compared the energy terms of the first four
rows of Table \ref{TabOrders} characterizing the \emph{physically linear }%
and isotropic constitutive law (\ref{GenCostLawIso}). However, it is of
interest here to determine the conditions under which the physical
linearization is a reasonable approximation of any real material behaviour.
It is expected that this is the case when the rotations are not too large.
In order to derive sharp bounds on the rotations, the quadratic constitutive
law (\ref{constQuad}) has to be considered, and the associated strain energy
must be computed. This leads to (see also (\ref{Pi-phys.lin}))
\begin{equation}
\begin{array}{l}
\mbox{\boldmath$\Pi$}^{\left( phys.non-lin.\right) }=\mbox{\boldmath$\Pi$}%
^{\left( phys.lin.\right) }+\left( O\left( \eta \right) +O\left( \eta
^{2p}\right) +O\left( \eta ^{1+p}\right) +O\left( \eta ^{2}\right) \right)
^{2} \\
\mathcal{F}=O\left( \eta ^{2}\right) +O\left( \eta ^{1+2p}\right) +O\left(
\eta ^{4p}\right) +\sum_{i=4}^{n_{nlin}}O\left( \eta ^{s_{i}}\right)%
\end{array}
\label{StrainEnergy-eta-p-Nlin}
\end{equation}%
i.e. some energy terms have to be added to those of the physically linear
case, as indicated in Table \ref{TabOrders}. Then, the same conditions as in
(\ref{condH1})-(\ref{Hb-rho12bis}) are considered, with $n$ substituted by $%
n_{nlin}$, in order to compute the conditions under which all the physically
non-linear terms are small compared with the dominant ones, either of order
2 or 4p. These conditions define the regions depicted in Figure \ref%
{Str-rot-domains-gen}b: they are the \emph{strain-rotation} \emph{domains}
where the physical linearization is admissible. An important difference with
respect to Figure \ref{Str-rot-domains-gen}a is that $\mathbb{H}_{b}$ and $%
\mathbb{H}_{c}$ are bounded by the vertical line $x\leq \frac{1}{2}\log
\zeta $, equivalent to $\eta ^{2p}\leq \zeta $, viz. the squared rotations,
not only the strains, must be small. It can be proven that this limitation
derives from the condition $\frac{\eta ^{6p}}{\eta ^{4p}}\leq \zeta $,
imposing that the term of order $6p$ associated with the physically
non-linear law remains small compared with the term of order $4p$, dominant
for $p\leq 1/2$.

The condition of having small Green-Lagrange strain reads: $\left\Vert
\mathbf{E}\right\Vert =O\left( \eta \right) +O\left( \eta ^{2p}\right)
+O\left( \eta ^{1+p}\right) +O\left( \eta ^{2}\right) \leq \zeta \ll 1$. It
easy to identify in Figure \ref{Str-rot-domains-gen} the domains in the
strain-rotation plane where the first, the second and the fourth term of $%
\mathbf{E}$ are less or equal to $\zeta $. It can be also proven that the
condition of having a small third term, i.e. $\eta ^{1+p}\leq \zeta $, is
fulfilled in the three sets $\mathbb{H}_{a}$, $\mathbb{H}_{b}$ and $\mathbb{H%
}_{c}$.

Other approximations retaining at least four energy terms are possible.
However, (a), (b) and (c) define situations often discussed in the
literature and for this reason the present analysis is restricted to them.
Case (b), collecting three terms, is formally more complex than the others
and is discussed in detail hereinafter.


\subsection{Numerical example: "exact" values of $\protect\eta $ and $p$ and
comparison of the exact and "simplified" energy maps}

Consider a problem of plain stress elasticity ($n_{\dim}$, the dimension of
the problem, is equal to $2$). The corresponding conditions on the
Piola-Kirchhoff tensor are $\Pi _{XZ}=\Pi _{YZ}=\Pi _{ZZ}=0$. A St.Venant-
Kirchhoff material is chosen with $\mbox{\boldmath$\Pi$}_{0i}=\mathbf{0}$
and $T=T_{0i}$. Then
\begin{equation}
\begin{array}{l}
\Psi =\Psi \left( \mathbf{E}\right) =\frac{1}{2}\mathbf{E}:\left( \tilde{%
\lambda}\mathbf{1}\otimes \mathbf{1+}2\mu \mathbf{I}\right) \mathbf{:E} \\
\mbox{\boldmath$\Pi$}=\mbox{\boldmath$\Pi$}\left( \mathbf{E}\right) =2\mu
\textrm{ }\mathbf{E+}\tilde{\lambda}\textrm{ }tr\left( \mathbf{E}\right) \mathbf{%
1=}\left[
\begin{array}{cc}
\Pi _{XX} & \Pi _{XY} \\
\Pi _{XY} & \Pi _{YY}%
\end{array}%
\right]%
\end{array}
\label{Const-Law_PlainStress}
\end{equation}%
where $\tilde{\lambda}=\frac{E\nu }{1-\nu ^{2}} \neq \lambda$ due to plain
stress assumption and
\begin{equation}
\mathbf{E}\left( \mathbf{U}\right) =\left[
\begin{array}{cc}
\frac{\partial \mathrm{u}}{\partial X} & \frac{1}{2}\left( \frac{\partial
\mathrm{u}}{\partial Y}+\frac{\partial \mathrm{v}}{\partial X}\right) \\
\frac{1}{2}\left( \frac{\partial \mathrm{u}}{\partial Y}+\frac{\partial
\mathrm{v}}{\partial X}\right) & \frac{\partial \mathrm{v}}{\partial Y}%
\end{array}%
\right] +\frac{1}{2}\left[
\begin{array}{cc}
\frac{\partial \mathrm{u}}{\partial X}\frac{\partial \mathrm{u}}{\partial X}+%
\frac{\partial \mathrm{v}}{\partial X}\frac{\partial \mathrm{v}}{\partial X}
& \frac{\partial \mathrm{u}}{\partial X}\frac{\partial \mathrm{u}}{\partial Y%
}+\frac{\partial \mathrm{v}}{\partial X}\frac{\partial \mathrm{v}}{\partial Y%
} \\
\frac{\partial \mathrm{u}}{\partial X}\frac{\partial \mathrm{u}}{\partial Y}+%
\frac{\partial \mathrm{v}}{\partial X}\frac{\partial \mathrm{v}}{\partial Y}
& \frac{\partial \mathrm{u}}{\partial Y}\frac{\partial \mathrm{u}}{\partial Y%
}+\frac{\partial \mathrm{v}}{\partial Y}\frac{\partial \mathrm{v}}{\partial Y%
}%
\end{array}%
\right]  \label{StrainMeas2D}
\end{equation}%
is the 2D Green-Lagrange strain with $\mathbf{U=}\left[ \mathrm{u,v}\right]
^{T}.\ $The relevant H-W functional is given by (\ref{Hu-Washizu-Gen}),
without the term depending on $\Lambda _{k}^{\ast }$ and with $\Psi $ given
by (\ref{Const-Law_PlainStress}-1). For a static problem with the external
volume force $\mathbf{f}_{0}=\left[ f_{X,0},f_{Y,0}\right] ^{T}$ and surface
force $\mathbf{g}_{0}=\left[ g_{X,0},g_{Y,0}\right] ^{T}$, the weak form of
the equilibrium equation reads:
\begin{equation}
\begin{array}{l}
\mathbf{R}\left( \mathbf{U},\delta \mathbf{U}\right) =-\int_{V_{00}}%
\mbox{\boldmath$\Pi$}\mathbf{\left( \mathbf{E}\left( U\right) \right) }%
:\delta \mathbf{E}\left( \mathbf{U,}\delta \mathbf{U}\right) \textrm{ }dV \\
+\int_{V_{00}}\mathbf{f}_{0}\mathbf{\cdot }\delta \mathbf{U}\textrm{ }%
dA+\int_{\partial V_{00,\sigma }}\mathbf{g}_{0}\mathbf{\cdot }\delta \mathbf{%
\ U}\textrm{ }dA=0\textrm{ \ \ \ \ \ \ for all }\delta \mathbf{U\in }\mathbb{V}%
\end{array}
\label{Rudu}
\end{equation}%
with $\delta \mathbf{U=}\left[ \delta \mathrm{u},\delta \mathrm{v}\right]
^{T}$ and $\delta \mathbf{E}\left( \mathbf{U,}\delta \mathbf{U}\right) $
defined by the right-hand side of Eq. (\ref{deltaEbar}). Eq. (\ref{Rudu}) is
a non-linear partial differential equation, which can be discretized by a
standard finite element method. The standard Newton algorithm has been
implemented in the code FreeFEM++ \citep{FreeFEM2003}: given the
displacement $\mathbf{U}_{n}$ at iteration $n$, the increment $\mathbf{w}$
is computed by
\[
\mathbf{w\in }\mathbb{U}\textrm{ such that \ }\mathbf{R}\left( \mathbf{U}%
_{n},\delta \mathbf{U}\right) +\delta \mathbf{R}\left( \mathbf{U}_{n},\delta
\mathbf{U,w}\right) \simeq \mathbf{0}\textrm{\ \ \ \ \ for all }\delta \mathbf{%
U\in }\mathbb{V}
\]%
with
\[
\begin{array}{l}
\delta \mathbf{R}\left( \mathbf{U}_{n},\delta \mathbf{U,w}\right)
=-\int_{\Omega }\mbox{\boldmath$\Pi$}\left( \delta \mathbf{E}\left( \mathbf{U%
}_{n}\mathbf{,w}\right) \right) :\delta \mathbf{E}\left( \mathbf{U}_{n}%
\mathbf{,}\delta \mathbf{U}\right) \textrm{ }d\Omega \\
\textrm{ \ \ \ \ \ \ \ \ \ \ \ \ \ \ \ \ \ \ \ \ \ \ }-\int_{\Omega }%
\mbox{\boldmath$\Pi$}\left( \mathbf{E}\left( \mathbf{U}_{n}\right) \right)
:\delta ^{2}\mathbf{E}\left( \delta \mathbf{U,w}\right) \textrm{ }d\Omega \\
\delta ^{2}\mathbf{E}\left( \delta \mathbf{U,w}\right) =\frac{1}{2}\nabla
\left( \delta \mathbf{U}\right) ^{T}\cdot \nabla \left( \mathbf{w}\right) +%
\frac{1}{2}\nabla \left( \mathbf{w}\right) ^{T}\cdot \nabla \left( \delta
\mathbf{U}\right)%
\end{array}%
\]%
Set $\mathbf{U}_{n+1}=\mathbf{U}_{n}+\mathbf{w}$. Repeat until $\frac{%
\left\Vert \mathbf{w}\right\Vert _{L_{2}\left( V_{00}\right) }}{\left\Vert
\mathbf{U}_{n}\right\Vert _{L_{2}\left( V_{00}\right) }}$ is small enough.

The structure examined in this example is a parallelepiped beam lying in the
$XY$-plane, the dimensions are $b=h=1cm$ and $L_{00}=50cm$. The beam is
clamped at both ends and the volume load is $\left[ f_{X,0},f_{Y,0}\right] =%
\left[ 0,-\left\vert f_{Y,0}\right\vert \right] $ $daN/cm^{3}$. Exploiting
the symmetry of the problem, only a half-beam is meshed, with the boundary
conditions $\left[ \mathrm{u}=0, \mathrm{v}=0\right] $ for $Y\in \left[
-h/2,h/2\right] $ and $X=0$ and $\mathrm{u}=0$ for $Y\in \left[ -h/2,h/2%
\right] $ and $X=L_{00}/2$. The material parameters (steel) read $E=2100000$
$daN/cm^{2},$ \ $\nu =0.28,$ \ \ $\mu =820312.5$ $daN/cm^{2},$ $\tilde{%
\lambda}=638020.8$ $daN/cm^{2}$. A mesh of triangular elements has been
chosen, with two elements inside every cell of a regular grid of 15x375
squares. The finite element space is of P1 type. The numerical simulations
give the results collected in Tables \ref{TabRes-num1} and \ref{TabRes-num2}%
, where $\mathrm{v}_{\max }=\mathrm{v} \left( X=L_{00}/2\right)$, $x=\log
_{10}\left( \eta ^{p}\right) ,$\ $y=\log _{10}\left( \eta \right) ,$ $p=%
\frac{x}{y}$ , $\rho _{2}=\frac{\eta ^{4p}}{\eta ^{2}}$, $\rho _{1}=\frac{%
\eta ^{1+2p}}{\eta ^{2}} $ and%
\begin{equation}
\begin{array}{l}
\eta =\frac{1}{\sqrt{V_{00}}}\left\Vert \mbox{\boldmath$\varepsilon$}%
\right\Vert _{L_{2}\left( V_{00}\right) }=\sqrt{\frac{\int_{V_{00}}\left(
\varepsilon _{XX}^{2}+\varepsilon _{YY}^{2}+2\varepsilon _{XY}^{2}\right) dV%
}{bhL_{00}/2}},\textrm{ \ \ \ }\eta
_{xx}=\sqrt{\frac{\int_{V_{00}}\varepsilon
_{XX}^{2}dV}{bhL_{00}/2}} \\
\eta ^{p}=\frac{1}{\sqrt{2V_{00}}}\left\Vert \mathbf{r}\right\Vert
_{L_{2}\left( V_{00}\right) }=\sqrt{\frac{\int_{V_{00}}r_{XY}^{2}dV}{%
bhL_{00}/2}}%
\end{array}
\label{ex1-num}
\end{equation}

\begin{figure}[tbp]
\begin{center}
\includegraphics[width=13cm]{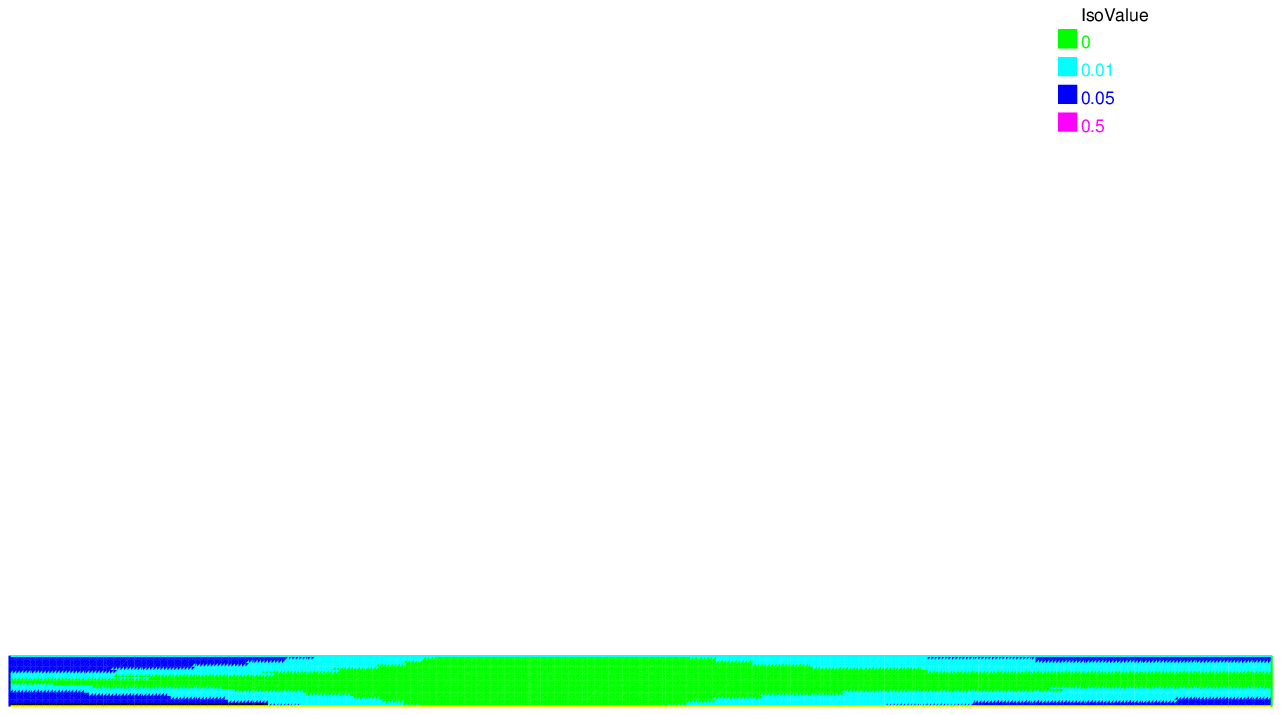}
\end{center}
\caption{Volume vertical force $f_{Y0}=-1$ $daN/cm^{3}$. Map of the strain
energy density $\Psi \left( \mathbf{E}\right) $ $[daN cm \ cm^{-3}]$, with $%
\mathbf{E=E}\left( \mathbf{U}\right) $.}
\label{Fig-en-map-fvol20}
\end{figure}

\begin{figure}[tbp]
\begin{center}
\includegraphics[width=13cm]{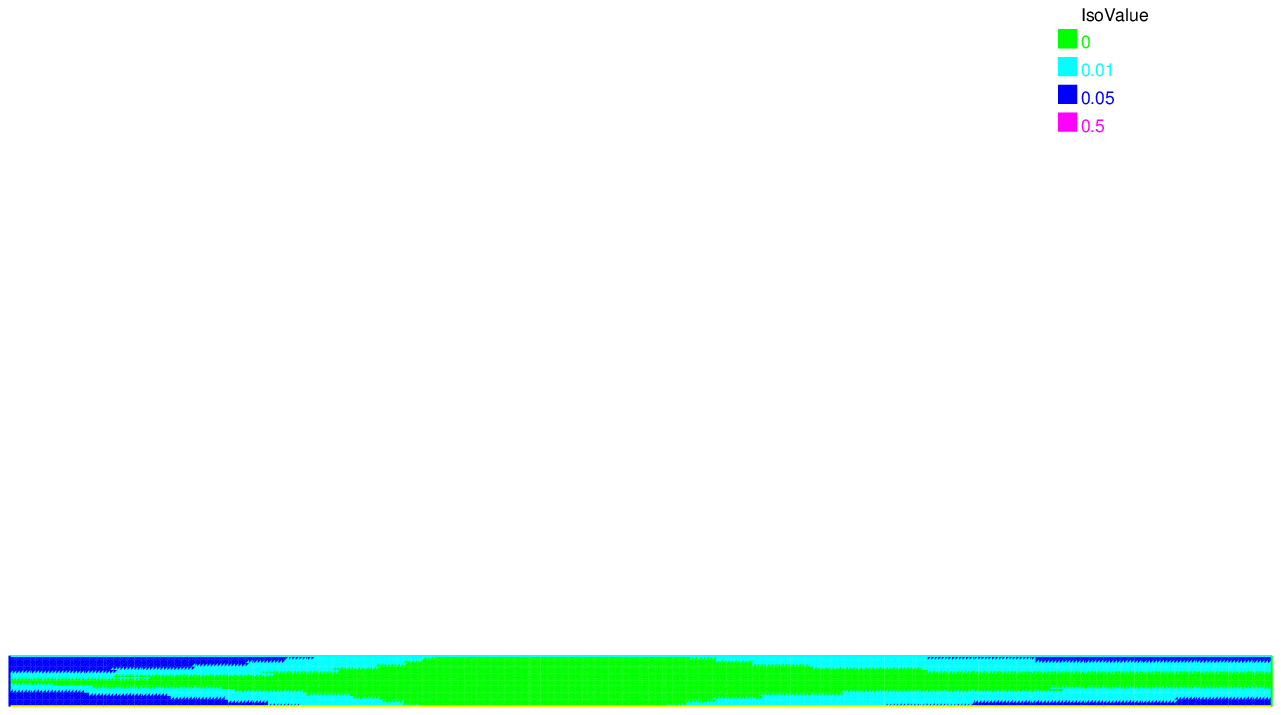}
\end{center}
\caption{Volume vertical force $f_{Y0}=-1$ $daN/cm^{3}$. Map of the \emph{%
approximated} strain energy density $\Psi \left( \mathbf{E}\right) $ $[daN
cm \ cm^{-3}]$, with $\mathbf{E=E}^{\left( b\right) }\left( \mathbf{U}%
\right) $). }
\label{Fig-en-mapAppr}
\end{figure}

The last column of Table \ref{TabRes-num2} provides a global estimation of
the energy error between the exact and the approximated energies $\mathcal{F=%
}\int \Psi \left( \mathbf{E}\right) dV$ and $\mathcal{F}^{\left( b\right) }%
\mathcal{=}\int \Psi \left( \mathbf{E}^{\left( b\right) }\right) dV$.
Observe the maximum absolute value of the strain $\varepsilon _{xx}$,
reported in the third column of Table \ref{TabRes-num1}: in the first four
cases it is less than $0.002,$ which is the limit elastic strain for a steel
having yielding stress approximately equal to $4200$ $daN/cm^{2}=420$ $MPa$.
For these situations, the material is truly physically linear. Conversely,
when $f_{Y0}=-6$ $daN/cm^{3}$ , see the last row of Table \ref{TabRes-num1},
the maximum absolute value of $\varepsilon _{xx}$ is larger than $0.002$.
Hence, the physical linearity is truly fulfilled only for steels having a
greater yielding stress. The $x-y$ coordinates of the \emph{strain-rotation
points} associated with each $f_{Y0}$ value are reported in the seventh and
eighth columns of Table \ref{TabRes-num1}. The corresponding graphical
representation is given in Figure \ref{Str-rot-domains-gen}-b, depicted
assuming $\zeta =0.01$. The map $\Psi \left( \mathbf{E}\right) $ , \emph{viz.%
} the strain energy density at the static equilibrium $\mathbf{U}_{0}$ is
shown in Figure \ref{Fig-en-map-fvol20}. The energy density corresponding to
the approximated strain energy $\Psi \left( \mathbf{E}^{\left( b\right)
}\right) $ is given in Figure \ref{Fig-en-mapAppr}. The relative difference
of energy density between the two cases is illustrated in Figure \ref%
{Fig-DiffEnergy}. With a surface load $\mathbf{g}_{0}=\left[ g_{X,0},g_{Y,0}%
\right] =\left[ 0,-\left\vert g_{Y,0}\right\vert \right] $ $daN/cm^{2}$ and $%
\mathbf{f}_{0}= \mathbf{0}$, the results are similar, as one can see from
Tables \ref{TabRes-num3}, \ref{TabRes-num4} and Figure \ref%
{Fig-Diff-Energ-surf1}. This confirms that $\eta $ and $p$ are not too
sensitive to the load distribution.

\section{Dissipative stress, Hu-Washizu functional and damping
pseudo-potential with stress constraints}

The linear law (\ref{GenCostLaw}) can be generalized by adding to $%
\mbox{\boldmath$\Pi$}^{nd}$ a dissipative term:
\begin{equation}
\mbox{\boldmath$\Pi$}=\mbox{\boldmath$\Pi$}^{nd}+\mbox{\boldmath$\Pi$}^{d}
\label{Pind+Pid}
\end{equation}%
The index $d$ indicates the dissipative part of the stress. Let
\begin{equation}
\begin{array}{l}
J^{d}\left( \mbox{\boldmath$\Pi$}^{d^{\ast }},\frac{d\mathbf{\tilde{E}}}{dt}%
^{\ast },\lambda _{k}^{\ast };\frac{d\mathbf{\bar{E}}}{dt}\right)
=\int_{V_{00}}\phi \left( \frac{d\mathbf{\tilde{E}}^{\ast }}{dt}\right) dV
\\
-\int_{V_{00}}\mbox{\boldmath$\Pi$}^{d^{\ast }}\mathbf{:}\left( \frac{d%
\mathbf{\tilde{E}}}{dt}^{\ast }-\frac{d\mathbf{\bar{E}}}{dt}\right)
dV-\int_{V_{00}}\sum_{k=1}^{n_{\Lambda }}\lambda _{k}^{\ast }\mathbf{R}_{k}:%
\mbox{\boldmath$\Pi$}^{d^{\ast }}dV%
\end{array}
\label{FuncDiss}
\end{equation}%
be the functional associated with a dissipative stress. It is assumed that
it depends on the dissipative stress $\mbox{\boldmath$\Pi$}^{d^{\ast }}\in
\mathbb{T}$, the generic strain flow $\frac{d\mathbf{\tilde{E}}}{dt}^{\ast
}\in \mathbb{T}$ and $\lambda _{k}^{\ast }\in \mathbb{F}$, i.e. the Lagrange
multipliers associated with the constraints imposed on $\mbox{\boldmath$\Pi$}%
^{d^{\ast }}$. Moreover, the \emph{actual} strain flow $\frac{d\mathbf{\bar{E%
}}}{dt}$ plays the role of additional parameter: for this reason it is
separated from the main variables by the semi-colon "$;$", instead of the
comma. The actual strain flow is computed from the problem associated with
the Hu-Washizu functional defined in Eq. (\ref{Hu-Washizu-Gen-diss}). The
scalar non-negative and convex function $\phi $ is called \emph{%
pseudo-potential} or \emph{dissipation potential.} A classical definition is
$\phi =\frac{1}{2}\frac{d\mathbf{\tilde{E}}}{dt}:\mathbf{F}:\frac{d\mathbf{\
\tilde{E}}}{dt}$ , i.e. a quadratic function. For an isotropic material, one
has $\mathbf{F=}\lambda _{d}\mathbf{1}\otimes \mathbf{1+}2\mu _{d}\mathbf{I}$
, where $\lambda _{d}$ and $\mu _{d}$ are analogous to the Lam\'{e}
constants $\lambda$ and $\mu$. The stationarity conditions imposed on (\ref%
{FuncDiss}) lead to following strong form expressions
\begin{equation}
\begin{array}{l}
a.\textrm{ \ \ }\forall k=1,n_{\Lambda }\textrm{ \ \ \ }\mathbf{R}_{k}:%
\mbox{\boldmath$\Pi$}^{d}=0\textrm{ \ \ \ \ \ \ \ \ \ \ \ \ \ \ \ \
\ \ \ \ in
}V_{00} \\
b.\textrm{ \ \ }\frac{d\mathbf{\tilde{E}}}{dt}=\frac{d\mathbf{\bar{E}}}{dt}%
-\sum_{k=1}^{n_{\Lambda }}\lambda _{k}\mathbf{R}_{k}\textrm{\ \ \ \
\ \ \ \ \
\ \ \ \ \ \ \ \ \ \ \ \ \ \ \ \ in }V_{00} \\
c.\textrm{ \ \ }\mbox{\boldmath$\Pi$}^{d}:=\left. \frac{\partial
\phi \left(
\frac{d\mathbf{\tilde{E}}^{\ast }}{dt}\right) }{\partial \frac{d\mathbf{%
\tilde{E}}^{\ast }}{dt}}\right\vert _{\frac{d\mathbf{\tilde{E}}}{dt}^{\ast }=%
\frac{d\mathbf{\tilde{E}}}{dt}}=\mathbf{F}:\frac{d\mathbf{\tilde{E}}}{dt}%
\textrm{\ \ \ \ \ \ \ \ \ \ in }V_{00}%
\end{array}
\label{StrongDiss1}
\end{equation}%
The first equation indicates the stress constraints imposed on $%
\mbox{\boldmath$\Pi$}^{d}$, the second one shows that the strain flow
governing the dissipative behaviour is not equal to the time derivative of
the strain when $\lambda _{k}\neq 0$. Finally, the third equation is the
constitutive law for the dissipative stress, obtained from the
pseudo-potential $\phi$. The constraints on the dissipative stress read
\begin{equation}
\mathbf{R}_{k}:\mbox{\boldmath$\Pi$}^{d}\mathbf{=R}_{k}:\left[ \mathbf{F}%
:\left( \frac{d\mathbf{E}}{dt}-\sum_{l=1}^{n_{\Lambda }}\frac{d\Lambda _{l}}{%
dt}\mathbf{R}_{l}-\sum_{l=1}^{n_{\Lambda }}\lambda _{l}\mathbf{R}_{l}\right) %
\right] =0
\end{equation}%
where $\Lambda _{l},$ $l=1,n_{\Lambda }$ are known from the analysis of the
non-dissipative part of the stress. Following the procedure indicated in the
Appendix,\ one can prove that $\mbox{\boldmath$\lambda$}=\left( \lambda
_{k}\right) $ , $k=1,n_{\Lambda }$ is the solution of a linear Gram system.
Therefore
\begin{equation}
\mbox{\boldmath$\lambda$}\mathbf{=}\mbox{\boldmath$\lambda$}^{\left(
1\right) }\mathbf{+}\mbox{\boldmath$\lambda$}^{\left( 2p\right) }\mathbf{+}%
\mbox{\boldmath$\lambda$}^{\left( 1+p\right) }\mathbf{+}%
\mbox{\boldmath$
\lambda$}^{\left( 2\right) }=O\left( \eta \right) +O\left( \eta ^{2p}\right)
+O\left( \eta ^{1+p}\right) +O\left( \eta ^{2}\right)  \label{lamDiss-orders}
\end{equation}
Using Eqs. (\ref{StrongDiss1}) and (\ref{lamDiss-orders}), one obtains
\begin{equation}
\mbox{\boldmath$\Pi$}^{d}=\mbox{\boldmath$\Pi$}^{d\left( 1\right) }+%
\mbox{\boldmath$\Pi$}^{d\left( 2p\right) }+\mbox{\boldmath$\Pi$}^{d\left(
1+p\right) }+\mbox{\boldmath$\Pi$}^{d\left( 2\right) }  \label{Pidiss}
\end{equation}%
where the following four terms of different orders are distinguished:
\begin{equation}
\begin{array}{l}
\mbox{\boldmath$\Pi$}^{d\left( 1\right) }=\mathbf{F}:\left( \frac{d%
\mbox{\boldmath$\varepsilon$}}{dt}-\sum_{k=1}^{n_{\Lambda }}\frac{d\Lambda
_{k}^{\left( 1\right) }}{dt}\mathbf{R}_{k}-\sum_{k=1}^{n_{\Lambda }}\lambda
_{k}^{\left( 1\right) }\mathbf{R}_{k}\right) \\
\mbox{\boldmath$\Pi$}^{d\left( 2p\right) }=\mathbf{F}:\left( \frac{1}{2}%
\frac{d\mathbf{r}^{T}}{dt}\cdot \mathbf{r+}\frac{1}{2}\mathbf{r}^{T}\cdot
\frac{d\mathbf{r}}{dt}-\sum_{k=1}^{n_{\Lambda }}\frac{d\Lambda _{k}^{\left(
2p\right) }}{dt}\mathbf{R}_{k}-\sum_{k=1}^{n_{\Lambda }}\lambda _{k}^{\left(
2p\right) }\mathbf{R}_{k}\right) \\
\mbox{\boldmath$\Pi$}^{d\left( 1+p\right) }=\mathbf{F}:\left( \frac{1}{2}%
\frac{d\mathbf{r}^{T}}{dt}\cdot \mbox{\boldmath
$\varepsilon$ }+\frac{1}{2}\mathbf{r}^{T}\cdot \frac{d%
\mbox{\boldmath
$\varepsilon$ }}{dt}+\frac{1}{2}\mbox{\boldmath $\varepsilon$ }^{T}\cdot
\frac{d\mathbf{r}}{dt}+\frac{1}{2}\frac{d\mbox{\boldmath $\varepsilon$ }^{T}%
}{dt}\cdot \mathbf{r}\right) \\
\textrm{ \ \ \ \ \ \ \ \ \ \ \ }-\mathbf{F}:\left( \sum_{k=1}^{n_{\Lambda }}%
\frac{d\Lambda _{k}^{\left( 1+p\right) }}{dt}\mathbf{R}_{k}+\sum_{k=1}^{n_{%
\Lambda }}\lambda _{k}^{\left( 1+p\right) }\mathbf{R}_{k}\right) \\
\mbox{\boldmath$\Pi$}^{d\left( 2\right) }=\mathbf{F}:\left( \frac{1}{2}\frac{%
d\mbox{\boldmath $\varepsilon$ }^{T}}{dt}\cdot
\mbox{\boldmath
$\varepsilon$ +}\frac{1}{2}\mbox{\boldmath $\varepsilon$ }\cdot \frac{d%
\mbox{\boldmath
$\varepsilon$ }^{T}}{dt}-\sum_{k=1}^{n_{\Lambda }}\frac{d\Lambda
_{k}^{\left( 2\right) }}{dt}\mathbf{R}_{k}-\sum_{k=1}^{n_{\Lambda }}\lambda
_{k}^{\left( 2\right) }\mathbf{R}_{k}\right)%
\end{array}
\label{PidissOrd}
\end{equation}

In order to define the non-dissipative part of the stress, as well as the
equilibrium equation, the following Hu-Washizu type functional is
introduced:
\begin{equation}
\begin{array}{l}
J_{H-W}\left( \mbox{\boldmath$\Pi$}^{nd^{\ast }},\mathbf{\mathbf{\bar{E}}%
^{\ast },U}^{\ast },\Lambda _{k}^{\ast };\mbox{\boldmath$\Pi$}^{d}\right)
=\int_{V_{00}}\Psi \left( \mathbf{\mathbf{\bar{E}}}^{\ast },T\right) \textrm{ }%
dV-\int_{V_{00}}\mbox{\boldmath$\Pi$}^{nd^{\ast }}\mathbf{:}\left( \mathbf{%
\mathbf{\bar{E}}}^{\ast }-\mathbf{E}\left( \mathbf{U}^{\ast }\right) \right)
dV \\
+\int_{V_{00}}\mbox{\boldmath$\Pi$}^{d}\mathbf{:E}\left( \mathbf{U}^{\ast
}\right) dV-\int_{V_{00}}\mathbf{f\cdot U}^{\ast }dV-\int_{\partial
V_{00,\sigma }}\mathbf{g\cdot U}^{\ast }dA \\
-\int_{\partial V_{00,u}}\left( \left[ \left( \mathbf{1}+\nabla \left(
\mathbf{U}^{\ast }\right) \right) \mathbf{\cdot }\left( \mbox{\boldmath$\Pi$}%
^{nd^{\ast }}+\mbox{\boldmath$\Pi$}^{d}\right) \right] \mathbf{\cdot N}%
\right) \mathbf{\cdot }\left( \mathbf{U^{\ast }-\bar{U}}\right) dA \\
-\int_{V_{00}}\sum_{k=1}^{n_{\Lambda }}\Lambda _{k}^{\ast }\mathbf{R}_{k}:%
\mbox{\boldmath$\Pi$}^{nd^{\ast }}dV%
\end{array}
\label{Hu-Washizu-Gen-diss}
\end{equation}%
Eq. (\ref{Hu-Washizu-Gen-diss}) should be compared with (\ref{Hu-Washizu-Gen}%
). An attentive reader can see that the functional depends on the
non-dissipative part of the stress $\mbox{\boldmath$\Pi$}^{nd^{\ast }}\in
\mathbb{T}$ instead of on the total stress $\mbox{\boldmath$\Pi$}^{\ast }$.
Moreover, an additional dependence on $\mbox{\boldmath$\Pi$}^{d}$ is
introduced, where $\mbox{\boldmath$\Pi$}^{d}$ is the stationary solution of (%
\ref{FuncDiss}). Stationarity imposed on $J_{H-W}$ leads to the strong form
expressions:
\begin{equation}
\begin{array}{l}
1.\textrm{ \ \ }\mathbf{R}_{k}:\mbox{\boldmath$\Pi$}^{nd}=0\textrm{ \ , \ \ \ }%
k=1,n_{\Lambda }\textrm{\ \ \ \ \ \ \ \ \ \ \ in }V_{00} \\
2.\textrm{ \ }\left\{
\begin{array}{l}
\mathbf{\bar{E}}=\mathbf{E}\left( \mathbf{U}\right) -\sum_{k=1}^{n_{\Lambda
}}\Lambda _{k}\mathbf{R}_{k}\textrm{\ \ \ \ \ \ \ \ in }V_{00} \\
\mathbf{U}\mathbf{=}\mathbf{\bar{U}}\textrm{\ \ \ \ \ \ \ \ \ \ \ \
\ \ \ \ \
\ \ \ \ \ \ \ \ \ \ \ \ \ \ \ \ \ \ \ \ \ \ on }\partial V_{00,u}%
\end{array}%
\right. \\
3.\textrm{ \ \ \ }\mbox{\boldmath$\Pi$}^{nd}=\left. \frac{\partial
\Psi \left(
\mathbf{\bar{E}}^{\ast },T\right) }{\partial \mathbf{\bar{E}}^{\ast }}%
\right\vert _{\mathbf{\bar{E}}^{\ast }=\mathbf{\bar{E}}}\textrm{\ \
\ \ \ \ \
\ \ \ \ \ in }V_{00}%
\end{array}
\label{StrongDiss2}
\end{equation}%
Moreover, recalling (\ref{Action}) and imposing stationarity in the
displacements, one obtains the following weak form equilibrium equation:
\[
\mathcal{W}_{i}\left( \mbox{\boldmath$\Pi$}^{nd},\mathbf{U,}\delta \mathbf{U}%
\right) +\mathcal{W}_{i}\left( \mbox{\boldmath$\Pi$}^{d},\mathbf{U,}\delta
\mathbf{U}\right) +\mathcal{W}_{e}\left( \mathbf{f},\mathbf{g,}\delta
\mathbf{U}\right) =\mathcal{W}_{a}\left( \mathbf{\ddot{U},}\delta \mathbf{U}%
\right) ,\textrm{\ \ }\forall \delta \mathbf{U}\in \mathbb{V}
\]%
where $\mathcal{W}_{i}\left( \mbox{\boldmath$\Pi$}^{nd},\mathbf{U,}\delta
\mathbf{U}^{\ast }\right) $ is defined in (\ref{Wdef}), $\mathcal{W}%
_{e}\left( \mathbf{f,g,}\delta \mathbf{U}\right) $ is also given in (\ref%
{Wdef}), with $\mbox{\boldmath$\Pi$}=\mbox{\boldmath$\Pi$}^{nd}+%
\mbox{\boldmath$\Pi$}^{d}$ and
\begin{equation}
\mathcal{W}_{i}^{d}=\mathcal{W}_{i}\left( \mbox{\boldmath$\Pi$}^{d},\mathbf{%
U,}\delta \mathbf{U}\right) =-\int_{V_{00}}\mbox{\boldmath$\Pi$}^{d}\mathbf{:%
}\delta \mathbf{E}\left( \mathbf{U,}\delta \mathbf{U}\right) dV  \label{Wid}
\end{equation}%
The virtual work of inertia forces is the same as in the previous
non-dissipative case. The dynamics of the system is ruled by (\ref{Strong0}%
), where $\mbox{\boldmath$\Pi$}$ is given by (\ref{Pind+Pid}). The case of
moderate rotations is obtained by just substituting $\mathbf{E}\left(
\mathbf{U}^{\ast }\right) $ with $\mathbf{E}^{\left( b\right) }\left(
\mathbf{U}^{\ast }\right) $ in Eqs. (\ref{FuncDiss}) and (\ref%
{Hu-Washizu-Gen-diss}). This corresponds to the substitutions $\mathcal{W}%
_{i}\rightarrow \mathcal{W}_{i}^{\left( b\right) }$ and $\mathcal{W}%
_{i}^{d}\rightarrow \mathcal{W}_{i}^{d\left( b\right) }$ , where (see Eq. (%
\ref{Wi2Wi1+2pWi4p}))%
\begin{equation}
\mathcal{W}_{i}^{d\left( b\right) }=\mathcal{W}_{i}^{\left( 2\right) }\left( %
\mbox{\boldmath$\Pi$}^{d},\mathbf{U,}\delta \mathbf{U}\right) +\mathcal{W}%
_{i}^{\left( 1+2p\right) }\left( \mbox{\boldmath$\Pi$}^{d},\mathbf{U,}\delta
\mathbf{U}\right) +\mathcal{W}_{i}^{\left( 4p\right) }\left( %
\mbox{\boldmath$\Pi$}^{d},\mathbf{U,}\delta \mathbf{U}\right)  \label{Widb}
\end{equation}

\section{Accounting for a static prestress}

As already discussed, when a static prestress due \emph{external} mechanical
and/or thermal loading occurs, the structure passes from the state $%
V_{00}=V_{0i}$ to a state $V_{0}$. It is interesting to write the equations
governing the equilibrium at the generic configuration $V_{1}$ as a function
of the unknown displacement $\mathbf{U}_{01}=\mathbf{U}_{1}-\mathbf{U}_{0}$,
expressing the motion with respect to $V_{0}$, as illustrated in Figure \ref%
{Config}. This can be easily done subtracting the equilibrium equations
established in the previous sections, written at $V_{1}$ and at $V_{0}$.
Both equilibrium conditions at $V_{0}$ and $V_{1}$ should be written,
together with the other expressions coming from the stationarity of the
relevant H-W type functional. For the sake of simplicity, only the weak form
of the dynamic equilibrium is reported in the analysis of this section. As
seen above, the virtual work of the internal forces is indicated by $%
\mathcal{W}_{i}$ in the general case, and by $\mathcal{W}_{i}^{\left(
b\right) }$ in the case of moderate rotations. All the equations of this
section are written as function of $\mathcal{W}_{i}$, and then refer to the
general case. However, the formal substitution of $\mathcal{W}_{i}^{\left(
b\right) }$ at the place of $\mathcal{W}_{i}$ gives the equations for the
moderate rotation case.

The static problem defining the prestressed configuration $V_{0}$ reads
\begin{equation}
\left\{
\begin{array}{l}
\textrm{Find }\mathbf{U}_{0}\mathbf{\in }\mathbb{U}\textrm{ such that \ for all }%
\delta \mathbf{U\in }\mathbb{V} \\
\mathcal{W}_{i}\left( \mbox{\boldmath$\Pi$}_{0}\mathbf{,U}_{0}\mathbf{,}%
\delta \mathbf{U}\right) +\mathcal{W}_{e}\left( \mathbf{f}_{0}\mathbf{,g}%
_{0},\delta \mathbf{U}\right) =0\textrm{\ ,\ \ }\mbox{\boldmath$\Pi$}_{0}%
\mathbf{=}\mbox{\boldmath$\Pi$}^{nd}\left( \mathbf{U}_{0},\Lambda
_{k,0}\right) \\
\mathbf{U}_{0}=\mathbf{\bar{U}}_{0}\textrm{ \ \ \ \ \ \ \ \ \ \ on
}\partial
V_{00,u}%
\end{array}%
\right.  \label{Wi-static}
\end{equation}%
where $\Lambda _{k,0}$ are the Lagrange multipliers associated with the
static problem and the prestress $\mbox{\boldmath$\Pi$}_{0}$ accounts for
the temperature field $T_{0}$. The dynamic problem defining the generic
configuration $V_{1}$ reads
\begin{equation}
\left\{
\begin{array}{l}
\textrm{Find }\mathbf{U}_{1}\mathbf{\in }\mathbb{U}\textrm{ such that \ for all }%
\delta \mathbf{U\in }\mathbb{V} \\
\mathcal{W}_{i}\left( \mbox{\boldmath$\Pi$}_{1}^{nd},\mathbf{U}_{1}\mathbf{,}%
\delta \mathbf{U}\right) +\mathcal{W}_{i}\left( \mbox{\boldmath$\Pi$}%
_{1}^{d},\mathbf{U}_{1}\mathbf{,}\delta \mathbf{U}\right) \\
+\mathcal{W}_{e}\left( \mathbf{f}_{0}+\mathbf{f}_{1}\left( t\right) \mathbf{%
,g}_{0}+\mathbf{g}_{1}\left( t\right) ,\delta \mathbf{U}\right) =\mathcal{W}%
_{a}\left( \mathbf{\ddot{U}}_{1}\mathbf{,}\delta \mathbf{U}\right) \\
\mbox{\boldmath$\Pi$}_{1}\mathbf{=\Pi }^{nd}\left( \mathbf{U}_{1},\Lambda
_{k,1}\right) +\mbox{\boldmath$\Pi$}^{d}\left( \mathbf{U}_{1},\lambda
_{k,1}\right) \\
\mathbf{U}_{1}=\mathbf{\bar{U}}_{0}\mathbf{+\bar{U}}_{01}\left( t\right)
\textrm{ \ \ \ \ \ \ \ \ \ \ \ \ on }\partial V_{00,u}%
\end{array}%
\right.  \label{Wi-dynamic}
\end{equation}%
where $\Lambda _{k,1}$ are the Lagrange multipliers computed for the dynamic
problem. The \emph{difference} between (\ref{Wi-dynamic}) and (\ref%
{Wi-static}) leads to
\begin{equation}
\left\{
\begin{array}{l}
\textrm{Find }\mathbf{U}_{01}\mathbf{\in }\mathbb{U}\textrm{ such
that\ \ for
all }\delta \mathbf{U\in }\mathbb{V} \\
\Delta \mathcal{W}_{i}\left( \mbox{\boldmath$\Pi$}_{0},\mbox{\boldmath$\Pi$}%
_{1}^{nd},\mathbf{U}_{01}\mathbf{,\mathbf{U}}_{0}\mathbf{\mathbf{,}}\delta
\mathbf{U}\right) +\mathcal{W}_{i}\left( \mbox{\boldmath$\Pi$}_{1}^{d},%
\mathbf{U}_{0}+\mathbf{U}_{01}\mathbf{,}\delta \mathbf{U}\right) \\
\textrm{ }+\mathcal{W}_{e}\left( \mathbf{f}_{1}\left( t\right) \mathbf{,g}%
_{1}\left( t\right) ,\delta \mathbf{U}\right) =\mathcal{W}_{a}\left( \mathbf{%
\ddot{U}}_{01}\mathbf{,}\delta \mathbf{U}\right) \textrm{\ \ \ \ \ \
\ \ \ \ \
\ \ \ \ \ \ \ \ \ \ } \\
\mbox{\boldmath$\Pi$}_{0}\mathbf{=}\mbox{\boldmath$\Pi$}\left( \mathbf{U}%
_{0},\Lambda _{k,0}\right) ,\textrm{ \ \ \ }\mbox{\boldmath$\Pi$}_{1}^{d}=%
\mbox{\boldmath$\Pi$}\left( \mathbf{U}_{0}+\mathbf{U}_{01},\lambda
_{k,1}\right) \\
\mbox{\boldmath$\Pi$}_{1}\mathbf{=}\mbox{\boldmath$\Pi$}^{nd}\left( \mathbf{U%
}_{0}+\mathbf{U}_{01},\Lambda _{k,1}\right) +\mbox{\boldmath$\Pi$}_{1}^{d}
\\
\mathbf{U}_{01}=\mathbf{\bar{U}}_{1}\left( t\right) \textrm{ \ \ \ \
\ \ \ \ \
\ \ \ \ \ \ \ on }\partial V_{00,u}%
\end{array}%
\right.  \label{Wi-difference}
\end{equation}%
with
\begin{equation}
\Delta \mathcal{W}_{i}\left( \mbox{\boldmath$\Pi$}_{0},\mbox{\boldmath$\Pi$}%
_{1}^{nd},\mathbf{U}_{01}\mathbf{,\mathbf{U}}_{0}\mathbf{\mathbf{,}}\delta
\mathbf{U}\right) =\mathcal{W}_{i}\left( \mbox{\boldmath$\Pi$}_{1}^{nd},%
\mathbf{U}_{0}+\mathbf{U}_{01}\mathbf{,}\delta \mathbf{U}\right) -\mathcal{W}%
_{i}\left( \mbox{\boldmath$\Pi$}_{0}\mathbf{,U}_{0}\mathbf{,}\delta \mathbf{U%
}\right)  \label{deltaWi}
\end{equation}%
knowing that $\mathbf{\ddot{U}}_{1}=\mathbf{\ddot{U}}_{01}$. Eq. (\ref%
{Wi-difference}) describes the dynamics around a statically prestressed
configuration for the general case. Observe that all the equations are
defined in the Lagrangian configuration $V_{00}=V_{0i}$, free of any
external prestress effect by definition. The case of moderate rotations is
retrieved introducing in the same equation $\mathcal{W}_{i}^{\left( b\right)
}$ instead of $\mathcal{W}_{i}$ and $\mathcal{W}_{i}^{d\left( b\right) }$
instead of $\mathcal{W}_{i}^{d}$. From Eq. (\ref{StrEn-3approx}) and
recalling (\ref{Wi_b}), (\ref{Pidiss}) and (\ref{PidissOrd}), Eq. (\ref%
{deltaWi}) becomes
\begin{equation}
\begin{array}{l}
-\Delta \mathcal{W}_{i}^{\left( b\right) }\left( \mbox{\boldmath$\Pi$}_{0},%
\mbox{\boldmath$\Pi$}_{1}^{nd},\mathbf{U}_{01}\mathbf{,\mathbf{U}}_{0}%
\mathbf{\mathbf{,}}\delta \mathbf{U}\right) \\
\mathcal{=}\int_{V_{00}}\left( \mbox{\boldmath$\Pi$}_{1}^{nd\left( 1\right)
}+\mbox{\boldmath$\Pi$}_{1}^{nd\left( 2p\right) }\right) \mathbf{:}\left(
\mathbf{\nabla }^{s}\left( \delta \mathbf{U}\right) +\delta %
\mbox{\boldmath$\chi$}\left( \mathbf{U}_{0}+\mathbf{U}_{01}\mathbf{,}\delta
\mathbf{U}\right) \right) dV \\
\textrm{ \ }-\int_{V_{00}}\left( \mbox{\boldmath$\Pi$}_{0}^{\left( 1\right) }+%
\mbox{\boldmath$\Pi$}_{0}^{\left( 2p\right) }\right) \mathbf{:}\left(
\mathbf{\nabla }^{s}\left( \delta \mathbf{U}\right) +\delta %
\mbox{\boldmath$\chi$}\left( \mathbf{U}_{0}\mathbf{,}\delta \mathbf{U}%
\right) \right) dV%
\end{array}
\label{Weak01}
\end{equation}

\section{Moderate rotations and Bernoulli-Navier kinematic assumptions:
undamped case}

\begin{figure}[tbp]
\begin{center}
\includegraphics[width=10cm]{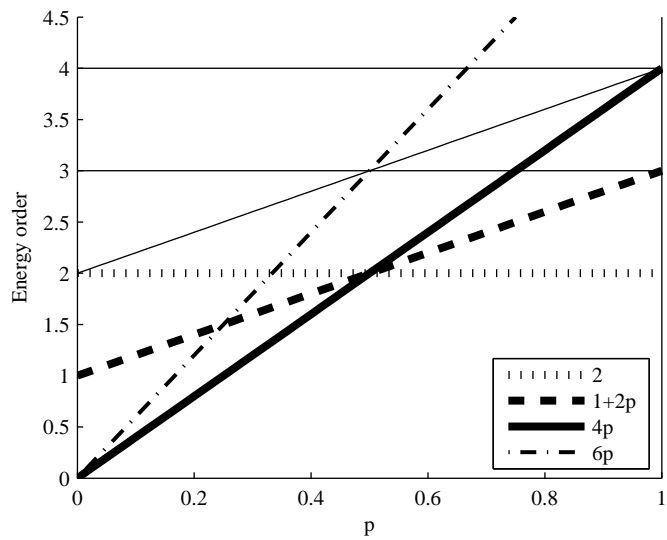}
\end{center}
\caption{Navier kinematic assumptions and orders of the strain energy terms.
The dominant terms ($2$, $1+2p$ and $4p$) are highlighted. The line
corresponding to the order $6p$ represents the most important energy term
associated with the non-linear constitutive law (\protect\ref{constQuad}).
Note that the terms of order $3+p$, $2+p$ and $1+3p $ no longer appear
(compare with Figure \protect\ref{Exponents}).}
\label{ExponentsEulBern}
\end{figure}

\begin{figure}[tbp]
\begin{center}
\includegraphics[width=14cm]{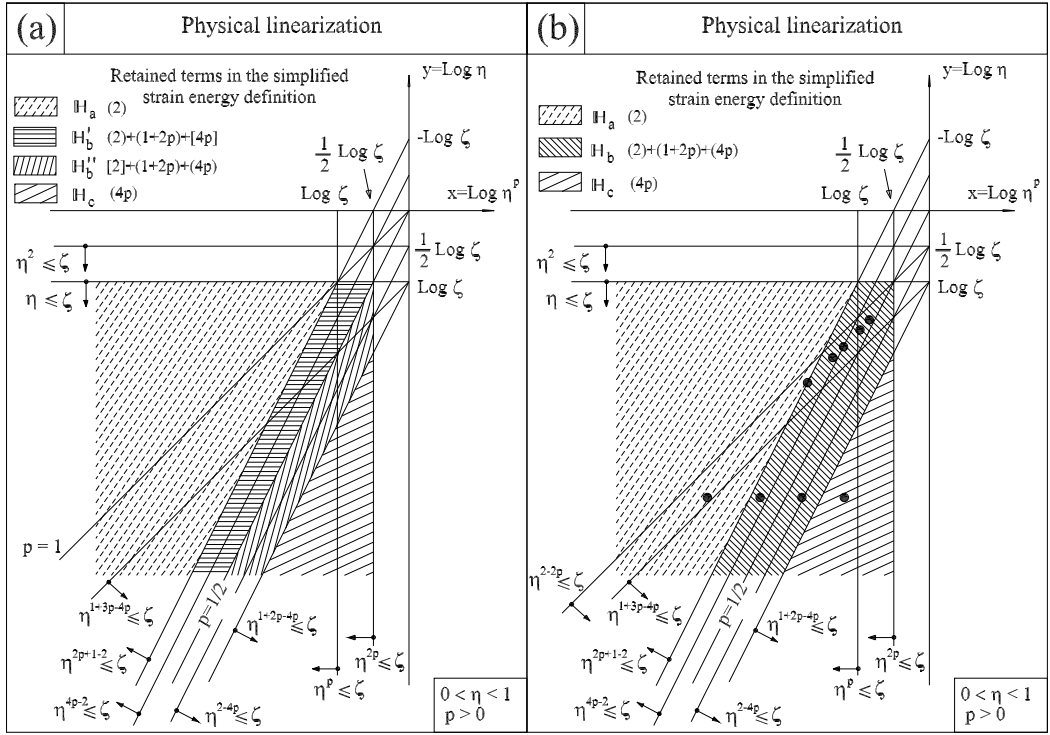}
\end{center}
\caption{Physical linearization and Navier kinematic assumptions. (a) Strain
rotation domains. (b) The strain-rotation domains, where $\mathbb{H}%
_{b}^{\prime }$ and $\mathbb{H}_{b}^{\prime \prime }$ are merged. The
"strain-rotation points" represent the deformed configurations of the
structures of the examples discussed in Section 8.2 .}
\label{Str-rot-Domains-Navier}
\end{figure}

In this section, the stationarity conditions (\ref{Strong123-gen}) with
suited stress constraints are used in conjunction with the strain measure (%
\ref{Eb-def}) for moderate rotations and the so-called Navier kinematic
assumptions for beams, in order to obtain the corresponding strong form of
the dynamic equilibrium equations. This analysis will enable a better
understanding of the general equations previously presented. In particular,
a simple way of estimating $\eta $ and $p$ is suggested with reference to
the example of a clamped-clamped beam. In the non deformed configuration $%
V_{00}=V_{0i}$, the beam axis coincides with the cartesian axis $X$ and the
beam motion is supposed to be limited to the plane $X-Y$. A quadratic
Helmholtz energy is adopted, leading to a linear constitutive law depending
on the tensors $\mathbf{D}$ and $\mathbf{A}$:
\begin{equation}
\mbox{\boldmath$\Pi$}=\mbox{\boldmath$\Pi$}_{0i}+\mathbf{D:\bar{E}}-\mathbf{A%
}\left( T-T_{0i}\right)  \label{Pi2}
\end{equation}%
with%
\begin{equation}
\mathbf{\bar{E}}=\mathbf{E}^{\left( b\right) }\left( \mathbf{U}\right)
-\sum_{k=1}^{n_{\Lambda }}\Lambda _{k}\mathbf{R}_{k}\mathbf{=}%
\mbox{\boldmath$\varepsilon$}\left( \mathbf{U}\right) +\mbox{\boldmath$\chi$}%
\left( \mathbf{U}\right) -\sum_{k=1}^{n_{\Lambda }}\Lambda _{k}\mathbf{R}_{k}
\label{Ecorr}
\end{equation}%
For an isotropic material, one has $\mathbf{D=}\frac{E}{\left( 1+\nu \right)
\left( 1-2\nu \right) }\left[ \nu \mathbf{1}\otimes \mathbf{1+}\left( 1-2\nu
\right) \mathbf{I}\right] $. Moreover, since $\mathbf{A=}\frac{\alpha E}{%
1-2\nu }\mathbf{1}$, one has $\mathbf{D} ^{-1}\mathbf{:A}\left(
T-T_{0i}\right) =\alpha \left( T-T_{0i}\right) \mathbf{1}$. The Navier
kinematic assumption reads
\begin{equation}
\mathbf{u}\mathbf{=}\left[ \mathrm{u}-Y\textrm{ }\mathrm{v}^{\prime },\mathrm{v%
},0\right] ^{T}  \label{KinHyp_Beam}
\end{equation}%
where $\mathrm{u}=\mathrm{u}\left( X,t\right) ,\mathrm{v}=\mathrm{v}\left(
X,t\right) $ are the X-and Y-displacement fields; the apex $^{\prime }$
indicates the derivation with respect to $X$. The strains $%
\mbox{\boldmath$\varepsilon$}\left( \mathbf{U}\right) $ and $%
\mbox{\boldmath$\chi$}\left( \mathbf{U}\right) $ read:
\[
\mbox{\boldmath$\varepsilon$}\left( \mathbf{U}\right) =\left[
\begin{array}{ccc}
\mathrm{u}^{\prime }-Y\mathrm{v}^{\prime \prime } & 0 & 0 \\
0 & 0 & 0 \\
0 & 0 & 0%
\end{array}%
\right] ,\textrm{ \ \ }\mbox{\boldmath$\chi$}\left( \mathbf{U}\right) =\frac{1%
}{2}\left[
\begin{array}{ccc}
\mathrm{v}^{\prime 2} & 0 & 0 \\
0 & \mathrm{v}^{\prime 2} & 0 \\
0 & 0 & 0%
\end{array}%
\right]
\]%
For the prestress $\mbox{\boldmath$\Pi$}_{0i}$, we assume:
\[
\mbox{\boldmath$\Pi$}_{0i}=\left[
\begin{array}{ccc}
\Pi _{0i,1} & 0 & 0 \\
0 & \Pi _{0i,2} & 0 \\
0 & 0 & \Pi _{0i,3}%
\end{array}%
\right] =\left[
\begin{array}{ccc}
\Pi _{0i}^{\left( x\right) }+Y\Pi _{0i,x}^{\left( b\right) } & 0 & 0 \\
0 & \Pi _{0i}^{\left( y\right) }+Y\Pi _{0i,y}^{\left( b\right) } & 0 \\
0 & 0 & \Pi _{0i}^{\left( z\right) }+Y\Pi _{0i,z}^{\left( b\right) }%
\end{array}%
\right]
\]%
Off-diagonal terms of $\mbox{\boldmath$\Pi$}_{0i}$ may not vanish.
Nonetheless, they have no influence on the following analysis, since they
are associated with zero virtual strain components in the virtual work
product. The thermal and load fields write
\[
\begin{array}{l}
\mathbf{f:}=\left[
\begin{array}{c}
f_{x}\left( X,t\right) -Y\textrm{ }f_{b}\left( X,t\right) \\
f_{y}\left( X,t\right) \\
0%
\end{array}%
\right] ,\textrm{ \ \ \ \ \ }\mathbf{g}=\left[
\begin{array}{c}
g_{x}\left( X,t\right) -Y\textrm{ }g_{b}\left( X,t\right) \\
g_{y}\left( X,t\right) \\
0%
\end{array}%
\right] \\
T-T_{0i}=\left[ T_{x}\left( X\right) -T_{0i,x}\left( X\right) \right] -Y%
\textrm{ }\left[ \gamma \left( X\right) -\gamma _{0i}\left( X\right)
\right]
=\Delta T_{x}-Y\textrm{ }\Delta \gamma%
\end{array}%
\]%
The stress constraints usually imposed to retrieve beam equations are $\Pi
_{YY}=\Pi _{ZZ}=0$, formally expressed by the conditions%
\begin{equation}
\mathbf{R}_{1}:\mbox{\boldmath$\Pi$}=\mathbf{R}_{2}:\mbox{\boldmath$\Pi$}=0%
\textrm{ \ with }\mathbf{R}_{1}=\left[
\begin{array}{ccc}
0 & 0 & 0 \\
0 & 1 & 0 \\
0 & 0 & 0%
\end{array}%
\right] \textrm{ ,\ }\mathbf{R}_{2}=\left[
\begin{array}{ccc}
0 & 0 & 0 \\
0 & 0 & 0 \\
0 & 0 & 1%
\end{array}%
\right]  \label{R1R2}
\end{equation}
Eqs. (\ref{etap-def}),(\ref{T-P0i-order}) and (\ref{etap-intervals}) written
for this case become%
\begin{equation}
\begin{array}{l}
\eta =\left( \frac{1}{V_{00}}\int_{V_{00}}\left[ \mathrm{u}^{\prime }-Y%
\mathrm{v}^{\prime \prime }\right] ^{2}\textrm{ }dV\right) ^{\frac{1}{2}%
}+\left( \frac{1}{V_{00}}\int_{V_{00}}3\left[ \alpha \left( \Delta T_{x}-Y%
\textrm{ }\Delta \gamma \right) \right] ^{2}\textrm{ }dV\right)
^{\frac{1}{2}}
\\
\textrm{ \ }=\left( \frac{1}{V_{00}}\int_{0}^{L_{00}}\left( A\left( \mathrm{u}%
^{\prime }\right) ^{2}+J\left( \mathrm{v}^{\prime \prime }\right)
^{2}\right) \textrm{ }dX\right) ^{\frac{1}{2}}+\left( \frac{3\alpha ^{2}}{%
V_{00}}\int_{0}^{L_{00}}\left( A\left( \Delta T_{x}\right) ^{2}+J\left(
\Delta \gamma \right) ^{2}\right) \textrm{ }dX\right) ^{\frac{1}{2}} \\
\eta ^{p}=\left( \frac{1}{V_{00}}\int_{V_{00}}\mathrm{v}^{\prime 2}\textrm{ }%
dV\right) ^{\frac{1}{2}}=\left( \frac{1}{V_{00}}\int_{0}^{L_{00}}A\mathrm{v}%
^{\prime 2}\textrm{ }dX\right) ^{\frac{1}{2}} \\
\mbox{\boldmath$\Pi$}_{0i}=O\left( \eta \right) +O\left( \eta ^{2p}\right)
\\
0<\eta <1\textrm{ \ \ \ \ \ \ \ \ \ }p>0%
\end{array}
\label{etap-def-Navier}
\end{equation}%
where $A$ is the area of the generic beam section; $J$ is the inertia moment
and $L_{00}$ is the beam length. Then, the same procedure as in the general
case can be applied here, in order to determine the strain-rotation domains $%
\mathbb{H}_{a}$, $\mathbb{H}_{b}$ and $\mathbb{H}_{c}$. As it is well-known,
the Navier-Bernoulli kinematic assumptions entails that all shear strains,
\emph{i.e.} the off-diagonal elements of $\mbox{\boldmath
$\varepsilon$}$, are equal to \emph{zero}. As a result, one can easily prove
that also the terms of order $3+p$, $2+p$ and $1+3p$ in the strain energy
\begin{equation}
\mathcal{F}=O\left( \eta ^{2}\right) +O\left( \eta ^{1+2p}\right) +O\left(
\eta ^{4p}\right) +\sum_{i=4}^{n_{NB}}O\left( \eta ^{s_{i}}\right)
\end{equation}%
become zero (see Figure \ref{ExponentsEulBern} and compare to Figure \ref%
{Exponents}). As a result, the strain-rotation domains are not the same as
in the general case, as illustrated in Figure \ref{Str-rot-Domains-Navier}.
The difference is highlighted by the small region excluded in the general
case and admitted by the Navier kinematic conditions. Inside this region,
the pertinence of the Navier assumptions (\ref{KinHyp_Beam}) should be
further investigated. Other kinematic assumptions, like for instance those
of Timoshenko, appear to be more sound.

\subsection{Strong form equations}

Eqs. (\ref{Pi2}), (\ref{R1R2}) \emph{and } (\ref{KinHyp_Beam}) lead to the
Lagrange multipliers
\begin{equation}
\begin{array}{l}
\Lambda _{1}=\frac{\Pi _{0i,2}\left( 1-\nu ^{2}\right) -\nu \left( 1+\nu
\right) \Pi _{0i,3}}{E}+\nu \left( \mathrm{u}^{\prime }-Y\mathrm{v}^{\prime
\prime }+\frac{1}{2}\mathrm{v}^{\prime 2}\right) \\
\textrm{ \ \ \ \ \ \ \ }+\frac{1}{2}\mathrm{v}^{\prime 2}-\alpha
\left( \Delta
T_{x}-Y\textrm{ }\Delta \gamma \right) \left( 1+\nu \right) \\
\Lambda _{2}=\frac{\Pi _{0i,3}\left( 1-\nu ^{2}\right) -\nu \left( 1+\nu
\right) \Pi _{0i,2}}{E}+\nu \left( \mathrm{u}^{\prime }-Y\mathrm{v}^{\prime
\prime }+\frac{1}{2}\mathrm{v}^{\prime 2}\right) -\alpha \left( \Delta
T_{x}-Y\textrm{ }\Delta \gamma \right) \left( 1+\nu \right)%
\end{array}
\label{LagrModer-Eulbeam}
\end{equation}%
from which (see (\ref{Ecorr})) $\bar{E}_{XX}=E_{XX}=\mathrm{u}^{\prime }-Y%
\mathrm{v}^{\prime \prime }+\frac{1}{2}\mathrm{v}^{\prime 2},$ $\bar{E}%
_{YY}=E_{YY}-\Lambda _{1}=\frac{1}{2}\mathrm{v}^{\prime 2}-\Lambda _{1},$ $%
\bar{E}_{ZZ}=E_{ZZ}-\Lambda _{2}=-\Lambda _{2}$ and
\[
\Pi _{XX}=\Pi _{0i,x}+Y\textrm{ }\Pi _{0i,b}+E\left( \mathrm{u}^{\prime }-Y%
\mathrm{v}^{\prime \prime }+\frac{1}{2}\mathrm{v}^{\prime 2}-\alpha
\left( \Delta T_{x}-Y\textrm{ }\Delta \gamma \right) \right)
\]%
where $\Pi _{0i,x}=\Pi _{0i}^{\left( x\right) }-\nu \left( \Pi
_{0i}^{\left( y\right) }+\Pi _{0i}^{\left( z\right) }\right) $ and
$\Pi _{0i,b}=\Pi _{0i,x}^{\left( b\right) }-\nu \left( \Pi
_{0i,y}^{\left( b\right) }+\Pi _{0i,z}^{\left( b\right) }\textrm{\
}\right) ,$ while the other stress components are zero. Moreover,
the virtual works read
\begin{equation}
\begin{array}{l}
\mathcal{W}_{i}^{\left( b\right) }\left( \mbox{\boldmath$\Pi$},\mathbf{U,}%
\delta \mathbf{U}\right) +\mathcal{W}_{e}\left( \mathbf{f,g,}\delta \mathbf{U%
}\right) =\mathcal{W}_{a}\left( \mathbf{\ddot{U},}\delta \mathbf{U}\right)
\\
\\
\mathcal{W}_{i}^{\left( b\right) }\left( \mbox{\boldmath$\Pi$},\mathbf{U,}%
\delta \mathbf{U}\right) =-\int_{V_{00}}\Pi _{XX}\left[ \delta \mathrm{u}%
^{\prime }-Y\delta \mathrm{v}^{\prime \prime }+\mathrm{v}^{\prime }\delta
\mathrm{v}^{\prime }\right] dV \\
\textrm{ \ \ \ }=\mathcal{W}_{i}^{\left( 2\right) }\left( \mbox{\boldmath$\Pi$}%
,\mathbf{U,}\delta \mathbf{U}\right) +\mathcal{W}_{i}^{\left( 1+2p\right)
}\left( \mbox{\boldmath$\Pi$},\mathbf{U,}\delta \mathbf{U}\right) +\mathcal{W%
}_{i}^{\left( 4p\right) }\left( \mbox{\boldmath$\Pi$},\mathbf{U,}\delta
\mathbf{U}\right) \\
-\mathcal{W}_{i}^{\left( 2\right) }\left( \mbox{\boldmath$\Pi$},\mathbf{U,}%
\delta \mathbf{U}\right) =\int_{0}^{L_{00}}\left[ EA\left( \mathrm{u}%
^{\prime }-\alpha \Delta T_{x}\right) \right] \delta \mathrm{u}^{\prime }%
\textrm{ }dX+\int_{0}^{L_{00}}\left[ EJ\left( \mathrm{v}^{\prime
\prime
}-\alpha \Delta \gamma \right) \right] \delta \mathrm{v}^{\prime \prime }%
\textrm{ }dX \\
-\mathcal{W}_{i}^{\left( 1+2p\right) }\left( \mbox{\boldmath$\Pi$},\mathbf{U,%
}\delta \mathbf{U}\right) =\int_{0}^{L_{00}}\left[ EA\frac{1}{2}\mathrm{v}%
^{\prime 2}\right] \delta \mathrm{u}^{\prime }\textrm{ }dX\textrm{\ }%
+\int_{0}^{L_{00}}\left[ EA\left( \mathrm{u}^{\prime }-\alpha \Delta
T_{x}\right) \right] \left[ \mathrm{v}^{\prime }\delta \mathrm{v}^{\prime }%
\right] dX \\
-\mathcal{W}_{i}^{\left( 4p\right) }\left( \mbox{\boldmath$\Pi$},\mathbf{U,}%
\delta \mathbf{U}\right) =\int_{0}^{L_{00}}\left[ EA\frac{1}{2}\mathrm{v}%
^{\prime 2}\right] \left[ \mathrm{v}^{\prime }\delta \mathrm{v}^{\prime }%
\right] dX%
\end{array}
\label{WiWeWa}
\end{equation}%
\[
\begin{array}{l}
\mathcal{W}_{e}\left( \mathbf{f,g,}\delta \mathbf{U}\right) =\int_{V_{00}}%
\left[ \left( f_{x}-Y\textrm{ }f_{b}\right) \left( \delta u-Y\delta \mathrm{v}%
^{\prime }\right) +f_{y}\textrm{ }\delta \mathrm{v}\right] dV \\
\textrm{ \ \ \ \ \ \ \ \ \ \ \ \ \ \ \ \ \ \ }+\int_{\partial V_{00,\sigma }}%
\left[ \left( g_{x}-Y\textrm{ }g_{b}\right) \left( \delta
\mathrm{u}-Y\delta
\mathrm{v}^{\prime }\right) +g_{y}\textrm{ }\delta \mathrm{v}\right] dA \\
\textrm{ \ \ \ \ \ \ \ \ \ \ \ \ \ \ \ \ \ \ }+\left[ R_{x}\delta \mathrm{u}%
+C_z\delta \mathrm{v}^{\prime }+R_{y}\delta \mathrm{v}\right]
_{X=0}^{X=L_{00}}%
\end{array}%
\]%
\[
\begin{array}{l}
\mathcal{W}_{a}\left( \mathbf{\ddot{U},}\delta \mathbf{U}\right)
=\int_{V_{00}}\rho _{00}\left[ \left( \mathrm{\ddot{u}}-Y\mathrm{\ddot{v}}
^{\prime }\right) \left( \delta \mathrm{u}-Y\delta \mathrm{v}^{\prime
}\right) +\mathrm{\ddot{v}}\delta \mathrm{v}\right] dV \\
\textrm{ \ \ \ \ \ \ \ \ \ \ \ \ \ \ \ \ }=\int_{0}^{L_{00}}\rho _{00}A\mathrm{%
\ \ \ddot{u}}\delta \mathrm{u}dX+\int_{0}^{L_{00}}\rho _{00}A\mathrm{\ddot{v}%
} \delta \mathrm{v}dX-\int_{0}^{L_{00}}\rho _{00}J\mathrm{\ddot{v}}^{\prime
\prime }\delta \mathrm{v}dX%
\end{array}%
\]%
where $R_{x}$ $R_{y}$ and $C_z$ are the reaction forces and the reaction
moment at the boundary $\partial V_{00,u}$. Observe the second and third
term in $\mathcal{W}_{a}$: 
it can be proven that the ratio $r_1$ between the third term (rotational
inertia) and the second term (translational inertia) reads $r_1=O\left(\frac{%
J}{AL_{00}^{2}}\right):=O\left(\frac{c h^2}{L_{00}^{2}}\right)$, where $h$
is the beam width. When the squared aspect ratio $\left(h/L_{00}\right)^2$
is small, $r_1$ is small too. Note that this ratio can be easily expressed
in terms of $\eta$ and $\eta^p$ when $\Delta T_x=\Delta \gamma=0$ and $%
u^{\prime}=O\left(\eta^{2p}\right)$. In this case, Eqs. (\ref%
{etap-def-Navier}-1,2) entail $r_1=O\left(\eta^{2p}\right)+O\left(%
\eta^{2-2p}\right)$. Hence, under these assumptions $\eta^{2p}<\zeta$ and $%
\eta ^{2-2p}<\zeta$ suffice to have $r_1$ small. The first condition is
always fulfilled due to physical linearization assumption, while the second
one is equivalent to $\log \eta ^{2-2p}<\log \zeta $ and $y<x+\log \zeta/2$
(see Figure \ref{Str-rot-Domains-Navier}) and is satisfied in a large
portion of $\mathbb{H}_{b}$. For simplicity, the rotational inertia is
always omitted hereinafter. 
The strong form equations corresponding to (\ref{WiWeWa}) are derived using
the standard procedure:
\begin{equation}
\left\{
\begin{array}{l}
\left[ A\Pi _{0i,x}+EA\left( \mathrm{u}^{\prime }+\frac{1}{2}\mathrm{v}%
^{\prime 2}-\alpha \Delta T_{x}\right) \right] ^{\prime }=-p_{x}+\rho _{00}A%
\mathrm{\ddot{u}} \\
\left( EJ\textrm{ }\left( \mathrm{v}^{\prime \prime }-\alpha \Delta
\gamma
\right) +J\Pi _{0i,b}\right) ^{\prime \prime } \\
-\left( \mathrm{v}^{\prime }\left[ A\Pi _{0i,x}+EA\left( \mathrm{u}^{\prime
}+\frac{1}{2}\mathrm{v}^{\prime 2}-\alpha \Delta T_{x}\right) \right]
\right) ^{\prime }=-q^{\prime }+p_{y}-\rho _{00}A\mathrm{\ddot{v}}%
\end{array}%
\right.
\end{equation}%
where $p_{x}=Af_{x}$ and $p_{y}=Af_{y}$ are the horizontal and vertical
loading per unit beam length, respectively; $q=Jf_{b}$ is a couple per unit
length. The boundary conditions of type $\partial V_{00,\sigma }$ involving
the external forces $P_{X}=Ag_{x}$ , $P_{Y}=Ag_{y}$ and the external couple $%
M=Jg_{b}$ at the ends of the beam, are not reported for brevity. At the
configuration $V_{00}=V_{0i}$, one has $\mathrm{u}=\mathrm{v}=\Delta
T_{x}=\Delta \gamma =0$, $\partial V_{00}=\partial V_{00,\sigma }$ with zero
loads, entailing $\left( A\Pi _{0i,x}\right) ^{\prime }=0$\ and\ $\left(
J\Pi _{0i,b}\right) ^{\prime \prime }=0$. This implies that $A\Pi _{0i,x}$
is constant along the length of the beam. Since zero force is applied on $%
\partial V_{00}$, \emph{i.e.} $P_{X}=Ag_{x}\left( 0\right) =0=A\Pi _{0i,x}$,
this constant is equal to zero and the same holds for $J\Pi _{0i,b}$. This
means that the initial \emph{self-equilibrated} stress is zero and the
equilibrium equations become
\begin{equation}
\left\{
\begin{array}{l}
-\left( EA\left( \mathrm{u}^{\prime }-\alpha \Delta T_{x}+\frac{1}{2}\mathrm{%
v}^{\prime 2}\right) \right) ^{\prime }=p_{x}-\rho _{00}A\mathrm{\ddot{u}}
\\
\left( EJ\mathrm{v}^{\prime \prime }-\alpha EJ\Delta \gamma \right) ^{\prime
\prime } \\
\ -\left( \mathrm{v}^{\prime }\left[ EA\left( \mathrm{u}^{\prime }+\frac{1}{2%
}\mathrm{v}^{\prime 2}-\alpha \Delta T_{x}\right) \right] \right) ^{\prime
}=-q^{\prime }+p_{y}-\rho _{00}A\mathrm{\ddot{v}}%
\end{array}%
\right.  \label{Navier-strong-moderate}
\end{equation}%
This is the general expression of the beam equation with temperature field.
The corresponding expression of the strain energy (see (\ref{StrEn-3approx})
and (\ref{Wi_b})) becomes
\begin{equation}
\begin{array}{l}
\mathcal{F}=\mathcal{F}^{(b)}=\mathcal{F}^{\left( 2\right) }+\mathcal{F}%
^{\left( 1+2p\right) }+\mathcal{F}^{\left( 4p\right) } \\
\mathcal{F}^{\left( 2\right) }=\int_{0}^{L_{00}}EA\left( \frac{1}{2}\mathrm{u%
}^{\prime }-\alpha \Delta T_{x}\right) \mathrm{u}^{\prime }\textrm{ }%
dX+\int_{0}^{L_{00}}EJ\left( \frac{1}{2}\mathrm{v}^{\prime \prime }-\alpha
\textrm{ }\Delta \gamma \right) \mathrm{v}^{\prime \prime }dX \\
\mathcal{F}^{\left( 1+2p\right) }=\int_{0}^{L_{00}}EA\left( \frac{1}{2}%
\mathrm{u}^{\prime }-\alpha \Delta T_{x}\right) \frac{1}{2}\mathrm{v}%
^{\prime 2}\textrm{ }dX+\frac{1}{2}\int_{0}^{L_{00}}EA\left( \frac{1}{2}%
\mathrm{v}^{\prime 2}\right) \mathrm{u}^{\prime }\textrm{ }dX \\
\mathcal{F}^{\left( 4p\right) }=\frac{1}{2}\int_{0}^{L_{00}}EA\left( \frac{1%
}{2}\mathrm{v}^{\prime 2}\right) ^{2}\textrm{ }dX%
\end{array}
\label{F-navier-gen}
\end{equation}

\subsection{The geometric interpretation of $\protect\eta $ and $\protect%
\eta ^{p}$}

In this Section, an interpretation of $\eta $ and $\eta ^{p}$ in terms of
suitable deflection and shape ratios, and as functions of the temperature
field is provided for the case of a homogeneous beam. A first example
concerns a beam with very small bending stiffness, \emph{i.e.} $J/A\simeq 0$%
. A static vertical load $F$ is applied at the midspan, where it induces a
transversal displacement $\mathrm{v}_{\max }$. Moreover, $\mathrm{\ddot{u}}=%
\mathrm{\ddot{v}}=p_{x}=p_{y}=q=0$ and an axial temperature field is
introduced. Then, Eq. (\ref{Navier-strong-moderate}) becomes
\begin{equation}
\left\{
\begin{array}{l}
EA\left( \mathrm{u}^{\prime }-\alpha \Delta T_{x}+\frac{1}{2}\mathrm{v}%
^{\prime 2}\right) =R_x \\
R_x\mathrm{v}^{\prime \prime }=0%
\end{array}%
\right.  \label{hing-Eq}
\end{equation}%
where $R_x$ is the constant horizontal reaction at $X=L_{00}$. Boundary
conditions write $\mathrm{u}\left( 0\right) =0$, $\mathrm{u}%
\left(L_{00}\right) =\mathrm{\bar{u}}\geq 0$, $\mathrm{v}\left( 0\right) =%
\mathrm{v}\left( L_{00}\right) =0$ and $\mathrm{v}^{\prime \prime }\left(
0\right) =\mathrm{v}^{\prime \prime }\left( L_{00}\right) =0$. Since $%
\mathrm{v}^{\prime }$ is piecewise constant, with a discontinuity at the
midspan, integration of the first equation in (\ref{hing-Eq}) yields
\[
\frac{R_x}{EA}=\frac{\mathrm{\bar{u}}}{L_{00}}+\frac{1}{2}\mathrm{v}^{\prime
2}-\frac{1}{L_{00}}\int_{0}^{L_{00}}\alpha \Delta T_{x}dX=\frac{\mathrm{\bar{%
u}}}{L_{00}}+\frac{1}{2}\mathrm{v}^{\prime 2}-\alpha \Delta \bar{T}_{x}
\]%
where $\Delta \bar{T}_{x}$ is the averaged temperature variation. It follows
$\mathrm{u}^{\prime }=\alpha \Delta T_{x}-\alpha \Delta \bar{T}_{x}+\frac{%
\mathrm{\bar{u}}}{L_{00}}$ i.e. $\mathrm{u}^{\prime }=\frac{\mathrm{\bar{u}}%
}{L_{00}}$ when the axial temperature field is constant, even if non-zero.
In this case $\Delta T_{x}=\Delta \bar{T}_{x}=const.$ and by using (\ref%
{etap-def-Navier}), one obtains
\begin{equation}
\eta =\eta _{\varepsilon }+\eta _{\Delta T}=\frac{\mathrm{\bar{u}}}{L_{00}}+%
\sqrt{3}\textrm{ }\alpha \left\vert \Delta T_{x}\right\vert \textrm{
\ \ \ \ and
\ \ \ \ }\eta ^{p}=\left\vert \mathrm{v}^{\prime }\right\vert =2\frac{%
\left\vert \mathrm{v}_{\max }\right\vert }{L_{00}}  \label{eta-expl}
\end{equation}%
which provide a simple interpretation of $\eta $ and\ $\eta ^{p}$ in terms
of temperature difference and of ratios between the maximum displacements
and the beam length. Hence, the strain-rotation domains of Figure \ref%
{Str-rot-Domains-Navier}, which depend on $\eta $ and $\eta ^{p}$, can also
be interpreted using these ratios. For instance, consider the case of zero
temperature field and given $\mathrm{\bar{u}}$ value, such that $\eta =\eta
_{\varepsilon }=\mathrm{\bar{u}/}L_{00}=10^{-8}$: the \emph{strain-rotation
points} corresponding to this situation and for different values of $%
\left\vert \mathrm{v}_{\max }\right\vert $\ are depicted in Figure \ref%
{Str-rot-Domains-Navier}-b: they have the same $y$-value (constant $\eta $)
and different $x$-values. The larger $\left\vert \mathrm{v}_{\max
}\right\vert $, the larger $\eta ^{p}$: then, according to the value of $%
\left\vert \mathrm{v}_{\max }\right\vert $ , the point representing the
structural state may belong to any of the sets $\mathbb{H}_{a},$ $\mathbb{H}%
_{b}$ or $\mathbb{H}_{c}$ and the relevant equilibrium equation is different
in each case. Since the strain-rotation domains depend on $\zeta $, Figure %
\ref{Str-rot-Domains-Navier} refers to the case $\zeta =0.01$. Observe in
addition that $A_{\alpha }:=\frac{1}{\sqrt{3}}\frac{\eta _{\Delta T}}{\eta
_{\varepsilon }}= \frac{\alpha \left\vert \Delta T_{x}\right\vert }{\frac{%
\mathrm{\bar{u}}}{L_{00}}}$ gives an estimate of the relative importance of
the thermal and mechanical strains.

Let us now consider a homogeneous beam with a distributed vertical static
load $p_y$ , a generic temperature field and with $\mathrm{\ddot{u}}=\mathrm{%
\ddot{v}}=p_{x}=q=0$. The same structure has been studied in the numerical
examples of Section 5.3. Eq. (\ref{Navier-strong-moderate}) becomes
\begin{equation}
\left\{
\begin{array}{l}
EA\left( \mathrm{u}^{\prime }-\alpha \Delta T_{x}+\frac{1}{2}\mathrm{v}%
^{\prime 2}\right) =R_{x} \\
EJ\left( \mathrm{v}^{\prime \prime \prime \prime }-\alpha \Delta \gamma
^{\prime \prime }\right) -R_{x}\mathrm{v}^{\prime \prime }=p_y%
\end{array}%
\right.  \label{Ex1strong}
\end{equation}%
For the $X$-direction, the boundary conditions (b.c.) are $\mathrm{u}\left(
0\right) =\mathrm{u}\left( L_{00}\right) =0$ and in the vertical direction
one has $\mathrm{v}\left( 0\right) =\mathrm{v}\left( L_{00}\right) =\mathrm{v%
}^{\prime }\left( 0\right) =\mathrm{v}^{\prime }\left( L_{00}\right) =0$.
Integrating the first equation with the b.c. at $X=0$ and $X=L_{00}$ leads
to
\begin{equation}
\frac{R_{x}}{EA}=-\frac{1}{L_{00}}\int_{0}^{L_{00}}\alpha \Delta T_{x}dX+%
\frac{1}{L_{00}}\int_{0}^{L_{00}}\frac{1}{2}\mathrm{\ v}^{\prime 2}dX
\label{Const}
\end{equation}%
It follows, according to (\ref{Ex1strong}-1)
\begin{equation}
\mathrm{u}^{\prime }-\alpha \Delta T_{x}=\frac{1}{L_{00}}\int_{0}^{L_{00}}%
\frac{1}{2}\mathrm{v}^{\prime 2}dX-\frac{1}{2}\mathrm{v}^{\prime 2}-\alpha
\Delta \bar{T}_{x}  \label{uprimDt}
\end{equation}%
where $\Delta \bar{T}_{x}$ has the same definition as in the previous
example. Assume that $\Delta T_{x}$ and $\Delta \gamma $ are constant and
substitute (\ref{uprimDt}) into the definition (\ref{etap-def-Navier}) of $%
\eta$. Hence
\[
\begin{array}{l}
\eta =\eta _{\varepsilon }+\eta _{\Delta T},\textrm{ \ \ }\eta
_{\varepsilon }=\left( \frac{1}{V_{00}}\int_{0}^{L_{00}}\left(
A\left( \frac{1}{2}\eta
^{2p}-\frac{1}{2}\mathrm{v}^{\prime 2}\right) ^{2}+J\left( \mathrm{v}%
^{\prime \prime }\right) ^{2}\right) \textrm{ }dX\right) ^{\frac{1}{2}} \\
\eta _{\Delta T}=\left( 3\alpha ^{2}\left( \left( \Delta T_{x}\right) ^{2}+%
\frac{J}{A}\left( \Delta \gamma \right) ^{2}\right) \right) ^{\frac{1}{2}}%
\end{array}%
\]%
In order to have a better understanding of the geometrical meaning of $\eta$%
, the solution $\mathrm{v}\left( X\right)$, depending on $p_y$, $\Delta
T_{x} $ and $\Delta \gamma $ should be analytically expressed. However, this
is not a simple task in general. Hence, accounting for the b.c., we assume
here that the deformed shape is approximately co-sinusoidal:
\begin{equation}
\mathrm{v}\left( X\right) =\frac{\mathrm{v}_{\max }}{2}\left[ 1-\cos \left(
\frac{2\pi X}{L_{00}}\right) \right]  \label{vcos}
\end{equation}%
Hence, by using the definitions (\ref{etap-def-Navier}), one obtains
\begin{equation}
\begin{array}{l}
\eta ^{p}=\frac{\pi }{\sqrt{2}}\frac{\left\vert \mathrm{v}_{\max
}\right\vert }{L_{00}}\textrm{ \ , \ \ \ \ \ \ \ \ \ }\eta =\eta
_{\varepsilon
}+\eta _{\Delta T} \\
\eta _{\varepsilon }=\frac{\pi ^{2}}{4\sqrt{2}}\frac{\left\vert \mathrm{v}%
_{\max }\right\vert }{L_{00}}\sqrt{\frac{\mathrm{v}_{\max }^{2}}{L_{00}^{2}}%
+64c\frac{h^{2}}{L_{00}^{2}}},\textrm{ \ \ \ }\eta _{\Delta
T}=\sqrt{3}\alpha
\sqrt{\Delta T_{x}^{2}+c\left( \Delta \gamma h\right) ^{2}}%
\end{array}
\label{etapetaex2}
\end{equation}%
where $c=\frac{J}{Ah^{2}}$, with $h$ the beam width. When the transversal
displacement is different from zero, the ratio between the thermal and
mechanical contributions in $\eta $ is equal to%
\begin{equation}
A_{\alpha }:=\frac{1}{\sqrt{3}}\frac{\eta _{\Delta T}}{\eta _{\varepsilon }}=%
\frac{\alpha }{\frac{\pi ^{2}}{4\sqrt{2}}\frac{\left\vert \mathrm{v}_{\max
}\right\vert }{L_{00}}}\sqrt{\frac{\Delta T_{x}^{2}+c\left( \Delta \gamma
h\right) ^{2}}{\frac{\mathrm{v}_{\max }^{2}}{L_{00}^{2}}+64c\frac{h^{2}}{%
L_{00}^{2}}}}  \label{TempInfl}
\end{equation}%
The interest of Eq. (\ref{etapetaex2}) is that it gives a geometrical
interpretation for the case of beams of the quantities\ $\eta $ and $\eta
^{p}$ defined in (\ref{etap-def}) for the general case and in (\ref%
{etap-def-Navier}) when the Navier-Bernoulli kinematic assumptions are
adopted. They can be easily related to geometrical ratios involving the
maximum deflection, the width and the length of the beam. These geometrical
ratios are known to be important for beam analysis, but they are clearly
related here to the tensorial quantities of a full 3D formulation of the
structural problem. When both maximum deflection and geometry of the beam
are known (or estimated), it is easy to find the corresponding point in
strain-rotation domains and to establish how many terms need to be taken
into account for the computation of the solution.

In order to find the explicit expression of the strain energy, note that $%
\Delta T_{x}$ constant entails from Eq. (\ref{uprimDt})
\begin{equation}
\mathrm{u}^{\prime }=\frac{1}{L_{00}}\int_{0}^{L_{00}}\frac{1}{2}\mathrm{v}%
^{\prime 2}dX-\frac{1}{2}\mathrm{v}^{\prime 2}=\frac{1}{2}\eta ^{2p}-\frac{1%
}{2}\mathrm{v}^{\prime 2}  \label{uprimv}
\end{equation}%
and $\delta \mathrm{u}^{\prime }=\eta ^{p}\delta \eta ^{p}-\mathrm{v}%
^{\prime }\delta \mathrm{v}^{\prime }$. Hence, using (\ref{vcos}) and the
expression of $\eta ^{p}$ given in (\ref{etapetaex2}), the strain energy
contributions (\ref{F-navier-gen}) read%
\[
\begin{array}{l}
\mathcal{F}^{\left( 2\right) }=EAL_{00}\left( \frac{\pi ^{4}}{64}\left(
\frac{\mathrm{v}_{\max }}{L_{00}}\right) ^{4}+\pi ^{4}c\left( \frac{h}{L_{00}%
}\right) ^{2}\left( \frac{\mathrm{v}_{\max }}{L_{00}}\right) ^{2}\right) \\
\mathcal{F}^{\left( 1+2p\right) }=EAL_{00}\left( -\frac{\pi ^{4}}{32}\left(
\frac{\mathrm{v}_{\max }}{L_{00}}\right) ^{4}-\frac{\pi ^{2}}{4}\alpha
\Delta T_{x}\left( \frac{\mathrm{v}_{\max }}{L_{00}}\right) ^{2}\right) \\
\mathcal{F}^{\left( 4p\right) }=EAL_{00}\frac{3\pi ^{4}}{64}\left( \frac{%
\mathrm{v}_{\max }}{L_{00}}\right) ^{4}%
\end{array}%
\]%
The ratio between the energy terms of order $4p$ and $2$ reads%
\[
\frac{\mathcal{F}^{\left( 4p\right) }}{\mathcal{F}^{\left( 2\right) }}=3%
\frac{\left( \frac{\mathrm{v}_{\max }}{h}\right) ^{2}}{64c+\left( \frac{%
\mathrm{v}_{\max }}{h}\right) ^{2}}=\frac{3}{8}\eta _{\varepsilon }^{4p-2}
\]%
$\allowbreak $and it only depends on the ratio $\mathrm{v}_{\max }/h$. The
virtual works of internal and external forces (see (\ref{WiWeWa})) read%
\[
\begin{array}{l}
-\mathcal{W}_{i}\left( \mbox{\boldmath $\Pi$},\mathbf{U,}\delta \mathbf{U}%
\right) =EAL_{00}\left( \frac{\pi ^{4}}{8}\left( \frac{\mathrm{v}_{\max }}{%
L_{00}}\right) ^{3}+2\pi ^{4}c\left( \frac{h}{L_{00}}\right) ^{2}\left(
\frac{\mathrm{v}_{\max }}{L_{00}}\right) -\frac{\pi ^{2}}{2}\alpha \Delta
T_{x}\left( \frac{\mathrm{v}_{\max }}{L_{00}}\right) \right) \frac{\delta
\mathrm{v}_{\max }}{L_{00}} \\
\mathcal{W}_{e}\left( \mathbf{f},\mathbf{g},\delta \mathbf{U}\right)
=\int_{0}^{L_{00}}p_y\frac{\delta \mathrm{v}_{\max }}{2}\left[ 1-\cos \left(
\frac{2\pi X}{L_{00}}\right) \right] dX=\frac{p_yL_{00}}{2}\delta \mathrm{v}%
_{\max }%
\end{array}%
\]%
with $p_y$ supposed constant, and by the virtual work principle, one obtains%
\begin{equation}
\left( 2\pi ^{4}c\left( \frac{h}{L_{00}}\right) ^{2}-\frac{\pi ^{2}}{2}%
\alpha \Delta T_{x}\right) \frac{\mathrm{v}_{\max }}{L_{00}}+\frac{\pi ^{4}}{%
8}\left( \frac{\mathrm{v}_{\max }}{L_{00}}\right) ^{3}=\frac{p_yL_{00}}{2EA}
\label{ex2plv}
\end{equation}%
which provides a simple nonlinear relationship between $\mathrm{v}_{\max }$
and $p_y$. Eq. (\ref{ex2plv}) does not depend on $\Delta \gamma ,$ since
this quantity is assumed constant on the beam length. Using Eq. (\ref{ex2plv}%
) with $\Delta T_{x}=0$ and supposing $h/L_{00}=0.01/0.5=0.02$, $c=1/12,$\ $%
A=1cm^{2}$ and $p_y=Af_{Y0}$ $daN/cm$ like in Section 5.3, one obtains the $%
\mathrm{v}_{\max }$ values collected in the first column of Table \ref%
{TabRes-beam1}. Moreover, by means of (\ref{etapetaex2}), it is easy to
compute $\eta =\eta _{\varepsilon }$ and $\eta ^{p}$. These values are good
estimations of the corresponding quantities computed from the numerical
tests (Table \ref{TabRes-num1}). The difference between $\eta $ estimated as
in (\ref{etapetaex2}) and in the numerical analysis is due to the shear
stress as well as the $\Pi _{YY}$ component. On the other hand, $\eta _{xx}$
issued by the numerical analysis and the analytical estimation of $\eta
=\eta _{\varepsilon }$ according to (\ref{etapetaex2}) are very close. The
corresponding co-ordinates in the strain-rotation domain are reported in the
fifth and the sixth column of Table \ref{TabRes-beam1} (see also Figure \ref%
{Str-rot-Domains-Navier}). The influence of a thermal field with $\alpha
=10^{-5}$ $%
{{}^\circ}%
C^{-1}$, $\Delta T_{x}=20%
{{}^\circ}%
C$ and $h\Delta \gamma =10%
{{}^\circ}%
C$ on the strain can be estimated as follows: Eq. (\ref{ex2plv}) with the
true temperature field is used to compute $\mathrm{v}_{\max }$, then $\eta
_{\varepsilon }$ and $\eta _{\Delta T}$ are separately evaluated, as well as
their ratio $A_{\alpha }$ (see the seventh column of Table \ref{TabRes-beam2}%
). A more useful comparison may be done by the ratio between $\mathrm{v}%
_{\max}$ of this last case and $\mathrm{v}_{\max }$ for the case $\Delta
T_{x}=0$ at the same level of $p_y$ (see the last column of Table \ref%
{TabRes-beam2}). This ratio correctly neglects the influence of $\Delta
\gamma$.

\subsection{Equilibrium around a prestressed configuration: strong form
equations}

By subtracting Eq. (\ref{Navier-strong-moderate}) written at the state $V_1$
from the same equation at the state $V_{0}$, one obtains
\begin{equation}
\left\{
\begin{array}{l}
-\left( EA\left( \mathrm{u}_{01}^{\prime }+\frac{1}{2}\mathrm{v}%
_{01}^{\prime 2}+\mathrm{v}_{01}^{\prime }\mathrm{v}_{0}^{\prime }\right)
\right) ^{\prime }=p_{1,x}-\rho _{00}A\mathrm{\ddot{u}}_{01} \\
\left( EJ\mathrm{v}_{01}^{\prime \prime }\right) ^{\prime \prime }-\left(
\mathrm{v}_{01}^{\prime }A\Pi _{0x}\right) ^{\prime } \\
\textrm{ }-\left( \left( \mathrm{v}_{01}^{\prime
}+\mathrm{v}_{0}^{\prime
}\right) \left[ EA\left( \mathrm{u}_{01}^{\prime }+\frac{1}{2}\mathrm{v}%
_{01}^{\prime 2}+\mathrm{v}_{01}^{\prime }\mathrm{v}_{0}^{\prime }\right) %
\right] \right) ^{\prime }=-q_{1}^{\prime }+p_{1,y}-\rho _{00}A\mathrm{\ddot{%
v}}_{01}%
\end{array}%
\right.  \label{EulmoderDiff}
\end{equation}%
where $\mathrm{u}_{0}$, $\mathrm{v}_{0}$ are the axial and transversal
displacements characterizing the statically prestressed configuration,
measured between $V_{00}$ and $V_{0}$; $T_{0,x}-T_{0i,x}$ is the axial
temperature field; $\Pi _{0,x}=EA\left( \mathrm{u}_{0}^{\prime }-\alpha
\left( T_{0,x}-T_{0i,x}\right) +\frac{1}{2}\mathrm{v}_{0}^{\prime 2}\right) $
is the static axial prestress; the unknowns $\mathrm{u}_{01}=\mathrm{u}_{1}-%
\mathrm{u}_{0}$ , $\mathrm{v} _{01}=\mathrm{v}_{1}-\mathrm{v}_{0}$ are the
displacements between the dynamic configuration $V_{1}$ and the static one.
The first and the second term of the bending equation (\ref{EulmoderDiff}-2)
are linear. The second one is related to the static configuration $V_{0}$ ($%
\mathrm{u}_{0}$, $\mathrm{v}_{0}^{\prime }$ and $T_{0,x}$): it is the effect
of the static prestress due to \emph{external} static loads, \emph{i.e.} it
is not self-equilibrated and is associated with a body configuration $V_{0}$
different from $V_{0i}=V_{00}$. The bending equation in (\ref{EulmoderDiff})
contains a term coupling $\mathrm{u}_{01}^{\prime }$ with the static and the
dynamic rotations$\mathrm{\ v}_{0}^{\prime }$ and $\mathrm{v}_{01}^{\prime }$%
. This term seems to be of paramount importance since experimental
investigations \citep{ft-ICSV2006} prove that pre-bending effect $\mathrm{v}%
_{0}^{\prime }$ play a crucial role, and not only prestress. When $\mathrm{u}%
_{01}^{\prime }$ is different from zero, this term cannot be neglected in
front of the static prestress contribution associated with $\Pi _{0x}$. This
situation occurs when a pulsating axial loading $P_X\left( t\right) $ is
applied at one end of the beam, inducing parametric resonance. In this case,
the first equation in (\ref{EulmoderDiff}) becomes
\[
EA\left( \mathrm{u}_{01}^{\prime }+\frac{1}{2}\mathrm{v}_{01}^{\prime 2}+%
\mathrm{v}_{01}^{\prime }\mathrm{v}_{0}^{\prime }\right) =-P_X\left(
t\right) +\int_{X}^{L_{00}}\rho _{00}A\mathrm{\ddot{u}}_{01}dX
\]%
where $p_{1x}=0$ by assumption. The influence of the axial inertia on the
solution is discussed, e.g., by \citet{Ribeiro2001}.


\section{Moderate rotations and Bernoulli-Navier kinematic assumptions:
viscous damping case}

The stress when linear viscous damping occurs is given by (see (\ref%
{Pind+Pid}), (\ref{StrongDiss1}) and (\ref{StrongDiss2})):
\[
\begin{array}{l}
\mbox{\boldmath$\Pi$}\mathbf{=}\mbox{\boldmath$\Pi$}^{nd}\mathbf{+} %
\mbox{\boldmath$\Pi$}^{d} \quad \quad \mbox{\boldmath$\Pi$}^{nd}=%
\mbox{\boldmath$\Pi$}_{0i}+\mathbf{D:}\left( \mathbf{E-}\sum_{k=1}^{n_{%
\Lambda }}\Lambda _{k}\mathbf{R}_{k}\right) \mathbf{-A}\textrm{
}\left(
T-T_{0i}\right) \\
\mbox{\boldmath$\Pi$}^{d}=\mathbf{F:}\frac{d\mathbf{\tilde{E}}}{dt}\mathbf{%
=F:}\left( \frac{d\mathbf{E}}{dt}-\sum_{k=1}^{n_{\Lambda }}\dot{\Lambda}_{k}%
\mathbf{R}_{k}\mathbf{-}\sum_{k=1}^{n_{\Lambda }}\lambda _{k}\mathbf{R}%
_{k}\right)%
\end{array}
\]
where the Lagrange multipliers can be computed imposing the constraints $%
\mathbf{R}_{k}:\mbox{\boldmath$\Pi$}^{nd}=0$ and\ $\mathbf{R}_{k}: %
\mbox{\boldmath$\Pi$}^{d}=0$ \ \ $k=1,2$ , where $\mathbf{R}_{1}$ and $%
\mathbf{R}_{2}$ are given in (\ref{R1R2}). Since the two parts of the stress
must be separately equal to zero, the Lagrange multipliers $\Lambda
_{1},\Lambda _{2}$ do not change with respect to the non-dissipative case,
and they are given in Eq. (\ref{LagrModer-Eulbeam}). Hence, their time
derivatives can be computed under the assumption, previously discussed, that
the time variations of the temperature field are very small compared to
those of displacements. Hence, $\dot{\Lambda}_{1}=\nu s+\mathrm{v}^{\prime }%
\mathrm{\dot{v}}^{\prime }$\ and $\dot{\Lambda}_{2}=\nu s$, with $s=\mathrm{%
\dot{u}}^{\prime }-Y\mathrm{\dot{v}}^{\prime \prime }+\mathrm{\dot{v}}%
^{\prime }\mathrm{v}^{\prime }$. As a result, one has
\[
\frac{d\mathbf{\tilde{E}}}{dt}=\frac{d\mathbf{\bar{E}}}{dt}\mathbf{-}%
\sum_{k=1}^{n_{\Lambda }}\lambda _{k}\mathbf{R}_{k}\mathbf{=}\left[
\begin{array}{ccc}
s & 0 & 0 \\
0 & -\nu s-\lambda _{1} & 0 \\
0 & 0 & -\nu s-\lambda _{2}%
\end{array}%
\right]
\]%
Let us assume that damping is proportional to the stiffness and to the mass,
\emph{viz}. $\mathbf{F}\mathbf{=}\alpha _{\xi }$ $\rho _{00}\mathbf{I+}\beta
_{\xi }\mathbf{D=}\alpha _{\xi }$ $\rho _{00}\mathbf{I+}\beta _{\xi }\left(
\lambda \mathbf{1}\otimes \mathbf{1+}2\mu \mathbf{I}\right) =\lambda _{d}%
\mathbf{1}\otimes \mathbf{1+}2\mu _{d}$\textbf{\ }$\mathbf{I,}$ with $%
\lambda _{d}=\beta _{\xi }\lambda $ and $2\mu _{d}=\alpha _{\xi }$ $\rho
_{00}+2\mu \beta _{\xi }$. Then, the constraints on the dissipative stress
read
\[
\begin{array}{c}
\left( \mathbf{F:}\frac{d\mathbf{\tilde{E}}}{dt}\right) _{YY}=\alpha _{\xi }%
\textrm{ }\rho _{00}\left( -\nu s-\lambda _{1}\right) +\beta _{\xi }\frac{E}{%
\left( 1+\nu \right) \left( 1-2\nu \right) }\left[ \left( 1-\nu \right)
\left( -\lambda _{1}\right) +\nu \left( -\lambda _{2}\right) \right] =0 \\
\left( \mathbf{F:}\frac{d\mathbf{\tilde{E}}}{dt}\right) _{ZZ}=\alpha _{\xi }%
\textrm{ }\rho _{00}\left( -\nu s-\lambda _{2}\right) +\beta _{\xi }\frac{E}{%
\left( 1+\nu \right) \left( 1-2\nu \right) }\left[ \nu \left( -\lambda
_{1}\right) +\left( 1-\nu \right) \left( -\lambda _{2}\right) \right] =0%
\end{array}%
\]%
It entails $\lambda _{1}=\lambda _{2}=\frac{-\alpha _{\xi }\textrm{
}\rho _{00}\nu }{\alpha _{\xi }\textrm{ }\rho _{00}+\beta _{\xi
}\frac{E}{\left( 1+\nu \right) \left( 1-2\nu \right) }}s$ , $\Pi
_{YY}^{d}=\Pi _{ZZ}^{d}=0$ and $\Pi _{XX}^{d}=\left(
\mathbf{F:}\frac{d\mathbf{\tilde{E}}}{dt}\right) _{XX}=c_{\xi }s$
with
\[
c_{\xi }=\frac{\left( \beta _{\xi }E\right) ^{2}+\alpha _{\xi
}\textrm{ }\rho _{00}\beta _{\xi }E\left( 2-\nu \right) +\left(
1+\nu \right) \left( 1-2\nu \right) \left( \alpha _{\xi }\textrm{
}\rho _{00}\right) ^{2}}{\beta _{\xi }E+\left( 1+\nu \right) \left(
1-2\nu \right) \alpha _{\xi }\textrm{ }\rho _{00}}
\]%
This expression relates the "structural" viscous damping coefficient $c_{\xi
}$ and the "material" parameters $\alpha _{\xi }$,$\beta _{\xi },\rho
_{00},E $ and $\nu $. The virtual work of dissipative forces becomes (see
Eq. (\ref{Wid}))
\[
\mathcal{W}^d_{i}=\mathcal{W}_{i}\left( \mbox{\boldmath$\Pi$}^{d},\mathbf{U,}%
\delta \mathbf{U}\right) =-\int_{V_{00}}\Pi _{XX}^{d}\textrm{
}\left( \delta \mathrm{u}^{\prime }-Y\delta \mathrm{v}^{\prime
\prime }+\mathrm{v}^{\prime }\delta \mathrm{v}^{\prime }\right)
\textrm{ }dV
\]%
and the strong form equations read
\begin{equation}
\left\{
\begin{array}{l}
-\left[ EA\left( \mathrm{u}^{\prime }-\alpha \left( T_{x}-T_{0i,x}\right) +%
\frac{1}{2}\mathrm{v}^{\prime 2}+c_{\xi }\left( \mathrm{\dot{u}}^{\prime }+%
\mathrm{v}^{\prime }\mathrm{\dot{v}}^{\prime }\right) \right) \right]
^{\prime }=p_{x}-\rho _{00}A\mathrm{\ddot{u}} \\
\left( EJ\textrm{ }\left( \mathrm{v}^{\prime \prime }-\alpha \left(
\gamma -\gamma _{0i}\right) +c_{\xi }\mathrm{\dot{v}}^{\prime \prime
}\right)
\right) ^{\prime \prime } \\
\textrm{ \ \ \ \ }-\left( \mathrm{v}^{\prime }\left[ EA\left( \mathrm{u}%
^{\prime }-\alpha \left( T_{x}-T_{0i,x}\right) +\frac{1}{2}\mathrm{v}%
^{\prime 2}+c_{\xi }\left( \mathrm{\dot{u}}^{\prime }+\mathrm{v}^{\prime }%
\mathrm{\dot{v}}^{\prime }\right) \right) \right] \right) ^{\prime
}=-q^{\prime }+p_{y}-\rho _{00}A\mathrm{\ddot{v}}%
\end{array}%
\right.  \label{Nav-damp}
\end{equation}%
When $\mathbf{F}\mathbf{=}\beta _{\xi }\mathbf{D}$, i.e. damping
proportional to the stiffness, one has $\lambda _{1}=\lambda _{2}=0$ and\ $%
c_{\xi }=\beta _{\xi }E$. Moreover, around a static configuration with
prestress $\Pi _{0x}=E\left( \mathrm{u}_{0}^{\prime }-\alpha \left(
T_{0x}-T_{0i,x}\right) +\frac{1}{2}\mathrm{v}_{0}^{\prime 2}\right) $, one
has
\begin{equation}
\left\{
\begin{array}{l}
-\left[ EA\left( \mathrm{u}_{01}^{\prime }+\frac{1}{2}\mathrm{v}%
_{01}^{\prime 2}+\mathrm{v}_{0}^{\prime }\mathrm{v}_{01}^{\prime }+c_{\xi
}\left( \mathrm{\dot{u}}_{01}^{\prime }+\left( \mathrm{v}_{0}^{\prime }+%
\mathrm{v}_{01}^{\prime }\right) \mathrm{\dot{v}}_{01}^{\prime }\right)
\right) \right] ^{\prime }=p_{1x}-\rho _{00}A\mathrm{\ddot{u}}_{01} \\
\left( EJ\textrm{ }\left( \mathrm{v}_{01}^{\prime \prime }+c_{\xi }\mathrm{%
\dot{v}}_{01}^{\prime \prime }\right) \right) ^{\prime \prime }-\left(
\mathrm{v}_{01}^{\prime }A\Pi _{0x}\right) ^{\prime } \\
-\left( \left( \mathrm{v}_{01}^{\prime }+\mathrm{v}_{0}^{\prime }\right) %
\left[ EA\left( \mathrm{u}_{01}^{\prime }+\frac{1}{2}\mathrm{v}_{01}^{\prime
2}+\mathrm{v}_{0}^{\prime }\mathrm{v}_{01}^{\prime }+c_{\xi }\left( \mathrm{%
\dot{u}}_{01}^{\prime }+\left( \mathrm{v}_{0}^{\prime }+\mathrm{v}%
_{01}^{\prime }\right) \mathrm{\dot{v}}_{01}^{\prime }\right) \right) \right]
\right) ^{\prime } \\
\textrm{ \ \ \ \ \ \ \ \ \ \ \ \ \ \ }=-q_{1}^{\prime }+p_{1y}-\rho _{00}A%
\mathrm{\ddot{v}}_{01}%
\end{array}%
\right.  \label{Nav-Damp01}
\end{equation}

\section{Conclusions}

Based on a Hu-Washizu functional accounting for stress constraints, an
original derivation of the dynamic equations for thin prestressed and
prestrained structures from continuum mechanics has been presented. The
assumption of small strains and moderate rotations, as well as the physical
linearization, have been formalized by the notion of strain-rotation domains
associated with new strain and rotation global measures. The corresponding
approximated expressions for the strain energy have been justified by 2D
non-linear finite element investigations. The presence of a generic
mechanical or thermal prestress and physically linear viscous damping has
been discussed in detail. In particular, damping is introduced by the
original definition of a pseudo-potential with stress constraints. Coupled
beam equations governing traction and bending for small strains and moderate
rotations have been derived by using the proposed general procedure.

\section{Appendix: Lagrange multipliers}

In this Section, we show how to compute the Lagrange multipliers. The stress
constraints write
\begin{equation}
\mathbf{R}_{k}\mathbf{:\Pi }=\mathbf{R}_{k}\mathbf{:}\left( %
\mbox{\boldmath$\Pi$}_{0i}+\mathbf{D:}\left( \mathbf{E}\left( \mathbf{U}%
\right) -\sum_{l=1}^{n_{\Lambda }}\Lambda _{l}\mathbf{R}_{l}\right) -\mathbf{%
A}\Delta T\right) \mathbf{=}0  \label{RkPi1}
\end{equation}%
with $k=1,n_{\Lambda }$. Eq. (\ref{RkPi1}) is equivalent to
\[
\sum_{k=1}^{n_{\Lambda }}\mathbf{R}_{k}\mathbf{:D:R}_{l}\textrm{ }\Lambda _{l}=%
\mathbf{R}_{k}\mathbf{:}\left( \mbox{\boldmath$\Pi$}_{0i}+\mathbf{D:E}\left(
\mathbf{U}\right) -\mathbf{A}\Delta T\right)
\]%
or $\mathbf{K\cdot \Lambda }=\mathbf{B,}$ where $\mbox{\boldmath$\Lambda$}%
\mathbf{=}\left( \Lambda _{i}\right) _{i=1,n\Lambda }$, $\mathbf{K=}\left(
k_{ij}\right) _{i,j=1,n_{\Lambda }}$, $\mathbf{B=}\left( B_{i}\right)
_{i=1,n\Lambda }$ and
\[
\begin{array}{cc}
k_{ij}=\mathbf{R}_{i}\mathbf{:D:R}_{j} \ , \ \  & B_{i}=\mathbf{R}_{i}%
\mathbf{:}\left( \mbox{\boldmath$\Pi$}_{0i}+\mathbf{D:E}\left(\mathbf{U}%
\right) -\mathbf{A}\Delta T\right)%
\end{array}%
\]%
By virtue of the symmetry and positivity of $\mathbf{D}$ it is easy to prove
that $\mathbf{K}$ is symmetric positive definite and therefore invertible,
provided that tensors $\mathbf{R}_{k}$ are linearly independent. Hence, the
Lagrange multipliers read $\mathbf{\Lambda =K}^{-1}\cdot \mathbf{B}$.
Then, the assumptions (\ref{etap-def}), (\ref{Ebar-orders}) and (\ref%
{T-P0i-order}) prove that $\mathbf{\Lambda =\Lambda }^{\left( 1\right) }+%
\mathbf{\Lambda }^{\left( 2p\right) }+\mathbf{\Lambda }^{\left( 1+p\right) }+%
\mathbf{\Lambda }^{\left( 2\right) }.$ In particular, the constraints (\ref%
{R1R2}) leads to
\[
\mathbf{K=}\left(%
\begin{array}{cc}
\lambda +2\mu & \lambda \\
\lambda & \lambda +2\mu%
\end{array}
\right)
\]
and
\[
\left[
\begin{array}{c}
\Lambda _{1} \\
\Lambda _{2}%
\end{array}%
\right] \mathbf{=}\frac{1}{\left( \lambda +2\mu \right) ^{2}-\lambda ^{2}}%
\left(
\begin{array}{cc}
\lambda +2\mu & -\lambda \\
-\lambda & \lambda +2\mu%
\end{array}%
\right) \cdot \left[
\begin{array}{c}
\left( \mbox{\boldmath$\Pi$}_{0i}+\mathbf{D:E}\left( \mathbf{U}\right) -%
\mathbf{A}\Delta T\right) _{YY} \\
\left( \mbox{\boldmath$\Pi$}_{0i}+\mathbf{D:E}\left( \mathbf{U}\right) -%
\mathbf{A}\Delta T\right) _{ZZ}%
\end{array}%
\right]
\]

For the dissipative stress, one has analogous constraints
\[
\mathbf{R}_{k}\mathbf{:\Pi }^{d}=\mathbf{R}_{k}\mathbf{:F:}\left( \frac{d%
\mathbf{E}\left( \mathbf{U}\right) }{dt}-\sum_{l=1}^{n_{\Lambda }}\frac{%
d\Lambda _{l}}{dt}\mathbf{R}_{l}-\sum_{l=1}^{n_{\Lambda }}\lambda _{l}%
\mathbf{R}_{l}\right) \mathbf{=}0
\]%
Then $\mbox{\boldmath$\lambda$}=\mathbf{\bar{K}} ^{-1}\cdot \mathbf{\bar{B}}=%
\mbox{\boldmath$\lambda$}^{\left( 1\right) }+\mbox{\boldmath$\lambda$}%
^{\left( 2p\right) }+\mbox{\boldmath$\lambda$}^{\left( 1+p\right) }+%
\mbox{\boldmath$\lambda$}^{\left( 2\right) },$ where $\mbox{\boldmath$%
\lambda$}=\left( \lambda _{i}\right) _{i=1,n\Lambda }$ and%
\[
\begin{array}{l}
\mathbf{\bar{K}=}\left( \bar{k}_{ij}\right) _{i,j=1,n_{\Lambda
}}\textrm{ ,\ \
\ \ \ }\bar{k}_{ij}=\mathbf{R}_{i}\mathbf{:F:R}_{j}\textrm{\textbf{\ }} \\
\mathbf{\bar{B}=}\left( \bar{B}_{i}\right) _{i=1,n\Lambda }\textrm{
,\ \ \ \ \ \ }\bar{B}_{i}=\mathbf{R}_{i}\mathbf{:F:}\left(
\frac{d\mathbf{E}\left(
\mathbf{U}\right) }{dt}-\sum_{l=1}^{n_{\Lambda }}\frac{d\Lambda _{l}}{dt}%
\mathbf{R}_{l}\right)%
\end{array}%
\]%
$\mathbf{\bar{K}}$ is symmetric positive definite. For a beam
\[
\bar{\mathbf{K}}=\left(%
\begin{array}{cc}
\lambda _{d}+2\mu _{d} & \lambda _{d} \\
\lambda _{d} & \lambda _{d}+2\mu _{d}%
\end{array}
\right)
\]
and
\[
\left[
\begin{array}{c}
\lambda _{1} \\
\lambda _{2}%
\end{array}%
\right] \mathbf{=}\frac{1}{\left( \lambda _{d}+2\mu _{d}\right) ^{2}-\lambda
_{d}^{2}}\left(
\begin{array}{cc}
\lambda _{d}+2\mu _{d} & -\lambda _{d} \\
-\lambda _{d} & \lambda _{d}+2\mu _{d}%
\end{array}%
\right) \cdot \left[
\begin{array}{c}
\left( \mathbf{F:}\left( \frac{d\mathbf{E}\left( \mathbf{U}\right) }{dt}%
-\sum_{l=1}^{n_{\Lambda }}\frac{d\Lambda _{l}}{dt}\mathbf{R}_{l}\right)
\right) _{YY} \\
\left( \mathbf{F:}\left( \frac{d\mathbf{E}\left( \mathbf{U}\right) }{dt}%
-\sum_{l=1}^{n_{\Lambda }}\frac{d\Lambda _{l}}{dt}\mathbf{R}_{l}\right)
\right) _{ZZ}%
\end{array}%
\right]
\]


\clearpage

\clearpage

\clearpage

\clearpage

\clearpage

\clearpage
\begin{figure}[tbp]
\begin{center}
\includegraphics[width=13cm]{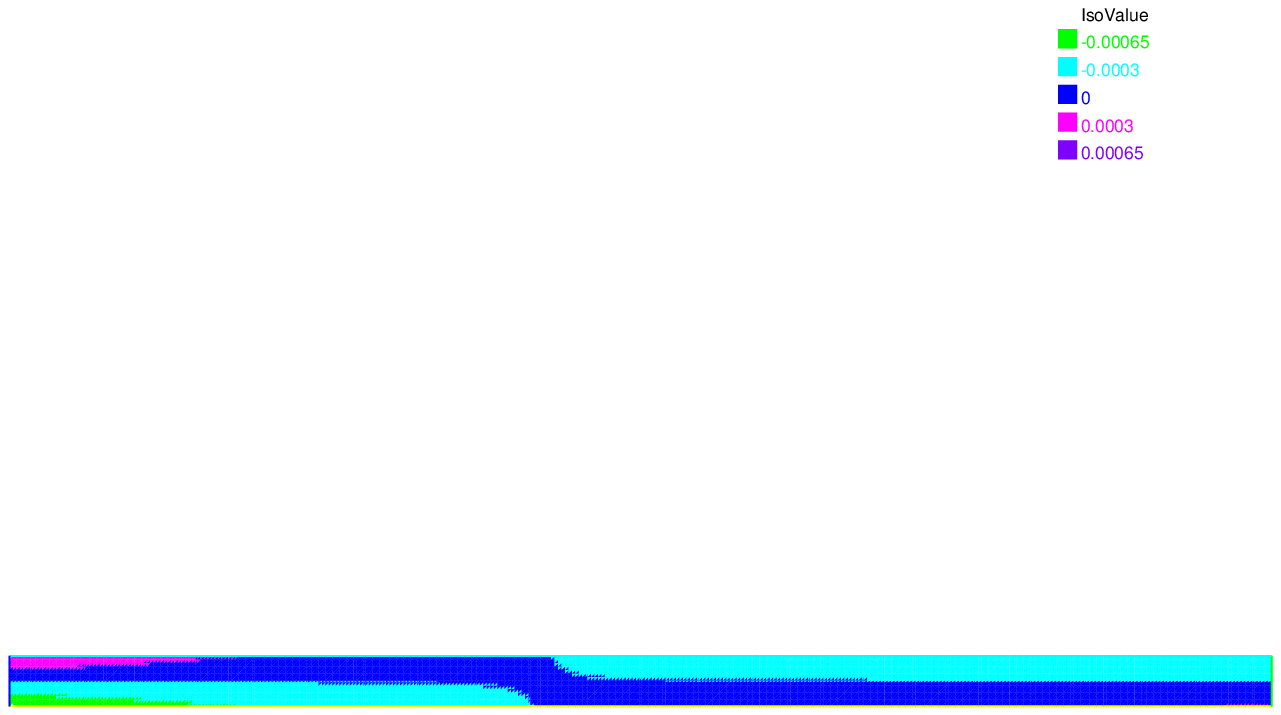}
\end{center}
\caption{Volume vertical force $f_{Y0}=-1$ $daN/cm^{3}$. Map of the relative
difference between the exact and the approximated strain energy density,
\emph{viz.} $\frac{\Psi \left( \mathbf{E}\left( \mathbf{U}\right) \right)
-\Psi \left( \mathbf{E}^{\left( b\right) }\left( \mathbf{U}\right) \right) }{%
\Psi \left( \mathbf{E}\left( \mathbf{U}\right) \right) }$.}
\label{Fig-DiffEnergy}
\end{figure}

\clearpage
\begin{figure}[tbp]
\begin{center}
\includegraphics[width=13cm]{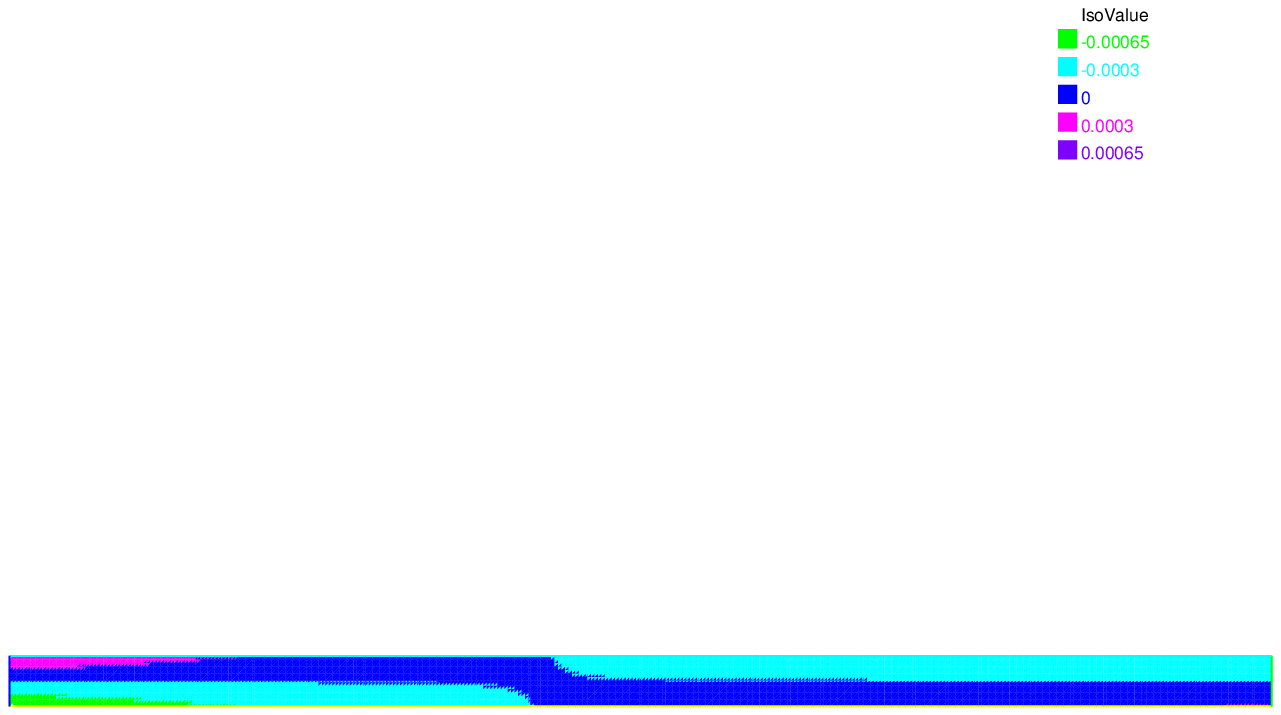}
\end{center}
\caption{Surface vertical force $g_{Y0}=-1$ $daN/cm^{2}$. Map of the
relative difference between the exact and the approximated strain energy
density, \emph{viz.} $\frac{\Psi \left( \mathbf{E}\left( \mathbf{U}\right)
\right) -\Psi \left( \mathbf{E}^{\left( b\right) }\left( \mathbf{U}\right)
\right) }{\Psi \left( \mathbf{E}\left( \mathbf{U}\right) \right) }$.}
\label{Fig-Diff-Energ-surf1}
\end{figure}

\clearpage

\clearpage



\clearpage
\begin{table}[tbp]
\begin{tabular}{ccccc}
\hline
& $V_{00}$ & $V_{0i}$ & $V_{0}$ & $V_{1}$ \\ \hline
$\mathbf{U}$ & $\mathbf{0}$ & $\mathbf{0}$ & $\mathbf{U}_{0}$ & $\mathbf{U}%
_{1}=\mathbf{U}_{0}+\mathbf{U}_{01}$ \\
$\mathbf{E}$ & $-$ & $\mathbf{0}$ & $\mathbf{E}\left( \mathbf{U}_{0}\right) $
& $\mathbf{E}\left( \mathbf{U}_{1}\right) $ \\
$\delta \mathbf{E}$ & $-$ & $\delta \mathbf{E}\left( \mathbf{0}\mathbf{,}%
\delta \mathbf{U}\right) =\delta \mbox{\boldmath $\varepsilon$ }$ & $\delta
\mathbf{E}\left( \mathbf{U}_{0}\mathbf{,}\delta \mathbf{U}\right) $ & $%
\delta \mathbf{E}\left( \mathbf{U}_{1}\mathbf{,}\delta \mathbf{U}\right) $
\\
$T$ & $-$ & $T_{0i}$ & $T_{0}$ & $T_{0}$ \\
$\mbox{\boldmath$\Pi$}$ & $-$ & $\mbox{\boldmath$\Pi$}_{0i}$ & ${\scriptsize
{\
\begin{array}{l}
\mbox{\boldmath$\Pi$}_{0}=\mbox{\boldmath$\Pi$}_{0i}+\mathbf{D:E}\left(
\mathbf{U}_{0}\right) \\
\ \ \ \ \ \ \ \ -\mathbf{A}\left( T_{0}-T_{0i}\right)%
\end{array}%
}}$ & {\scriptsize {$%
\begin{array}{l}
\mbox{\boldmath$\Pi$}_{0i}+\mathbf{D:}\mathbf{E}\left( \mathbf{U}_{1}\right)
\\
\ \ \ -\mathbf{A}\left( T_{0}-T_{0i}\right)%
\end{array}%
$}} \\
$\mathbf{f}$ & $-$ & $\mathbf{0}$ & $\mathbf{f}_{0}$ & $\mathbf{f}_{0}+%
\mathbf{f}_{1}(t)$ \\
$\mathbf{g}$ & $-$ & $\mathbf{0}$ & $\mathbf{g}_{0}$ & $\mathbf{g}_{0}+%
\mathbf{g}_{1}(t)$ \\
$\mathbf{\bar{U}}$ & $\mathbf{0}$ & $\mathbf{0}$ & $\mathbf{\bar{U}}_{0}$ & $%
\mathbf{\bar{U}}_{0}+\mathbf{\bar{U}}_{1}(t)$ \\
$\mathbf{\ddot{U}}$ & $-$ & $\mathbf{0}$ & $\mathbf{0}$ & $\mathbf{\ \ \ddot{%
U}}_{1}=\mathbf{\ddot{U}}_{01}$%
\end{tabular}%
\caption{Displacements, strains, temperatures and stresses associated to
each structural configuration. $\mathbf{U}_{0i}=\mathbf{\bar{U}}_{0i}=%
\mathbf{0}$ in the analyses of this paper, since no geometric imperfection
is considered. }
\label{TabConf}
\end{table}

\clearpage
\begin{table}[tbp]
\begin{tabular}{|c||cccc|}
\hline
$\left(\mbox{\boldmath $\Pi$}\right) \diagdown \left( \mathbf{E}\right)$ & $%
\mbox{\boldmath $\varepsilon$}$ & \multicolumn{1}{|c}{$\mathbf{\ r}^{T}\cdot
\mathbf{r} $} & \multicolumn{1}{|c}{$%
\begin{array}{c}
\left( \mbox{\boldmath $\varepsilon$}^{T}\ \cdot \mathbf{r}\right) \\
\left( \mathbf{r}^{T}\cdot \mbox{\boldmath $\varepsilon$}\right)%
\end{array}
$} & \multicolumn{1}{|c|}{$\mbox{\boldmath
$\varepsilon$}^{T}\cdot \mbox{\boldmath $\varepsilon$}$} \\ \hline\hline
$\mbox{\boldmath $\varepsilon$},\alpha \Delta T$ & $2$ & $1+2p$ &
\multicolumn{1}{|c}{$2+p$} & \multicolumn{1}{|c|}{$3$} \\ \cline{1-1}
$\mathbf{r}^{T}\cdot \mathbf{r}$ & $1+2p$ & $4p$ & \multicolumn{1}{|c}{$1+3p$%
} & \multicolumn{1}{|c|}{$2+2p$} \\ \cline{1-3}
$%
\begin{array}{c}
\mbox{\boldmath $\varepsilon$}^{T}\cdot \mathbf{r} \\
\mathbf{r}^{T}\cdot \mbox{\boldmath $\varepsilon$}%
\end{array}
$ & $2+p$ & $1+3p$ & $2+2p$ & \multicolumn{1}{|c|}{$3+p$} \\ \cline{1-4}
$\mbox{\boldmath $\varepsilon$}^{T}\cdot \mbox{\boldmath
$\varepsilon$}$ & $3$ & $2+2p$ & $3+p $ & $4$ \\ \hline\hline
$\mbox{\boldmath $\varepsilon$}\cdot \mbox{\boldmath
$\varepsilon$},\left( \alpha \Delta T\right) ^{2}$ & $3$ & $2+2p$ & $3+p$ & $%
4$ \\ \hline
$\left( \mathbf{r}^{T}\cdot \mathbf{r}\right) \cdot \left( \mathbf{r}%
^{T}\cdot \mathbf{r}\right) $ & $1+4p$ & $6p $ & $1+5p$ & $2+4p$ \\ \hline
$%
\begin{array}{c}
\left( \mbox{\boldmath $\varepsilon$}^{T}\cdot \mathbf{r}\right) \cdot
\left( \mbox{\boldmath $\varepsilon$}^{T}\cdot \mathbf{r}\right) \\
\left( \mathbf{r}^{T}\cdot \mbox{\boldmath $\varepsilon$}\right) \cdot
\left( \mathbf{r}^{T}\cdot \mbox{\boldmath $\varepsilon$}\right)%
\end{array}
$ & $3+2p$ & $2+4p$ & $3+3p$ & $4+2p$ \\ \hline
$\left( \mbox{\boldmath $\varepsilon$}^{T}\cdot
\mbox{\boldmath
$\varepsilon$}\right) \cdot \left( \mbox{\boldmath
$\varepsilon$}^{T}\cdot \mbox{\boldmath $\varepsilon$}\right) $ & $5$ & $%
4+2p $ & $5+p$ & $6$ \\ \hline
$\left( \mbox{\boldmath
$\varepsilon$}\right) \cdot \left( \mathbf{r}^{T}\cdot \mathbf{r}\right)
,\alpha \Delta T\left( \mathbf{r}^{T}\cdot \mathbf{r}\right) $ & $2+2p$ & $%
1+4p$ & $2+3p$ & $3+2p$ \\ \hline
$%
\begin{array}{c}
\mbox{\boldmath $\varepsilon$}\cdot \left( \mbox{\boldmath
$\varepsilon$} ^{T}\cdot \mathbf{r}\right) ,\alpha \Delta T\left(
\mbox{\boldmath
$\varepsilon$} ^{T}\cdot \mathbf{r}\right) \\
\mbox{\boldmath $\varepsilon$}\cdot \left( \mathbf{r}^{T}\cdot %
\mbox{\boldmath $\varepsilon$}\right) ,\alpha \Delta T\left( \mathbf{r}%
^{T}\cdot \mbox{\boldmath
$\varepsilon$}\right)%
\end{array}%
$ & $3+p$ & $2+3p$ & $3+2p$ & $4+p$ \\ \hline
$\mbox{\boldmath
$\varepsilon$}\cdot \mbox{\boldmath $\varepsilon$}^{T}\cdot
\mbox{\boldmath
$\varepsilon$},\alpha \Delta T$ $\mbox{\boldmath
$\varepsilon$}^{T}\cdot \mbox{\boldmath $\varepsilon$}$ & $4$ & $3+2p$ & $%
4+p $ & $5$ \\ \hline
$%
\begin{array}{c}
\left( \mathbf{r}^{T}\cdot \mathbf{r}\right) \cdot \left(
\mbox{\boldmath
$\varepsilon$} ^{T}\cdot \mathbf{r}\right) \\
\left( \mathbf{r}^{T}\cdot \mathbf{r}\right) \cdot \left( \mathbf{r}%
^{T}\cdot \mbox{\boldmath $\varepsilon$}\right)%
\end{array}
$ & $2+3p$ & $1+5p$ & $2+4p$ & $3+3p$ \\ \hline
$\left( \mathbf{r}^{T}\cdot \mathbf{r}\right) \cdot \left(
\mbox{\boldmath
$\varepsilon$}^{T}\cdot \mbox{\boldmath $\varepsilon$}\right) $ & $3+2p$ & $%
2+4p$ & $3+3p$ & $4+2p$ \\ \hline
$%
\begin{array}{c}
\left( \mbox{\boldmath $\varepsilon$}^{T}\cdot \mathbf{r}\right) \cdot
\left( \mbox{\boldmath
$\varepsilon$}^{T}\cdot \mbox{\boldmath $\varepsilon$}\right) \\
\left( \mathbf{r}^{T}\cdot \mbox{\boldmath $\varepsilon$}\right) \cdot
\left( \mbox{\boldmath $\varepsilon$}^{T}\cdot
\mbox{\boldmath
$\varepsilon$}\right)%
\end{array}
$ & $4+p$ & $3+3p$ & $4+2p$ & $5+p$ \\ \hline
\end{tabular}%
\caption{Orders of the terms appearing in the strain energy expressions (%
\protect\ref{strain-energy-complete})-(\protect\ref{StrainEnergy-eta-p}) and
(\protect\ref{StrainEnergy-eta-p-Nlin}-2), representing the physically
linear and non-linear case, respectively.}
\label{TabOrders}
\end{table}

\clearpage
\begin{table}[tbp]
\begin{tabular}{ccccccccc}
\hline
$\left\vert f_{Y,0}\right\vert $ & $\left\vert \mathrm{v}_{\max }\right\vert
$ & $\max \left\vert \varepsilon _{xx}\right\vert $ & $\eta _{xx}$ & $\eta $
& $\eta ^{p}$ & $x$ & $y$ & $p$ \\
$\lbrack daN/cm^{3}]$ & $[cm]$ &  &  &  &  &  &  &  \\ \hline
$0.1$ & $0.0092$ & $6.46e-5$ & $1.52e-5$ & $1.59e-5$ & $4.05e-4$ & $-3.4$ & $%
-4.8$ & $0.71$ \\
$0.5$ & $0.0459$ & $3.20e-4$ & $7.58e-5$ & $7.93e-5$ & $2.02e-3$ & $-2.7$ & $%
-4.1$ & $0.66$ \\
$1$ & $0.0914$ & $6.34e-4$ & $1.51e-4$ & $1.58e-4$ & $4.03e-3$ & $-2.39$ & $%
-3.8$ & $0.63$ \\
$3$ & $0.263$ & $1.85e-3$ & $4.37e-4$ & $4.66e-4$ & $1.157e-2$ & $-1.94$ & $%
-3.33$ & $0.58$ \\
$6$ & $0.4745$ & $3.58e-3$ & $8.03e-4$ & $8.96e-4$ & $2.09e-2$ & $-1.68$ & $%
-3.05$ & $0.55$ \\ \hline
\end{tabular}%
\caption{Clamped-clamped beam (2D plain stress analysis). Numerical results
for different values of the vertical volume load $f_{Y0}$.}
\label{TabRes-num1}
\end{table}

\clearpage
\begin{table}[tbp]
\begin{tabular}{cccccc}
\hline
$\left\vert f_{Y,0}\right\vert $ & $\rho _{1}=\eta ^{2p-1}$ & $\rho
_{2}=\eta ^{4p-2}$ & $\mathcal{F}$ & $\mathcal{F}^{\left( b\right) }$ & $%
\frac{\mathcal{F}-\mathcal{F}^{\left( b\right) }}{\mathcal{F}}$ \\
$\lbrack daN/cm^{3}]$ &  &  & $[daN$ $cm]$ & $[daN$ $cm]$ &  \\ \hline
$0.1$ & $0.0103$ & $0.00011$ & $0.00613653$ & $0.00613653$ & $\sim 0$ \\
$0.5$ & $0.0515$ & $0.00265$ & $0.153092$ & $0.153091$ & $6.5e-6$ \\
$1$ & $0.1025$ & $0.01052$ & $0.608433$ & $0.60842$ & $2.1e-5$ \\
$3$ & $0.2873$ & $0.0825$ & $5.15335$ & $5.15244$ & $1.8e-4$ \\
$6$ & $0.4875$ & $0.238$ & $17.8727$ & $17.8626$ & $5.7e-4$ \\ \hline
\end{tabular}%
\caption{Clamped-clamped beam (2D plain stress analysis). Energies for
different values of the vertical volume load $f_{Y0}$.}
\label{TabRes-num2}
\end{table}

\clearpage
\begin{table}[tbp]
\begin{tabular}{ccccccccc}
\hline
$\left\vert g_{Y,0}\right\vert $ & $\left\vert \mathrm{v}_{\max }\right\vert
$ & $\max \left\vert \varepsilon _{xx}\right\vert $ & $\eta _{xx}$ & $\eta $
& $\eta ^{p}$ & $x$ & $y$ & $p$ \\
$\lbrack daN/cm^{2}]$ & $[cm]$ &  &  &  &  &  &  &  \\ \hline
$0.1$ & $0.0092$ & $6.46e-5$ & $1.52e-5$ & $1.59e-5$ & $4.052e-4$ & $-3.39$
& $-4.80$ & $0.71$ \\
$0.5$ & $0.0459$ & $3.20e-4$ & $7.58e-5$ & $7.93e-5$ & $2.02e-3$ & $-2.69$ &
$-4.10$ & $0.66$ \\
$1$ & $0.09142$ & $6.34e-4$ & $1.51e-4$ & $1.58e-4$ & $4.029e-3$ & $-2.39$ &
$-3.80$ & $0.63$ \\
$3$ & $0.263$ & $1.85e-3$ & $4.37e-4$ & $4.67e-4$ & $1.158e-2$ & $-1.94$ & $%
-3.33$ & $0.58$ \\
$6$ & $0.475$ & $3.58e-3$ & $8.04e-4$ & $8.98e-4$ & $2.092e-2$ & $-1.68$ & $%
-3.05$ & $0.55$ \\ \hline
\end{tabular}%
\caption{Clamped-clamped beam (2D plain stress analysis). Numerical results
for different values of the vertical surface load $g_{Y0}$.}
\label{TabRes-num3}
\end{table}

\clearpage
\begin{table}[tbp]
\begin{tabular}{cccccc}
\hline
$\left\vert g_{Y,0}\right\vert $ & $\rho _{1}=\eta ^{2p-1}$ & $\rho
_{2}=\eta ^{4p-2}$ & $\mathcal{F}$ & $\mathcal{F}^{\left( b\right) }$ & $%
\frac{\mathcal{F-F}^{\left( b\right) }}{\mathcal{F}}$ \\
$\lbrack daN/cm^{2}]$ &  &  & $[daN$ $cm]$ & $[daN$ $cm]$ &  \\ \hline
$0.1$ & $0.01035$ & $0.00011$ & $0.0061368$ & $0.0061368$ & $\sim 0$ \\
$0.5$ & $0.0515$ & $0.00265$ & $0.153124$ & $0.153123$ & $\allowbreak
6.\,\allowbreak 53e-6$ \\
$1$ & $0.1025$ & $0.0105$ & $0.608689$ & $0.608676$ & $\allowbreak
2.\,\allowbreak 14e-5$ \\
$3$ & $0.2870$ & $0.0824$ & $5.15922$ & $5.15831$ & $1.\,\allowbreak 76e-4$
\\
$6$ & $0.4876$ & $0.2377$ & $17.9055$ & $17.8953$ & $\allowbreak
5.\,\allowbreak 70e-4$ \\ \hline
\end{tabular}%
\caption{Clamped-clamped beam (2D plain stress analysis). Energies for
different values of the vertical surface load $g_{Y0}$.}
\label{TabRes-num4}
\end{table}

\clearpage
\begin{table}[tbp]
\begin{tabular}{ccccccc}
\hline
$\left\vert p_y\right\vert $ & $\left\vert \mathrm{v}_{\max ,\Delta
T=0}\right\vert $ & $\eta $ & $\eta ^{p}$ & $x$ & $y$ & $p$ \\
$\lbrack daN/cm]$ & $[cm]$ &  &  &  &  &  \\ \hline
$0.1$ & $0.009165$ & $1.\,\allowbreak 477\,e-5$ & $4.\,\allowbreak 072e-4$ &
$\allowbreak -3.\,\allowbreak 39$ & $-4.\,\allowbreak 83$ & $\allowbreak
0.702$ \\
$0.5$ & $0.04576$ & $7.\,\allowbreak 377\,e-5$ & $2.\,\allowbreak 033\,e-3$
& $-2.\,\allowbreak 69$ & $-4.\,\allowbreak 13$ & $0.651\,$ \\
$1$ & $0.0911$ & $1.\,\allowbreak 469\,e-4$ & $4.\,\allowbreak 047\,e-3$ & $%
-2.\,\allowbreak 39$ & $-3.\,\allowbreak 83$ & $0.624$ \\
$3$ & $0.2616$ & $4.\,\allowbreak 243\,e-4$ & $1.\,\allowbreak 162\,e-2$ & $%
-1.\,\allowbreak 935$ & $\allowbreak -3.\,\allowbreak 37$ & $0.574$ \\
$6$ & $0.4714$ & $7.\,\allowbreak 754\,e-4$ & $2.\,\allowbreak 094\,e-2$ & $%
-1.\,\allowbreak 68$ & $-3.\,\allowbreak 11$ & $0.54$ \\ \hline
\end{tabular}%
\caption{Clamped-clamped beam with co-sinusoidal deformed shape. Numerical
results for different values of the vertical load $p_{y}$.}
\label{TabRes-beam1}
\end{table}

\clearpage
\begin{table}[tbp]
\begin{tabular}{cccccccc}
\hline
$\left\vert p_y\right\vert $ & $\left\vert \mathrm{v}_{\max }\right\vert $ &
$\eta _{\varepsilon }$ & $\eta _{\Delta T}$ & $\eta $ & $\eta ^{p}$ & $%
A_{\alpha }$ & $\frac{\left\vert \mathrm{v}_{\max }\right\vert }{\left\vert
\mathrm{v}_{\max ,\Delta T=0}\right\vert }$ \\
$\lbrack daN/cm]$ & $[cm]$ &  &  &  &  &  &  \\ \hline
$0.1$ & $0.010808$ & $1.\,\allowbreak 742e-5$ & $6.\,\allowbreak 083e-4$ & $%
6.\,\allowbreak 26\,0e-4$ & $4.\,\allowbreak 802e-4$ & $34.\,\allowbreak 9$
& $\allowbreak 1.\,\allowbreak 180$ \\
$0.5$ & $0.053906$ & $8.\,\allowbreak 690\,e-5$ & $6.\,\allowbreak 083e-4$ &
$6.\,\allowbreak 952e-4$ & $2.\,\allowbreak 395\,e-3$ & $7.0$ & $%
1.\,\allowbreak 178$ \\
$1$ & $0.1070$ & $1.\,\allowbreak 726\,e-4$ & $6.\,\allowbreak 083e-4$ & $%
7.\,\allowbreak 809\,e-4$ & $4.\,\allowbreak 754e-3$ & $3.\,\allowbreak 52\,$
& $1.\,\allowbreak 174$ \\
$3$ & $0.3003$ & $4.\,\allowbreak 8801e-4$ & $6.\,\allowbreak 083e-4$ & $%
1.\,\allowbreak 096\,e-3$ & $1.\,\allowbreak 334e-2$ & $1.\,\allowbreak 25$
& $1.\,\allowbreak 148$ \\
$6$ & $0.52242$ & $8.\,\allowbreak 632\,e-4$ & $6.\,\allowbreak 083e-4$ & $%
1.\,\allowbreak 472e-3$ & $2.\,\allowbreak 321\,e-2$ & $0.705$ & $%
1.\,\allowbreak 108$ \\ \hline
\end{tabular}%
\caption{Clamped-clamped beam with co-sinusoidal deformed shape. The
vertical load is the same as in Table \protect\ref{TabRes-num1}. Moreover, a
thermal field with $\Delta T_{x}=20
{{}^\circ}%
C$ and $h\Delta \protect\gamma =10%
{{}^\circ}%
C$ has been added.}
\label{TabRes-beam2}
\end{table}

\end{document}